\documentclass[useAMS,usenatbib]{mn2e}

\title[Simulations of tidally induced structure in M51]
{Simulations of the grand design galaxy M51: a case study for analysing tidally induced spiral structure}
\author[C. L. Dobbs, C. Theis, J. E Pringle, M. R. Bate]
{C. L. Dobbs\thanks{E-mail:
dobbs@astro.ex.ac.uk}$^1$, C. Theis$^2$, J. E Pringle$^3$ \& M. R. Bate$^1$\\
$^1$ School of Physics, University of Exeter, 
Stocker Road, Exeter, EX4 4QL \\
$^2$ Institute of Astronomy, University of Vienna, T\"urkenschanzstr.\ 17, 1180, Austria  \\
$^3$ Institute of Astronomy, Madingley Road, Cambridge, CB3 0HA \\}
\usepackage{amssymb}
\usepackage{amsmath}
\usepackage{graphicx}
\usepackage{multirow}
\usepackage{epsfig}

\begin{document}
\date{\today}

\pagerange{\pageref{firstpage}--\pageref{lastpage}} \pubyear{0000}

\maketitle

\label{firstpage}
 \begin{abstract}
   We present hydrodynamical models of the grand design spiral M51
   (NGC~5194), and its interaction with its companion
   NGC~5195. Despite the simplicity of our models, our simulations
   capture the present day spiral structure of M51 remarkably well,
   and even reproduce details such as a kink along one spiral arm, and
   spiral arm bifurcations. We investigate the offset between the
   stellar and gaseous spiral arms, and find at most times (including the
   present day) there is no offset between the stars and gas to within
   our error bars. We also compare our simulations with recent
   observational analysis of M51. We compute the pattern speed versus
   radius, and like the observations, find no single global pattern
   speed. We also show that the spiral arms cannot be fitted well by
   logarithmic spirals. We interpret these findings as evidence that
   M51 does not exhibit a quasi-steady density wave, as would be
   predicted by density wave theory. The internal structure of M51
   derives from the complicated and dynamical interaction with its
   companion, resulting in spiral arms showing considerable structure
   in the form of short-lived kinks and bifurcations. Rather than
   trying to model such galaxies in terms of global spiral modes with
   fixed pattern speeds, it is more realistic to start from a picture
   in which the spiral arms, while not being simple material arms, are
   the result of tidally induced kinematic density `waves' or density
   patterns, which wind up slowly over time.
\end{abstract}


\section{Introduction}

Nearby galaxies such as M51 provide an ideal basis to examine
molecular cloud and star formation, both observationally and
theoretically. Whilst the last decade has seen a huge advance in high
resolution hydrodynamic simulations of galaxies
(e.g. \citealt{Wada1999,deAvillez2005,Shetty2006,Dobbs2008,Tasker2009,Agertz2009}),
such calculations are rarely designed to model specific galaxies,
e.g. M33, M81, M51 where we are now obtaining detailed CO,
H$\alpha$ and with Herschel, FIR observations
(e.g. \citealt{Blitz2007,Tamburro2008,Koda2009}). However several
groups have successfully computed orbits of nearby interacting
galaxies, including M51 \citep{Salo2000,Theis2003} and M81
\citep{Yun1999} using N-body codes. We take the results from one such
calculation \citep{Theis2003} as the basis for modelling the gas dynamics of M51.

\subsection{Spiral structure}

Spiral galaxies exhibit a variety of morphologies, from flocculent to
grand design. Their structure and dynamics are thought to depend on
whether the spiral arms are due to small scale gravitational
instabilities in the stars and/or gas, or larger scale perturbations
caused by global density waves, tidal interactions or bars
(e.g. \citealt{Toomre1977,Lin1985,Elme1991,Elmegreen1995}).

Flocculent galaxies are thought to occur when local gravitational
instabilities in the gas lead to multiple short arm segments
\citep{Elmegreen1991,Elmegreen1995,Bottema2003,Li2005b}. When
instabilities are present in both the gas and stars, longer spiral
arms may develop (e.g. \citealt{Sellwood1984,Elmegreen1993}).

In addition to flocculent spirals, grand design galaxies, predominantly with a
symmetric 2 or 4 armed spiral pattern, constitute around 50 per cent of
spiral galaxies \citep{Elmegreen1982}. Traditionally `density wave
theory' has been used to provide an explanation of the spiral patterns
in these galaxies, where the stars and gas are assumed to exhibit
quasi-stationary standing wave patterns \citep{Lin1964}. The spiral
arms occur where stars are aligned at particular points of their
orbits \citep{Toomre1977}. For both types of spiral, swing amplification may be important either by increasing the amplitude of local perturbations \citep{Julian1966,Toomre1981,Sellwood1984}, or by providing a feedback mechanism to maintain wave packets propagating through the disc \citep{Mark1976,Goldreich1978,Toomre1981}.

However, an alternative picture is that grand design spiral structure
may be caused predominantly by tides, driven by internal bars or by
interactions with other galaxies
\citep{Chamberlin1901,Toomre1972,Tully1974,Oh2008}. In fact,
\citet{Kormendy1979} proposed that, unless their rotation curves have
special properties, all non-barred grand design galaxies must be the
result of interactions with nearby galaxies. This view is strengthened
by the simulations by \citet{Bottema2003}.  

The first comprehensive
investigation of the degree to which specific grand design spirals
could be modelled by tidal interactions was presented by
\citet{Toomre1972}.  However the limitations of the simple analysis of
\citet{Toomre1972} were such that tidal features could not propagate
radially and so could be induced only in the outer parts of galaxies.
Later calculations by \citet{Hernquist1990} of tidal interactions
indicated that with higher resolution, and when including a fully
consistent gravitational model, more realistic rotation curves and
self-gravity of the stars, the spiral structure extends to much
smaller radii and is longer lived. \citet{Hernquist1990} found that tidally induced waves are 
amplified by the swing amplification mechanism, as predicted by 
\citet{Toomre1981}. Furthermore, the spiral structure
is more easily seen and so prolonged when gas is present
\citep{Sundelius1987,Bertin1989,Chak2008}. Nevertheless, the tidal
encounter could produce the stimulus for a strong density wave
perturbation, as originally hypothesised by \citet{Toomre1972}, and
discussed in more detail for M51 by \citet{Tully1974} and \citet{Elmegreen1989c}.

\subsection{Molecular clouds and spurs}

For flocculent galaxies, the formation of molecular clouds is most
likely intrinsically linked to the formation of spiral structure -
gravitational instabilities in the gas lead to both the spiral arms
and molecular clouds \citep{Elmegreen1995}. Recently computer
simulations have become capable of modeling the hydrodynamics of such
galaxies. Several show the formation of giant molecular clouds in this
manner \citep{Wada1999,Wada2002,Li2005b,Tasker2006,Robertson2008},
although collisions between clouds may also contribute
\citep{Tasker2009}. For grand design spirals, GMCs basically form in
the same way, but long-lived spiral perturbations force the gas into
the spiral arms periodically, and for longer. Consequently, the growth
of instabilities are more confined to, and more dependent on the
conditions in the spiral arm, e.g. density, degree of
shear. Furthermore collisional formation of molecular clouds can play
a much more important role as the mean free path between collisions
decreases \citep{Dobbs2008}.

In addition to molecular cloud complexes along the arms, interarm
spurs are clearly visible in many spiral galaxies
\citep{Elmegreen1980,LaVigne2006} and are particularly distinguishable
in M51 \citep{Corder2008}. One possibility is that interarm spurs
correspond to features in the underlying stellar distribution
\citep{Julian1966,Elmegreen1980,Byrd1984} and are thus due to
gravitational instabilities in the stars. More recent simulations,
which assume a static potential, have also shown the formation of
spurs purely from the gaseous component of the disc. In this case,
GMCs form in the spiral arms (e.g. by agglomeration
\citep{Dobbs2006,Dobbs2008} or gravitational instabilities in the gas
\citep{Balbus1988,Kim2002,Shetty2006, Dobbs2008}) and are then sheared
into spurs as they leave the arms. In the presence of some underlying
driving frequency (a global pattern speed) longer gaseous branches may
also develop at certain radii due to resonances in the disc
\citep{Patsis1997,Chak2003,Yanez2008}. In addition, spurs may be
associated with stochastic star formation
\citep{Elmegreen1980,Feitzinger1982}, but simulations by
\citet{Shetty2008} indicate that stellar feedback tends to disrupt
large spurs in the disc.

\subsection{Previous models of M51}

M51 in particular is considered the hallmark for grand design galaxies
and as such is a prime candidate for the application of density wave
theory. However there are clear indications of a departure from
standard density wave theory. \citet{Shetty2007} show huge variations
in the velocities of gas in the disc, apparently showing large net
radial mass fluxes at some radii, suggesting the spiral structure of
the galaxy is not in a steady state. Furthermore \citet{Meidt2008}
find multiple pattern speeds in M51 (or a radial dependence of pattern
speed), indicative that the pattern is both radially and time
dependent.

There have been numerous numerical models of M51,
(e.g. \citealt{Toomre1972,Hernquist1990}), but the most thorough are
by \citet{Salo2000,Salo2000b}. \citet{Salo2000} use a least squares
technique to find the orbits of the galaxy--galaxy interaction which
produce the observed structure. They show that a multiple, rather than
single, encounter can produce velocities in the HI tail in agreement
with observations. They also indicate that tidal perturbations
overwhelm any pre-existing spiral structure. \citet{Salo2000b}
perform higher resolution simulations, though only with a stellar
disc, to investigate inner spiral structure. Similar to
\citet{Toomre1969}, they propose that the inner spirals are a
consequence of tidal waves propagating to the centre of the disc.

In order to analyse the extended parameter space, \citet{Theis2003} 
instead employ a genetic algorithm to determine the 
orbit of M51 and NGC 5195. Based on a much larger number of simulations,
they corroborate the results of \citet{Salo2000} i.e.\ they
find the orbit involves multiple encounters.  Both \citet{Salo2000} and 
\citet{Theis2003} require that NGC 5195 lies on a bound orbit to match the spatial and velocity structure of M51.

\subsection{The current paper}

So far, simulations of grand design galaxies have largely implicitly
assumed density wave theory, invoking a global mode with a fixed
pattern speed by applying a steady rotating spiral potential
\citep{Kim2002,DBP2006,Dobbs2008}. However if grand design galaxies
arise from interactions or bars, it is unclear whether the density
wave scenario invoked by \citet{Lin1964} is appropriate in many galaxies. 
Thus it is important to
study the gas dynamics, and formation of molecular clouds in the
context of realistically induced spiral perturbations.  For the case
of a bar--driven grand design spiral, the spiral is believed to be
long-lasting, thus the idea of a spiral structure with fixed pattern
speed may well be applicable. This is less likely to be true for
interacting galaxies, where there is a dynamical interaction and a
constantly changing spiral structure.

In this paper we study the dynamics in an interacting system, using
orbital data for M51 provided by \citet{Theis2003}. Whilst
\citet{Salo2000b} mainly concentrated on the stellar structure and
dynamics, we focus on the gas. \citet{Salo2000} do include gas in
lower resolution simulations, though they use a sticky particle method
to evolve the gas particles. We instead use a Lagrangian hydrodynamics
code, \textsc{SPH} (Smoothed Particle Hydrodynamics) to model a
dynamic halo, disc (which contains gas and stars) and bulge. We
describe the evolution of the spiral arms, the generation of
substructure and the velocities in the gas. We consider whether the
evolution of the disc is significantly different from the previous
simulations which assumed a static potential and therefore a rigidly
rotating density wave.

\section{Computational details}

We model M51 and its interaction with NGC~5195 using \textsc{SPH}, a
Lagrangian fluids code. The code is based on an original version by
Benz \citep{Benz1990}, but has since been subject to significant
modifications, including individual timesteps and sink particles
\citep{Bate1995,Batesph1995}. More recently magnetic fields have been
included \citep{Price2005,PB2007}, although these are not used in the
current paper.

Unlike GADGET \citep{Springel2001,Springel2005}, which has previously
been applied to simulations of interacting galaxies
(e.g. \citealt{Cox2006,Jachym2007}), our code has predominantly been
used for simulations of star formation
(e.g. \citealt{Bate2003,Price2008,Bate2009a,Bate2009b}). Even for
calculations of a galactic disc \citep{Dobbs2008,DGCK2008,DB2008}, we
restricted our models to the gaseous component of the disc. For the
simulations in this paper, we model the stellar disc, bulge and halo,
so have therefore adapted the code to include 2 types of particle,
gaseous and stellar.

All particles have variable smoothing lengths, the smoothing length
and density solved iteratively according to
\begin{equation}
h=\nu \bigg(\frac{m}{\rho}\bigg)^{1/3}
\end{equation}
\citep{PM2007}. Here $\rho$ is the density, $m$ the mass of the
particle, and $\nu$ is a dimensionless parameter set to 1.2 in order
that each particle has $\sim$60 neighbours. We calculate smoothing lengths 
for the gas and star particles separately. For both types of
particle, $h$ sets the gravitational softening length, and for gaseous
particles it is also the SPH smoothing length. Only gas particles are
subject to pressure and viscous forces. Artificial viscosity is
included to treat shocks, using the standard parameters, $\alpha=1$
and $\beta=2$ \citep{Monaghan1985}.

\subsection{The orbit of M51 and NGC 5195}

In order to reproduce spiral structure similar to that observed in
M51, we model a galaxy representing M51 and its companion galaxy
NGC~5195. We assign the initial positions and velocities of the two
galaxies according to the results from N-body calculations by
\citet{Theis2003}. \citet{Theis2003} used \textsc{MINGA}, a restricted
N-body code combined with a genetic algorithm code, to determine the
orbit of M51 and NGC~5195.  In their calculations, the interacting
galaxy NGC~5195 is represented by a single point mass, which has one
third the mass of M51. The test particles otherwise comprise the disc and
bulge of M51 - the halo is represented by a potential. They generated
spatial and velocity maps from the outcome of each N-body calculation
to compare with observed maps of HI. The genetic algorithm code is
used to constrain the parameters of the orbit, and find the best fit
to the observed data. The best-fit model corresponds to a highly
elliptical orbit, with two passages of NGC~5195 through the plane of
the disc of M51. This model provided the initial velocities and
positions for the \textsc{SPH} calculations at a time of $\sim300$ Myr
prior to their current position. At this point in the N-body model,
the two galaxies are separated by approximately 24~kpc. The resulting
orbit of the two galaxies in the SPH calculations is shown in Fig.~1.
The initial positions and velocities of the two galaxies are listed in Table~1.
\subsubsection{Dynamical friction}

A difference between the simulations presented here and those produced
using the \textsc{MINGA} code is that we use a live halo, whereas
\citet{Theis2003} use a potential. Consequently the orbit of the SPH
calculations begins to deviate from that derived from the
\textsc{MINGA} calculations with time. \citet{Petsch2008} have since
implemented dynamical friction into an improved \textsc{MINGA} code,
which takes into account the effect of the companion on the halo. They
also reran the calculations of \citet{Theis2003} using dynamical
friction. Unfortunately they have not yet calculated a new orbit for
M51, which includes the effects of dynamical friction. 

We did however run calculations of the interaction without gas, to compare with
unpublished purely stellar-dynamical calculations
 of Harfst \& Theis (private communication). Although Harfst \& Theis
modelled NGC~5195 as
well as M51 (whilst here we adopt a point mass for NGC~5195), we found
that after 300 Myr, the difference in the position of the companion,
and the structure of the disc was negligible for the \textsc{MINGA}
and \textsc{SPH} codes. After 500 Myr the structure of M51 was still
the same, although the companion galaxy had shifted about $25^o$
further in its orbit for the \textsc{SPH} code.

\subsection{Initial setup of M51 and NGC 5195}

We determined the initial distribution of particles in our model of
M51 by using the mkkd95 program \citep{Kuijken1995}. This is a
publicly available program from the NEMO stellar dynamical software
package \citep{Teuben1995}. In the mkkd95 program we set psi0=-7.65
and ra=0.142 (which determine the size of the halo), md=1.36 (the mass
of the disc), router=3 (the radius where the disc begins truncate) and
drtrunc=0.25 (the distance over which the disc truncates). The
remaining parameters (including those for the bulge) were set as the
defaults. We selected these parameters to produce a mass and
maximum radial extent for each component similar to those of \citet{Theis2003}, 
although we adopted a slightly less extended halo based on the rotation curve shown in \citet{Sofue1999}.
However although the radius and mass chosen for the halo were based on the observed rotation curve, our rotation curve was not as flat as \citet{Sofue1999}, peaking at around 275 rather than 250 km s$^{-1}$. Furthermore the maximum velocity increases slightly during the interaction. Thus in retrospect we ideally would have needed to start with a more extended halo. 
The main difference this makes to our simulation is that the pattern has rotated further and is therefore slightly out of phase compared to the observed structure of M51 (see Sections~4.1 and 4.2). 

The properties of our model of M51 are listed in Table~2, as well as the
number of particles in each component. 
We chose to place the most resolution in the disc (and in particular
the gas). Thus we used 1 million particles for the disc, $6\times
10^4$ for the halo and $4\times 10^4$ for the bulge, giving a total of
1.1 million. We then assigned SPH particles the velocities, positions
and masses outputted from the mkkd95 program. 

We work in a Cartesian coordinate system in which the $x$-- and
$y$--axes lie in the plane of the sky, and the $z$--axis lies towards
us along the line of sight. In order to end up with the observed
orientation of M51 we start with our model galaxy in the plane of the
sky and then perform two rotations, first one of $20^o$ clockwise
about the $y$--axis, and then one $10^o$ clockwise about the
$x$--axis. Before adding the perturbing galaxy,
we first ran a simulation with stars only, to ensure that the galaxy
had a stable configuration.

Lastly we placed the two galaxies at the relative positions, and with
the respective velocities to reproduce their orbit, as determined by
\citet{Theis2003}. The initial positions and velocities (with respect to our Cartesian grid) 
of the two galaxies are those shown in Table~1. Similar to \citet{Theis2003} we model the
interaction by designating a single particle as the companion galaxy
NGC~5195, which has a third of the mass of M51. The softening of the point mass is 
treated in the same way as the other particles in the simulation. \citet{Salo2000b}
similarly only assigned 1 particle to NGC 5195, although their first
calculations modelled the companion galaxy consistently
\citep{Salo2000}.

\begin{table}
\centering
\begin{tabular}{r|c|c|c}
 \hline 
& & M51 & NGC 5195 \\
\hline
Initial & $x$ & 4.91 & -17 \\ 
position & $y$ & 1.89 & -6.55 \\ 
(kpc) & $z$ & 0.95 & -3.30 \\ 
\hline
Initial & $v_x$ & -1.46 & 5.86 \\ 
velocity & $v_y$ & 0.68 & -2.44 \\ 
(km s$^{-1}$) & $v_z$ & -3.26 & 15.6 \\ 
\hline
\end{tabular}
\caption{The initial positions and velocities for M51 and NGC 5195 are listed. These are extracted from the N-body calculations performed by \citet{Theis2003}.}
\label{orbit}
\end{table}

\begin{table}
\centering
\begin{tabular}{c|c|c|c|c|c|c}
 \hline 
Component & Mass  & Radial Extent & No.  \\
& (M$_{\odot}$) &  (kpc) & particles \\
 \hline
M51: Disc & 5.9 $\times 10^{10}$ & 15 & $10^6$ \\
M51: Halo & 1.45  $\times 10^{11}$ & 20 & $6\times 10^4$  \\
M51: Bulge & 5.25 $\times 10^{9}$ & 4.8 & $4\times 10^4$ \\
NGC 5195 & 7.07 $\times 10^{10}$ & -  & 1 \\
\hline
\end{tabular}
\caption{The mass, radial extent and number of particles is listed for each component of the M51 galaxy (the radial extent is the distance to the edge of each component, although all fall off with radius). The companion galaxy NGC 5195 is represented by a single point mass in the simulation, thus has no physical radius.}
\label{runs}
\end{table}

\begin{table}
\centering
\begin{tabular}{c|c|c|c|c|c|c}
 \hline 
Model & Gas mass & Temperature & $min(Q_{\rm g})$ \\
& (\% of disc) & K  & \\
 \hline
A & 1 & $10^4$ & 30 \\
B & 10 & $10^4$ & 3 \\
C & 0.1 & 100 & 30 \\
\hline
\end{tabular}
\caption{The properties of the gas component are listed for the 3 models presented in this paper. The final column is the minimum value of the Toomre stability parameter for the gas (see Section~2.2.2 and Fig.~2).}
\label{runs}
\end{table}
\begin{figure}
\centerline{
\includegraphics[bb=150 0 550 800,scale=0.36]{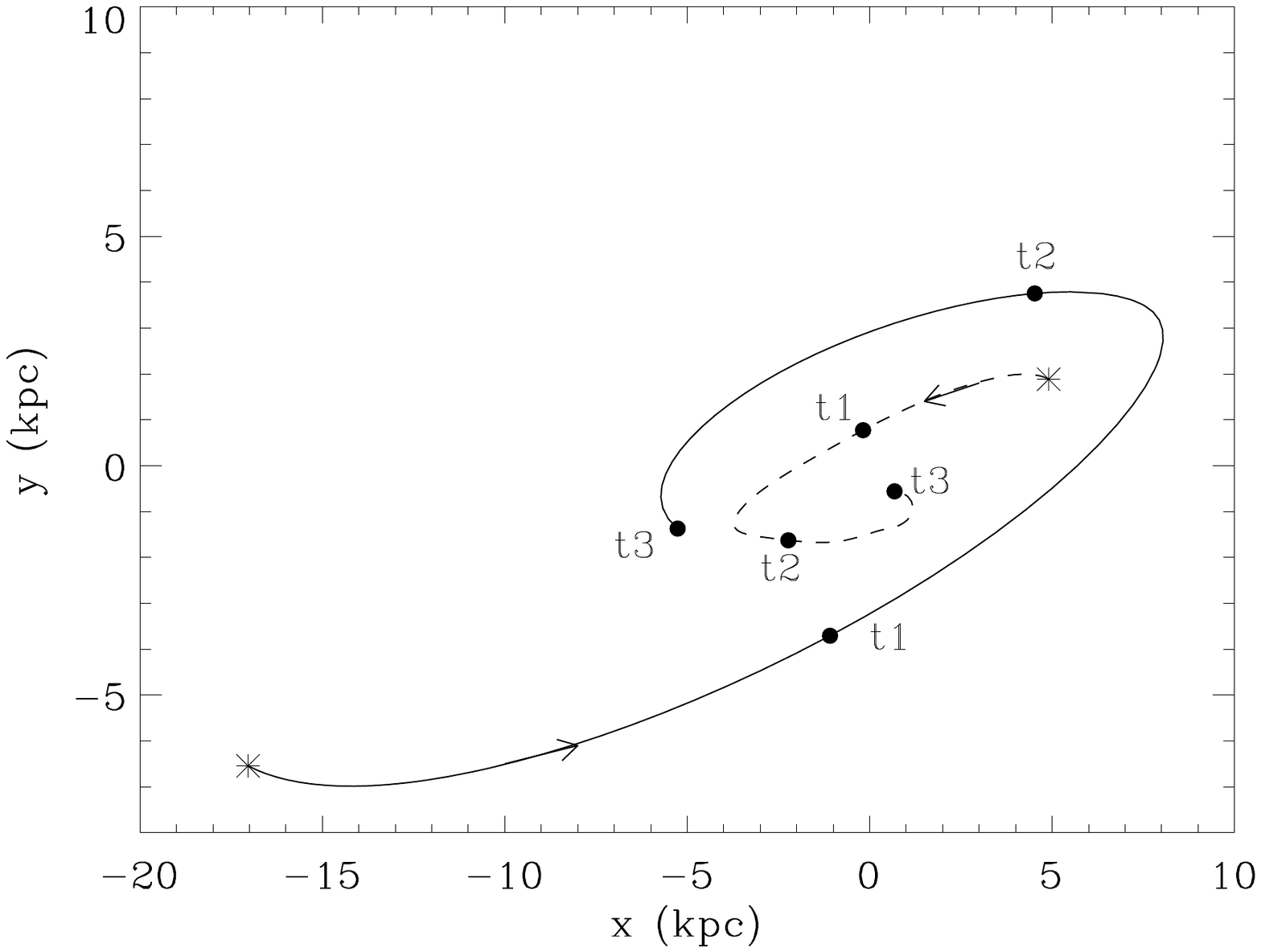}}
\centerline{
\includegraphics[bb=150 350 550 420,scale=0.36]{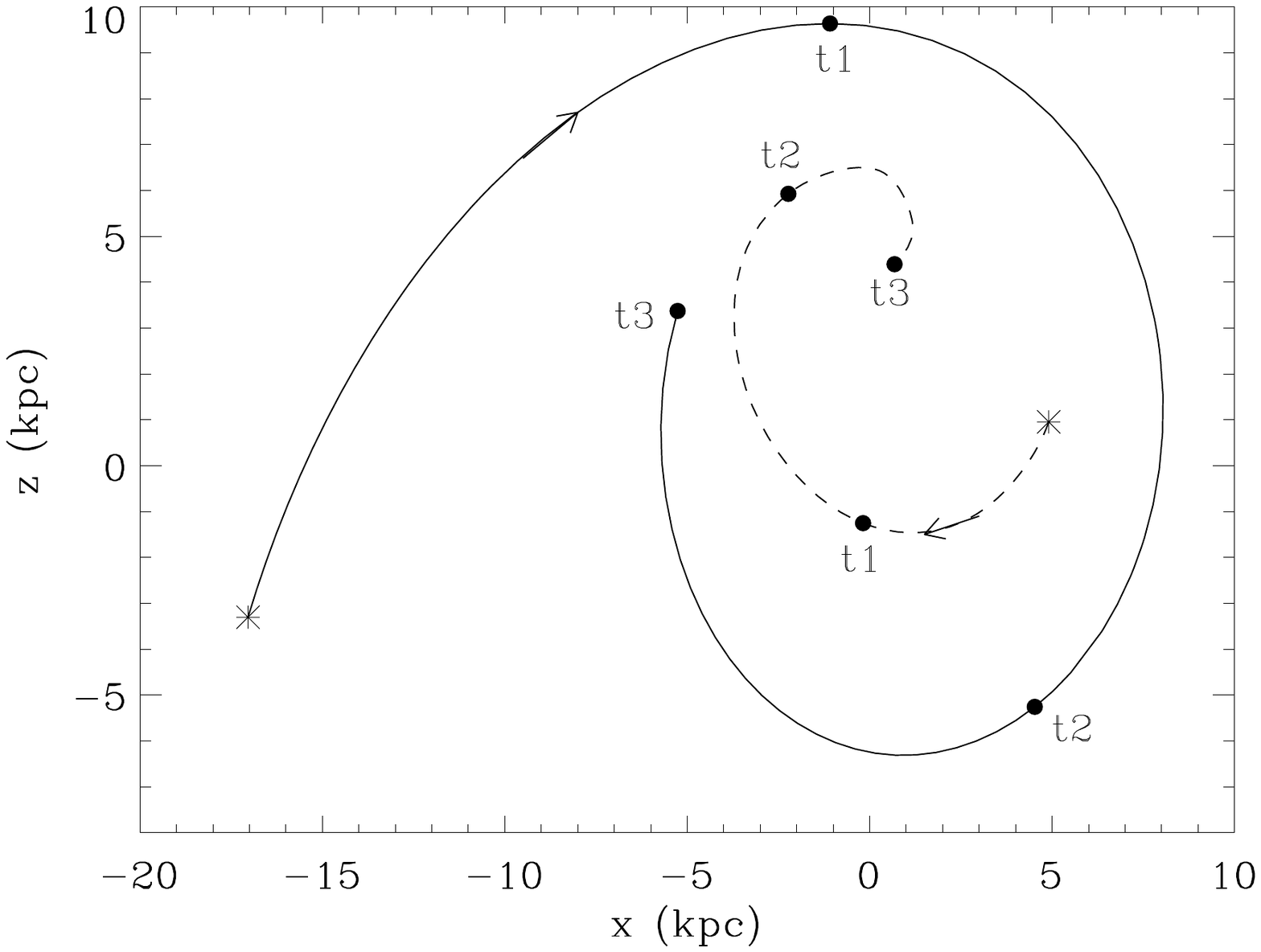}}
\caption{Orbit of M51 (dotted line) and companion galaxy NGC 5195
  (solid line) in the $xy$-- (top) and $yz$-- (lower) planes. The
  \rlap{$\times$}+ symbol denotes the initial positions of the
  galaxies and t1, t2, t3 denote times of 100, 200 and 300 Myr. Recall
  that the $z$--axis lies along the line of sight (with positive $z$
  pointing towards the observer, while the $x$-- and $y$--axes lie
  in the plane of the sky.}
\end{figure} 

\subsubsection{The gaseous component of the disc}
We performed 3 calculations with different conditions for the gaseous
component of the disc. In all cases we set 900,000 random particles in
the disc as gas particles - thus leaving 100,000 stellar particles in
the disc. In model A, this implied a mass resolution of 180
$M_{\odot}$ per particle for the gas and 2.4 $\times 10^6 M_{\odot}$
per particle for the stellar disc. The average smoothing length for
the gas particles is 100 pc.  We did not change the positions of the
gas particles, rather we assumed they would settle in equilibrium in
the direction perpendicular to the disc (see Section~3). 
We then set masses and temperatures for the gas particles, as shown in Table~3.

We chose the gas mass to be 0.1, 1 or 10 per cent of the disc mass,
and the temperature as either 100 or $10^4$ K. For these calculations,
we take a simple approach and assume the gas to be isothermal.  The
cases where the gas mass is 0.1 or 1 per cent are unrealistically low
for a disc galaxy. However such a low mass largely prevents
gravitational collapse in the gas. Gravitational collapse slows down
the calculations, and requires the inclusion of sink
particles. Furthermore in order to carry out a global comparison of
the gas distribution between models and observations, we are mainly
interested in the structure induced in the gas in response to the
interaction, as opposed to the instabilities in the gas. For the warm
($10^4$ K) gas though, which is less susceptible to gravitational
instabilities, we also ran a model where 10 per cent of the disc is
gas (by mass).

We initially ran a calculation with a temperature of 100 K, setting 1\% 
of the disc as gas. However this gave rise to widespread
self-gravitational collapse throughout the disc. For this reason we
restarted the calculation but taking 0.1\% of the mass of the disc to
be gas, which was more stable.  In addition to running models with an
interaction, we also ran a calculation for a galaxy in isolation. For
this, we used our fiducial mode (A) with 1\% warm gas.

\subsubsection{Stability of the disc}

To clarify the gravitational stability of the disc, we computed the
Toomre stability parameter for the gas ($Q_{\rm g}$) and stars ($Q_{\rm s}$). We
took $Q_{\rm g}=c_{\rm s} \kappa / \pi G \Sigma$ and $Q_{\rm s}=\sigma_{\rm R}
  \kappa / 3.36 G \Sigma$ where $c_{\rm s}$ is the sound speed, $\kappa$ is
the epicyclic frequency, $\Sigma$ the surface density and $\sigma_{\rm R}$
the radial velocity dispersion \citep{Toomre1964,Goldreich1965,Binney}. 
A detailed analysis for the Toomre criterion in a two-fluid system is described in \citet{Jog1984} and \citep{Romeo1992}, but we adopt the limiting case where one component is stable for our simple analysis here. We show
$Q_{\rm g}$ and $Q_{\rm s}$ versus radius for the least stable case (model B) in
Fig.~2. We also state the minimum value of $Q_{\rm g}$ in Table~3. As $Q_{\rm s}$
is $\approx 1$ at larger radii, the stars are unstable to perturbations,
and as we show in Section~ 3.1.1, multiple spiral arms develop even in
the absence of a tidal interaction. For the model where 10\% of the
disc is gas, $Q_{\rm g}$ is as low as 3, hence the gas is only marginally
stable. This is the least stable case -- $Q_{\rm g}$ is a factor of 10 higher
in models A and C.
\begin{figure}
\centerline{
\includegraphics[bb=150 350 500 800,scale=0.42]{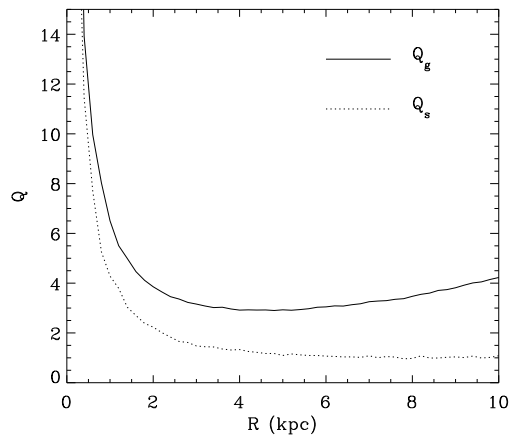}}
\caption{$Q_{\rm s}$ and $Q_{\rm g}$ are plotted against radius for model B, where
  10\% of the disc is warm ($10^4$ K) gas. $Q_{\rm s}$ is the same for all
  the models as the stellar distribution does not change. $Q_{\rm g}$ is a
  factor of 10 higher for models A and C, which are more stable.}
\end{figure}  

\subsubsection{Sink particles}

Even with low surface densities, we found it necessary for
computational reasons to allow the formation of sink particles. Sink
particles replaced regions with gas densities exceeding $10^{-12}$ g
cm$^{-3}$, with accretion radii of 10 pc. Clearly we do not resolve
the Jeans length up to this density with our resolution -- the
inclusion of sink particles merely allows the calculations to
continue. In our main model (A) only 5 sink particles are formed
during the simulation.
\begin{figure}
\centerline{
\includegraphics[bb=100 20 500 480,scale=0.4]{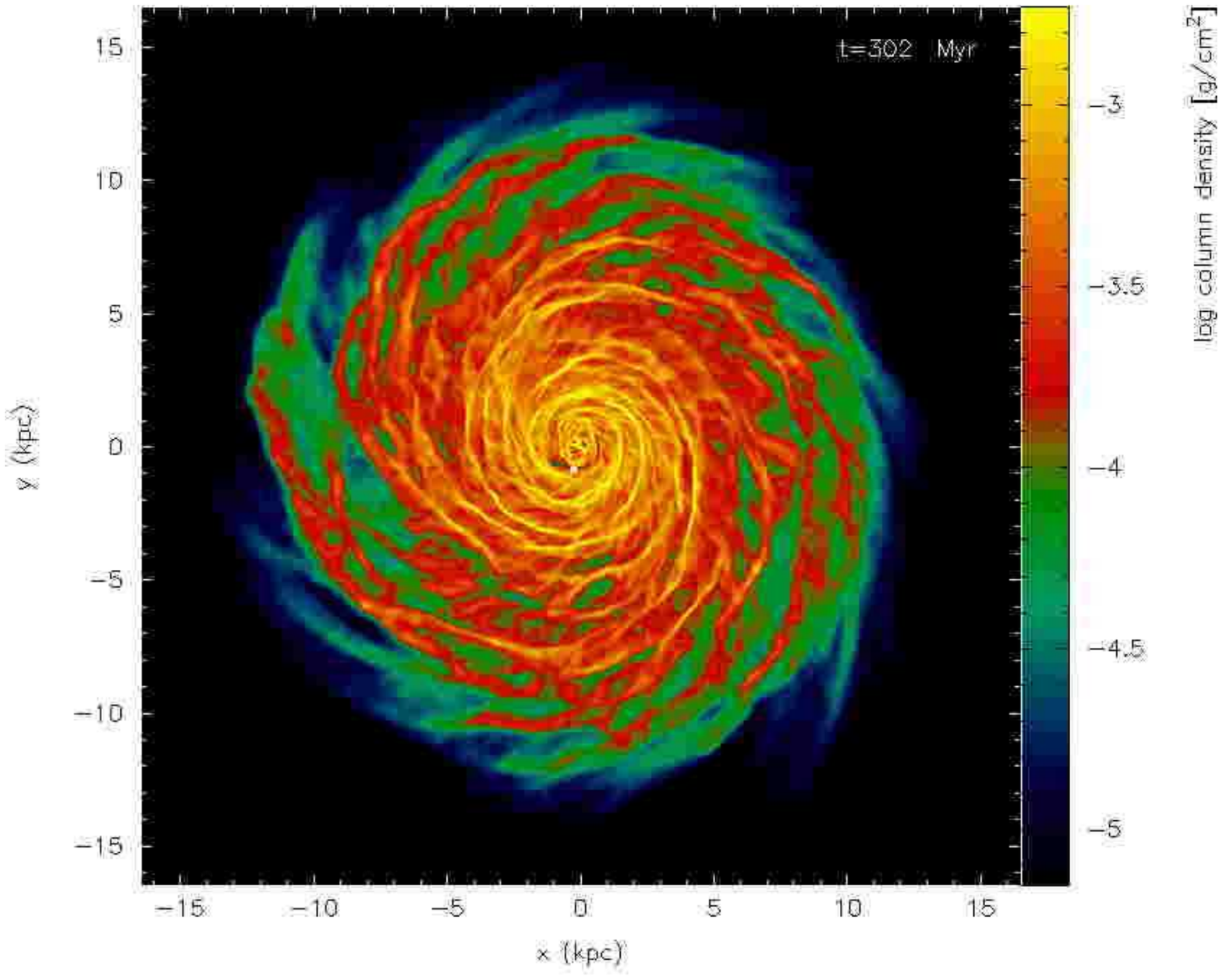}}
\caption{The gas column density is shown for an isolated galaxy, using
  the galaxy from model A with 1\% warm gas. The time is 300 Myr,
  though the nature of the spiral structure does not change significantly between
  100 and 300 Myr. The combined stellar+gaseous disc is unstable to
  gravitational perturbations (with $Q_{\rm s} < 2$), hence a multi-armed
  flocculent spiral pattern develops.}
\end{figure} 

\begin{figure*}
\centerline{
\includegraphics[scale=0.32,bb=20 100 500 500]{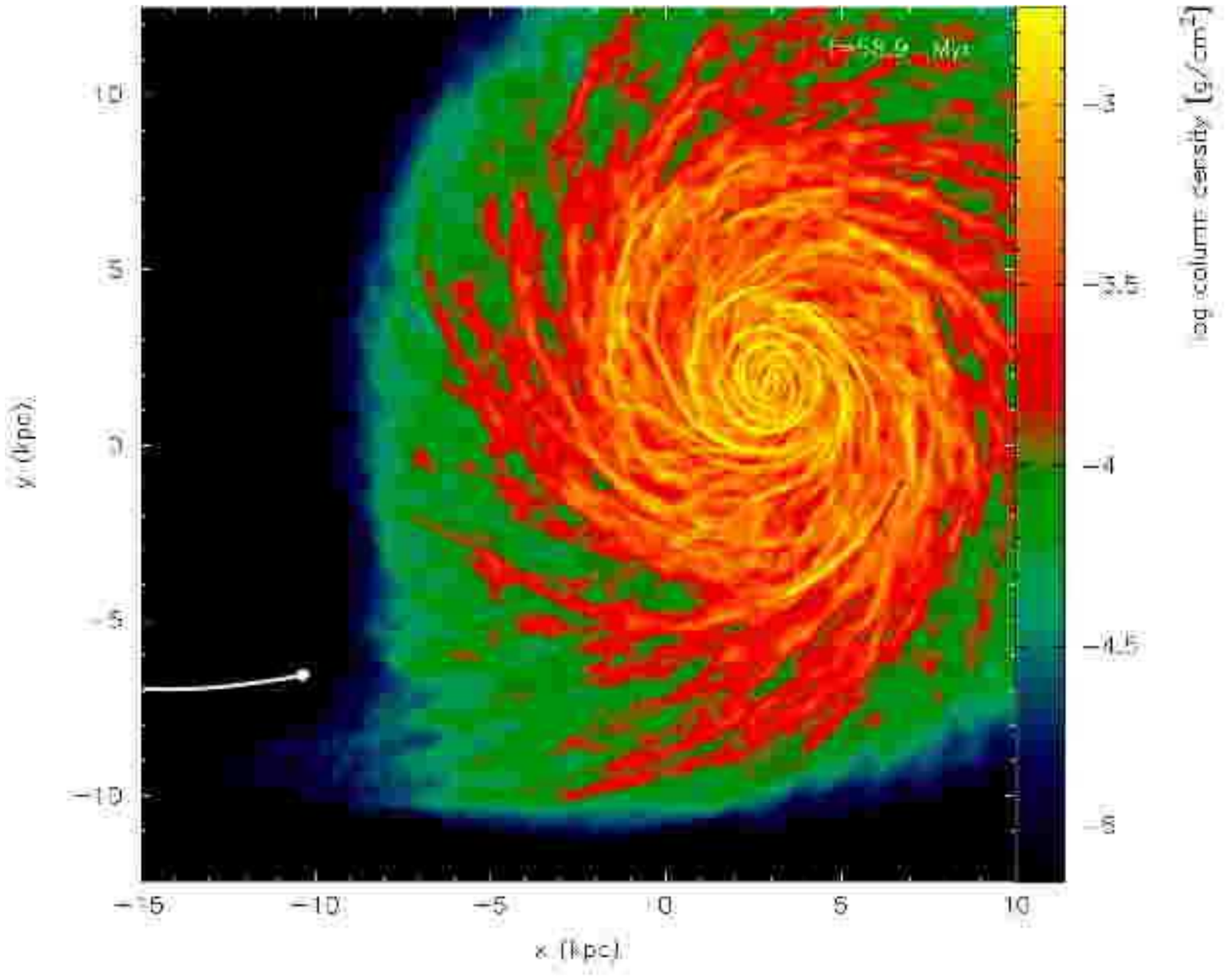} 
\includegraphics[scale=0.32,bb=20 100 500 500]{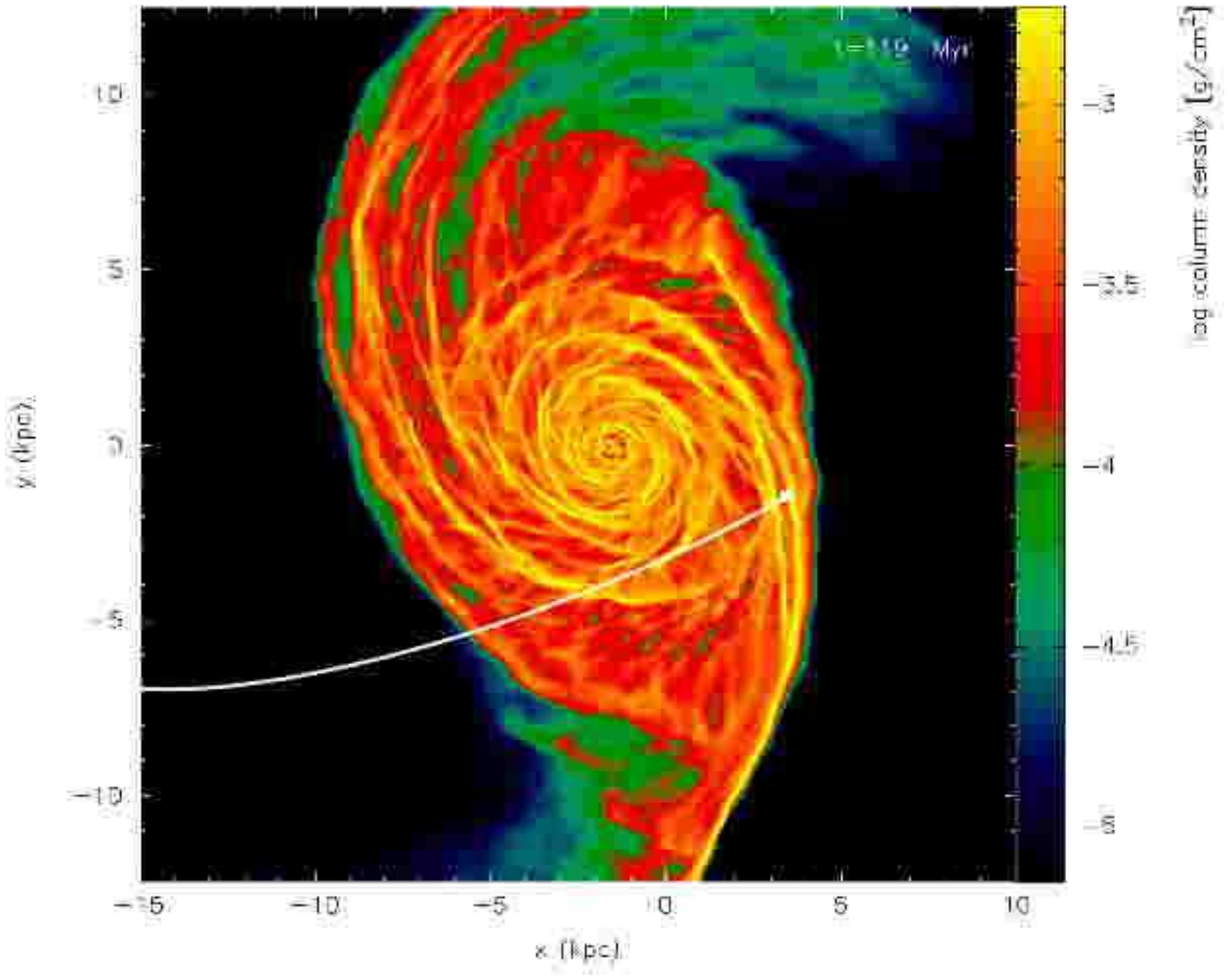} 
\includegraphics[scale=0.32,bb=20 100 500 500]{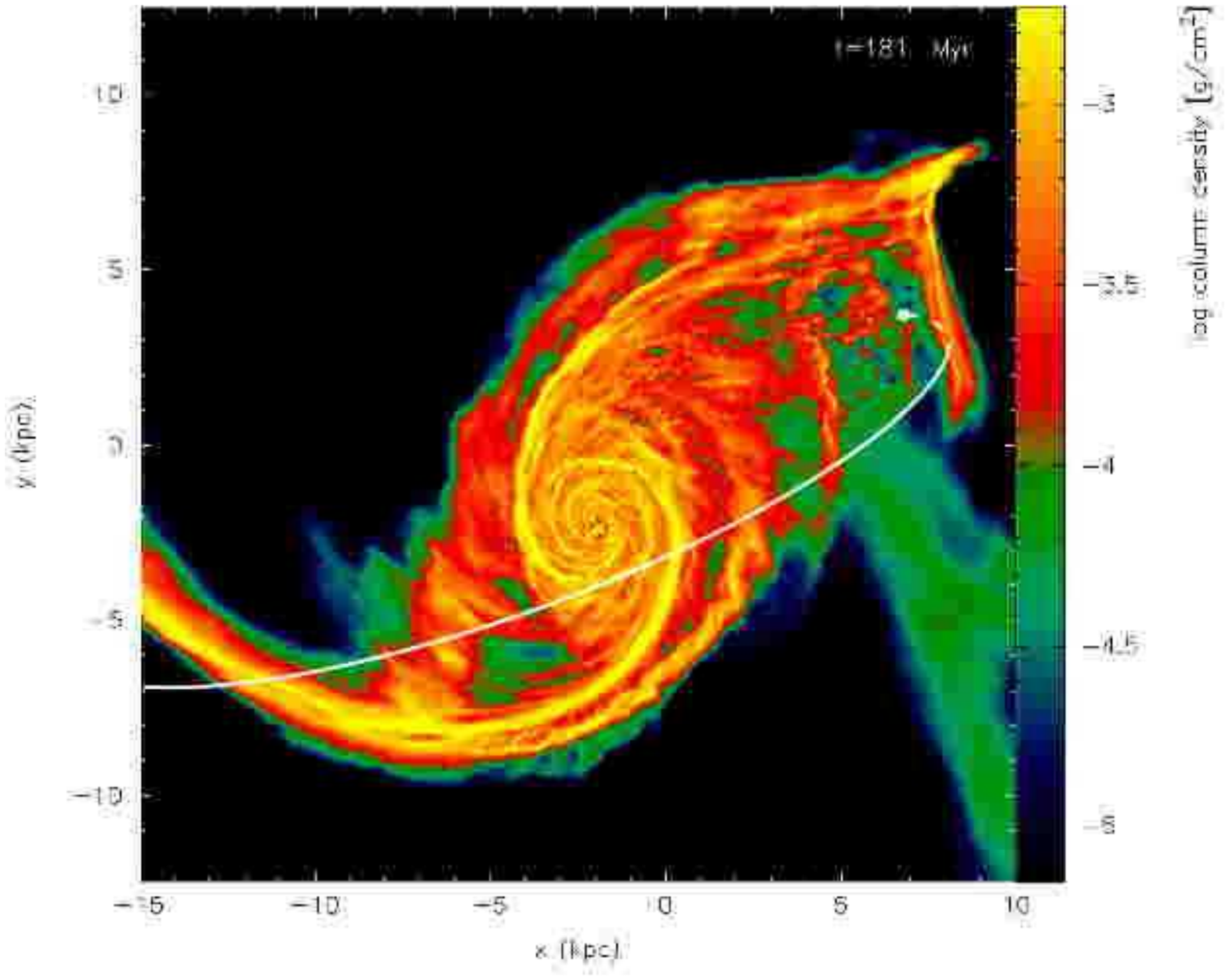}}
\centerline{
\includegraphics[scale=0.32,bb=20 0 500 550]{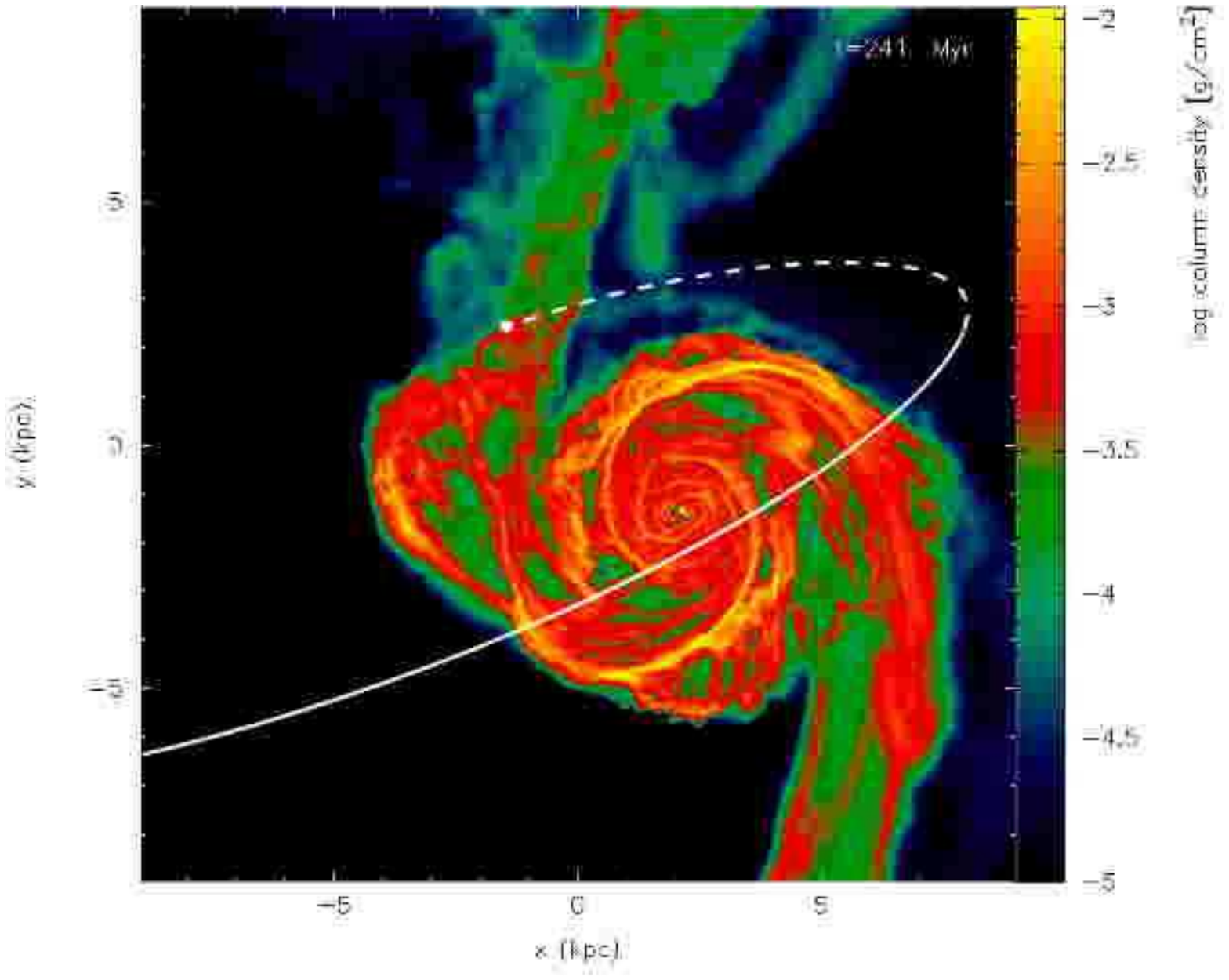} 
\includegraphics[scale=0.32,bb=20 0 500 550]{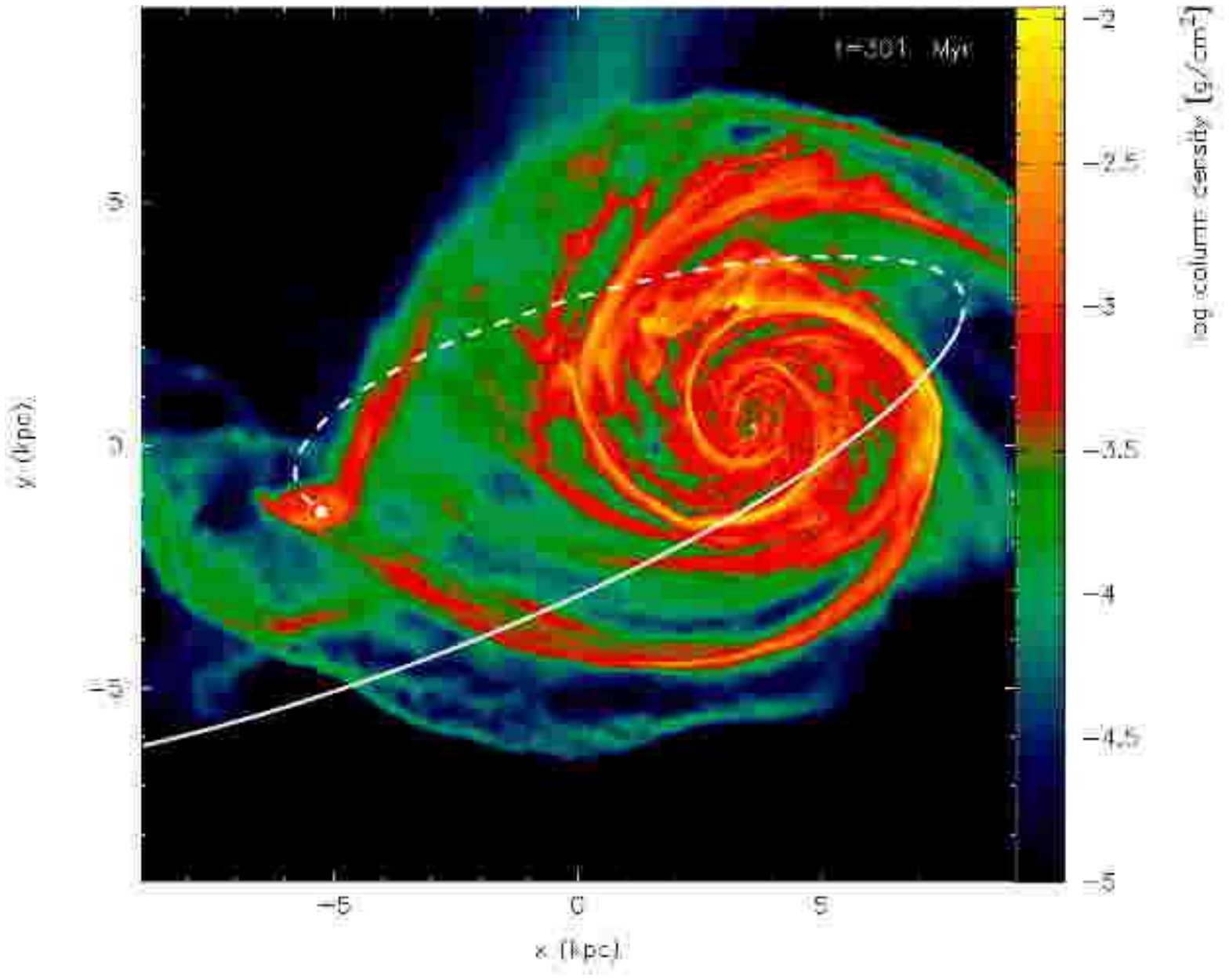}
\includegraphics[scale=0.32,bb=20 0 500 550]{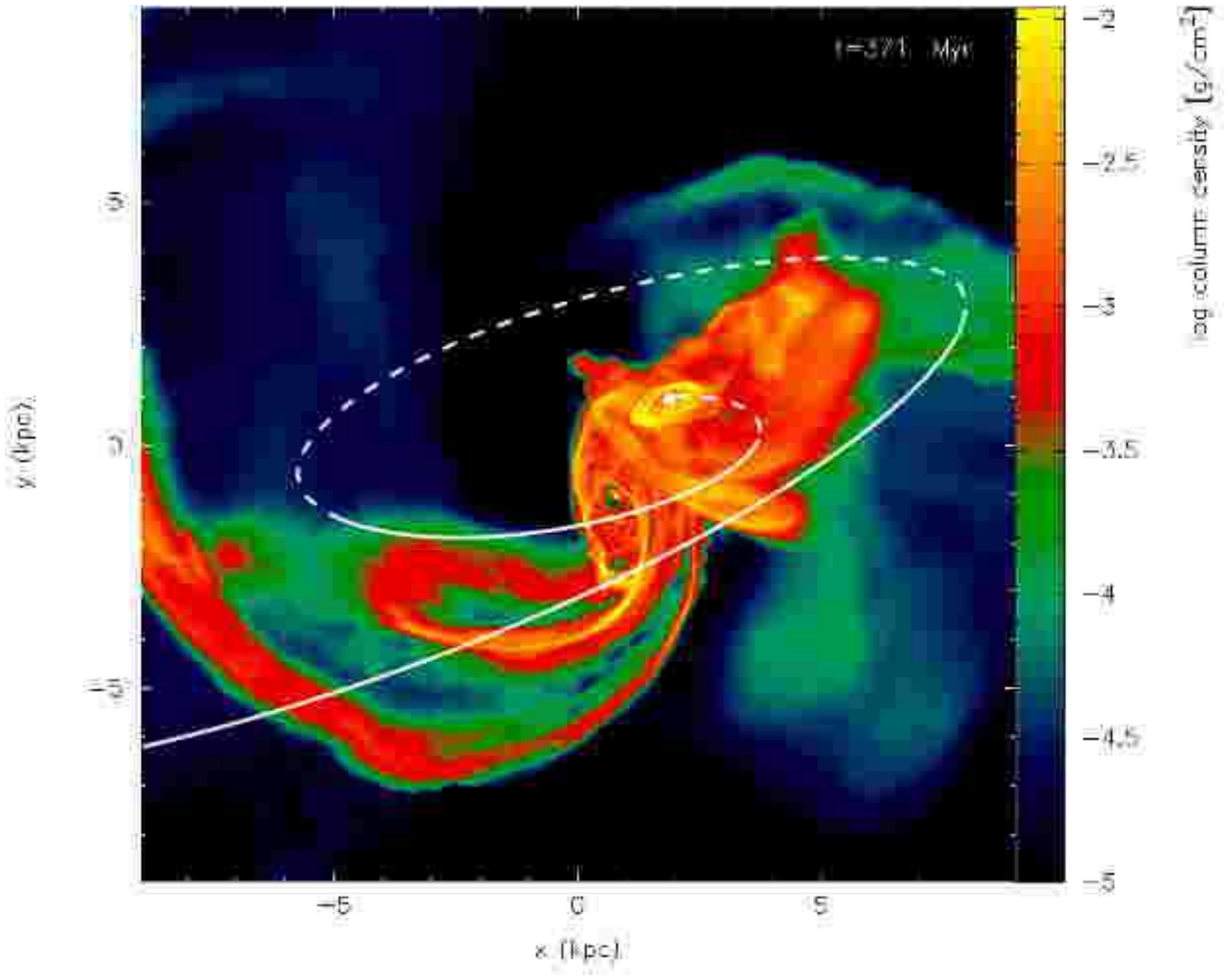}}
\caption{Column density plots show the time evolution of the simulated
  interaction of M51 and NGC~5195, at times of 60, 120, 180, 240, 300
  (corresponding approximately to the present day) and 371 Myr. These
  plots only show the gas, which represents 1\% of the disc by mass,
  and has a temperature of $10^4$ K (model A). We model a galaxy
  representing M51, whilst the galaxy NGC~5195 (a point mass) is
  indicated by the white spot. Sink particles are otherwise omitted
  from the figures (see text). The orbit of the companion galaxy is
  also shown on the panels (the dashed section indicates that the
  companion is behind the M51 galaxy). The galaxy undergoes a
  transition from a flocculent spiral to a grand design spiral during
  the course of the interaction. At the last time frame (371 Myr), the
  two galaxies are in the process of merging. Note, both the spatial
  and density scales differ in the lower 3 plots.}
\end{figure*} 

\begin{figure}
\centerline{
\includegraphics[scale=0.44,bb=0 -20 500 480]{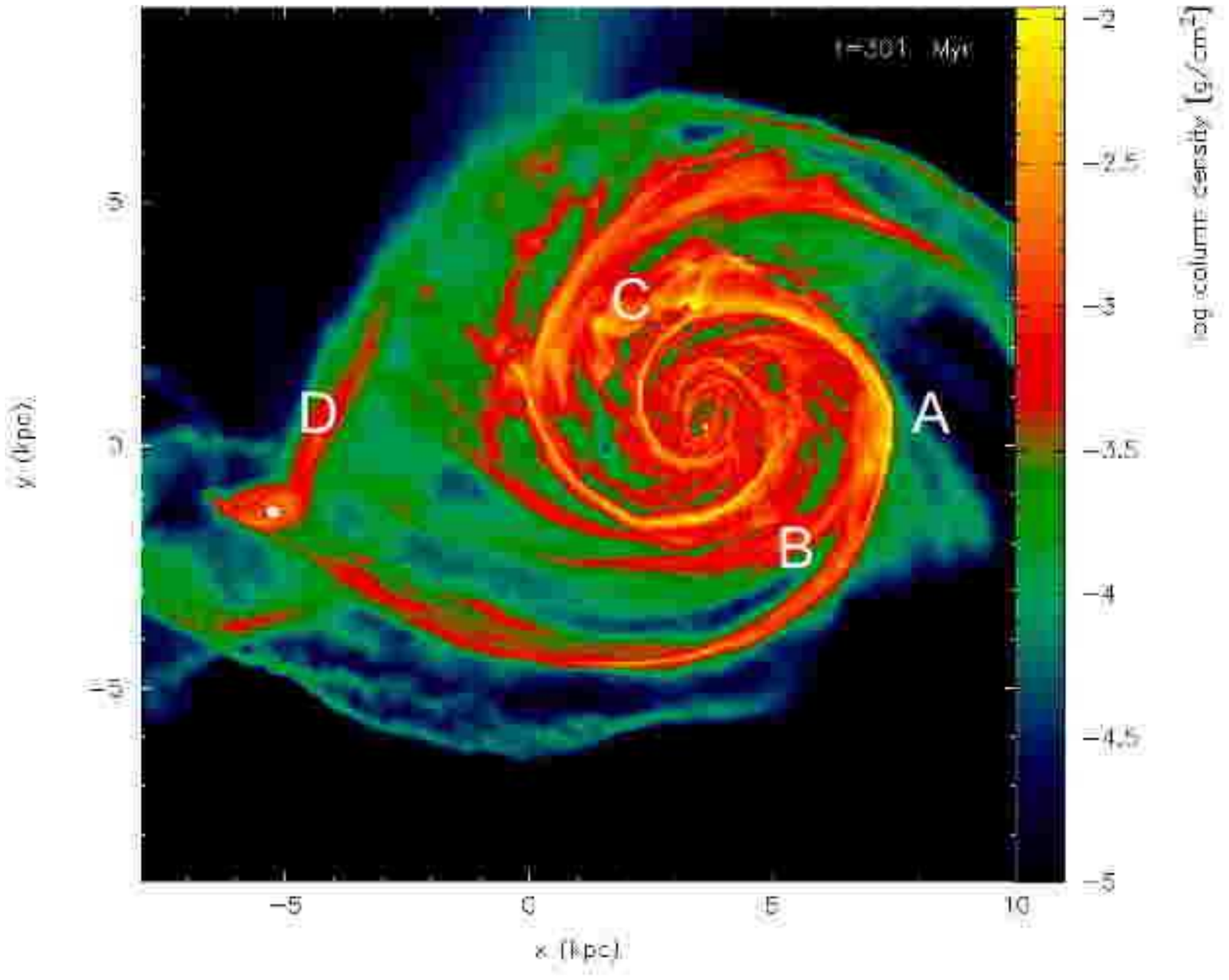}}
\vspace{10pt}
\centerline{
\includegraphics[scale=0.33,bb=0 -20 500 480]{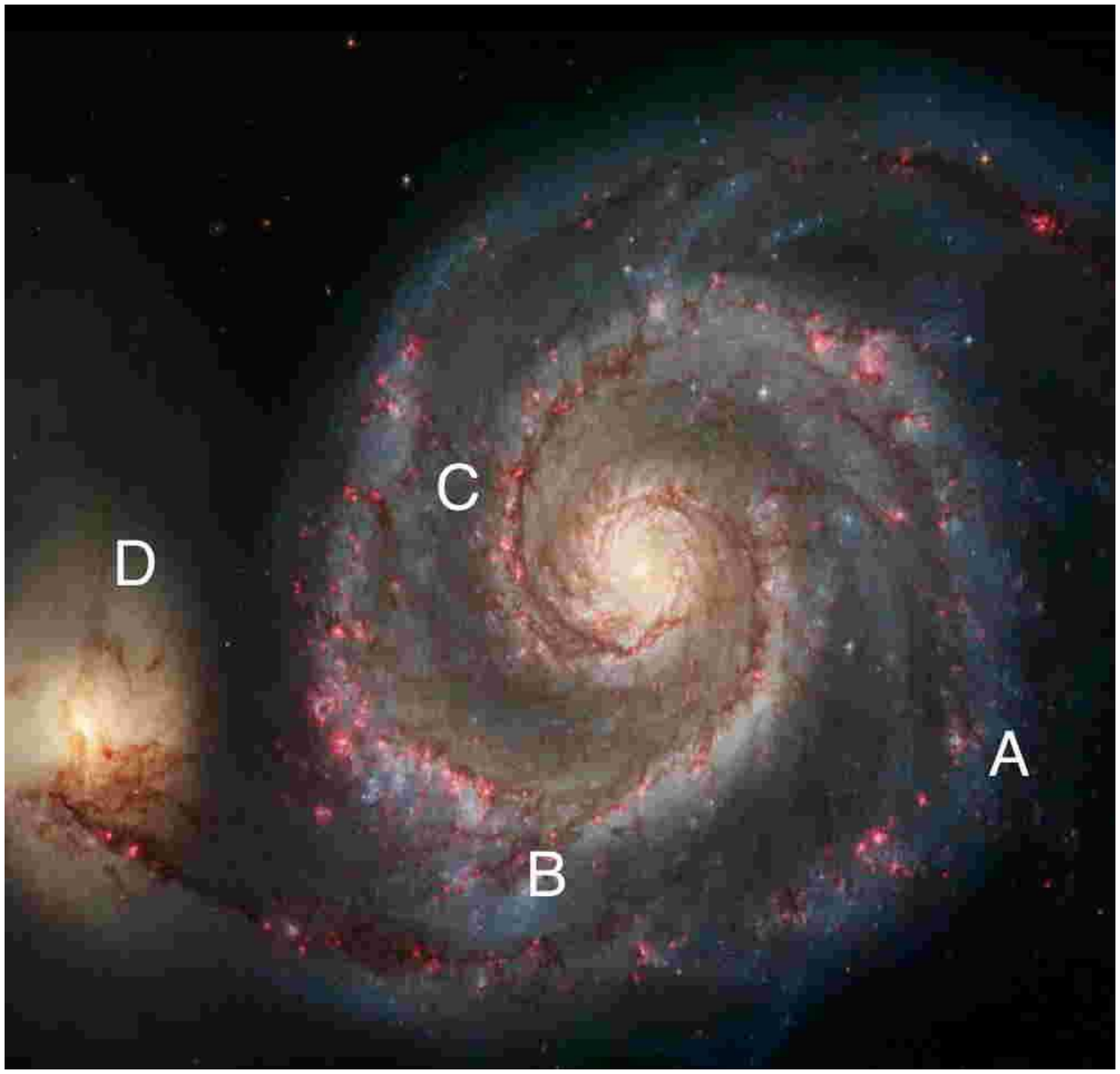}}
\caption{The top panel is a column density plot from our simulation
  with 1\% warm gas after 300 Myr, whilst the lower panel is a HST
  image of M51 produced by NASA, ESA, S. Beckwith (STScI), and The Hubble Heritage Team (STScI/AURA). The spiral structure
  e.g. shape and pitch angle of the spiral arms appears very similar
  to the actual M51. In addition we mark on several features (A - a
  kink in the spiral arm, B - a branch, C - interarm gas, D - a stream
  of gas extending from the companion) which appear in both the
  simulation and the observations. }
\end{figure}

\begin{figure}
\vspace{10pt}
\centerline{
\includegraphics[scale=0.22,bb=0 -20 850 650]{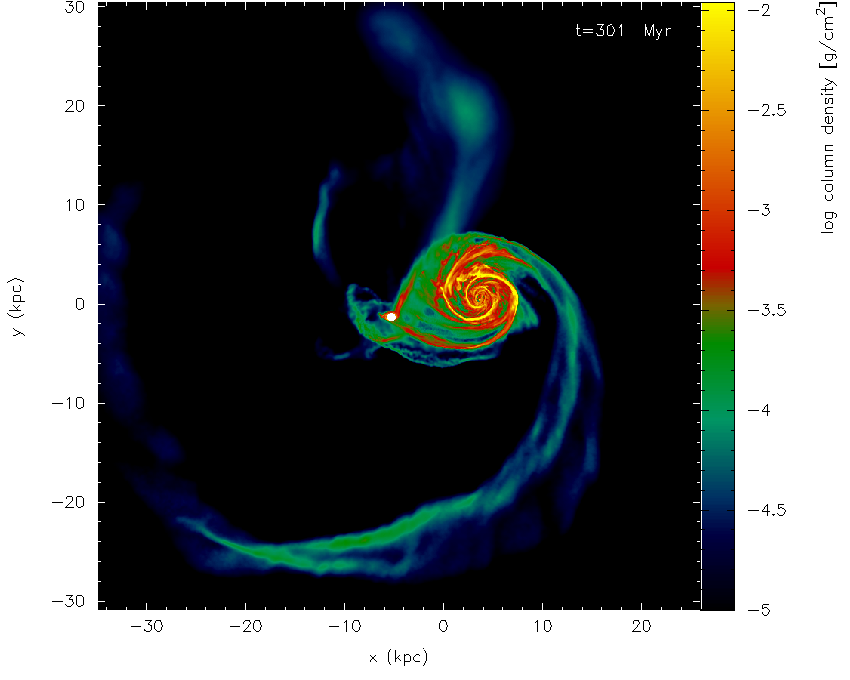}}
\centerline{
\includegraphics[scale=0.28,bb=0 -20 520 500]{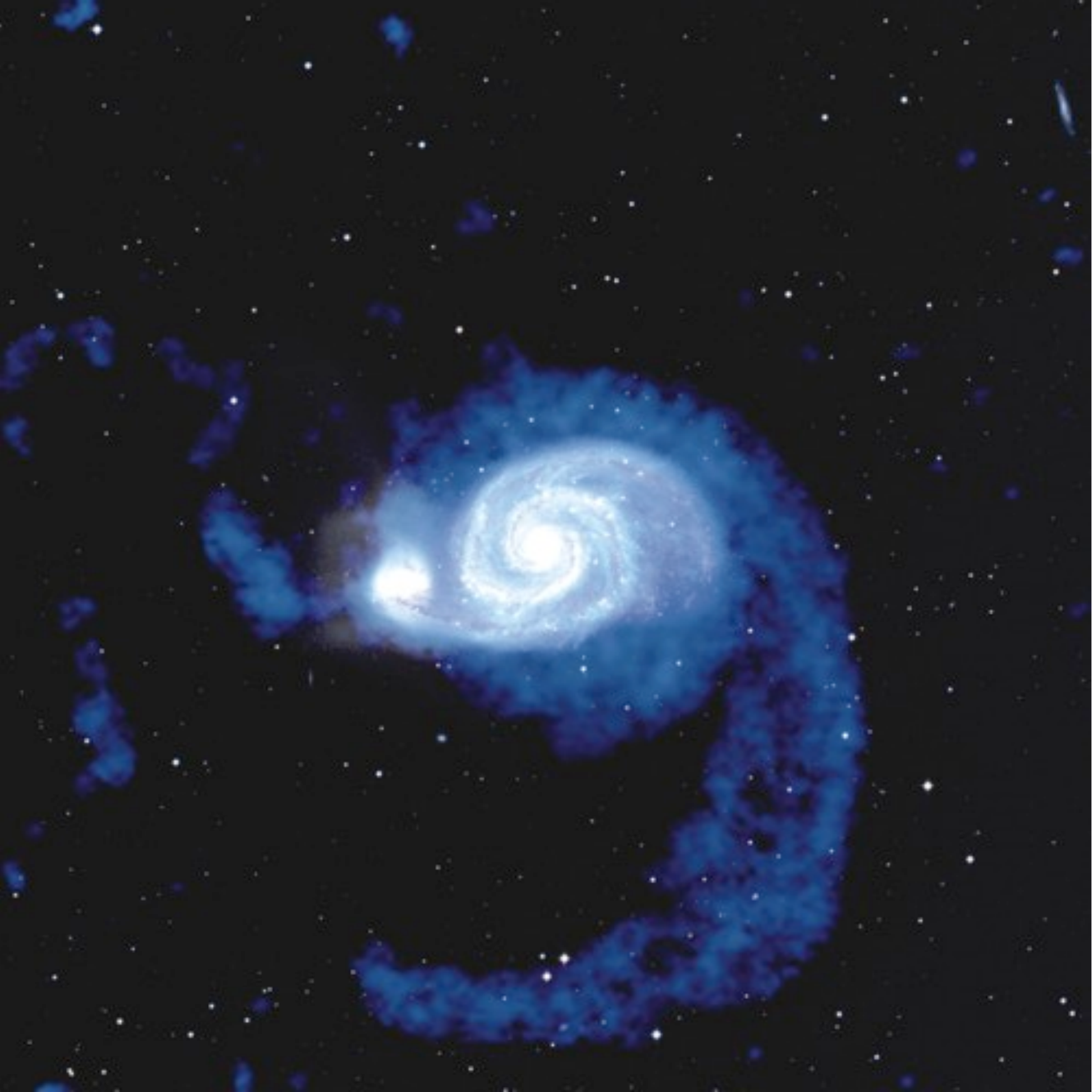}}
\caption{The top panel shows a column density plot from our fiducial
  simulation after 300 Myr but zoomed out to larger scales. There is a
  tidal tail of low density gas which roughly matches the HI
  observations by \citet{Rots1990}, lower panel. Image courtesy of NRAO/AUI/NSF.}
\end{figure}

\section{Isolated galaxy}

Before considering the results of the models where we have two
interacting galaxies, we first consider the case where we model an
isolated galaxy. This galaxy is essentially the same as that set up
for model A, except the galaxy is not inclined to the line of sight,
and is centred at Cartesian coordinates (0,0,0). We show in Fig.~3 the
column density of the gaseous component after 300 Myr. The structure
at this time is typical of the structure present at any time after 100 Myr
- there are multiple (relatively long) gaseous spiral arms, but no
obvious spiral mode. The combined stellar+gaseous disc is evidently
susceptible to gravitational perturbations (as also indicated by the
low $Q_{\rm s}$ parameter in Fig.~2). However the gas tends to be locally
stable (since we chose such a low gas mass) to runaway gravitational
collapse, and only one sink particle forms over the length of the
simulation, 300 Myr. The overall appearance is one of a typical
flocculent galaxy. 

As described previously, we did not set the gas particles in 
vertical equilibrium initially. In the isolated case however, we find that the
gas settles in the $z$ direction within 30 Myr, with a typical scale height of 
around 200 pc. 
This is comfortably before the interaction occurs in our M51 models 
described next.

\section{M51 model}

We describe in this Section the structure of the disc in
simulations of M51 and its orbit with NGC~5195. We provide a more
detailed comparison between our results and the structure of M51 in
Section~4.1.1, a comparison between models A, B and C in Section~4.1.2, 
and a Fourier analysis of the spiral components in
Section~4.1.3. We also investigate any offset between the stellar and
gaseous arms in Section~4.2, and compute the pattern speed in
Section~4.3. Finally in Section~5 we make comparisons between our
results and the observations of \citet{Shetty2007}.
\subsection{Structure of the disc}

We show the evolution of the gas disc of model A (with 1\% warm gas)
in Fig.~4, where the interacting galaxy NGC~5195 is visible as a white
dot. After 60 Myr (first panel), the interaction is not far advanced,
and the structure resembles the flocculent spiral structure seen in
the isolated case (Fig.~3). By 120 Myr (second panel), we start to see
a two-armed spiral pattern develop at larger radii ($R >$ 5 kpc).  At
later times, the two-armed structure extends to much smaller ($R \sim 2$
kpc) radii. The spiral pattern evidently changes with time, but it
evolves more slowly between 240 and 300 Myr.  During the simulation, the
companion galaxy becomes increasingly bound to the M51 galaxy, and at
the latest time (370 Myr), the two galaxies are beginning to merge.

The time of 300 Myr corresponds approximately with the present day in
\citet{Theis2003}. The orbit in this simulation deviates from the one
obtained by \citet{Theis2003} because our active halo permits
dynamical friction and and therefore the model does not exactly
reproduce the current positions of M51 and NGC~5195. Nevertheless the
tightly wound spiral arms and overall morphology strongly resemble
that of M51, as we will discuss in Section~4.1.1.

Although not shown in Fig.~4, to avoid confusion with the companion
galaxy, the gas disc also contains sink particles. These have been
inserted at regions which acquire sufficiently high densities that
would otherwise halt the calculation. However for model A, the disc is
still relatively stable and only 5 such particles form over the course
of the calculation.

Interestingly, there appears to be substructure in the gas at all
stages of the interaction. Initially the substructure is flocculent,
with many shorter segments of spiral arms, due to gravitational
instabilities primarily affecting the gas \citep{Toomre1964}. By 180
Myr (top right panel), the galaxy contains a dominant 2--armed spiral
pattern. Some remaining flocculent structure is seen towards the top
right of the panel ($x=5$ kpc, $y=5$ kpc), which has not yet passed
through the tidal arms. However there are also bifurcations of the
spiral arms, spurs and material shearing away from the spiral arms,
which we discuss further in later parts of the paper.

\subsubsection{Detailed analysis of structure and comparison with M51}

In Fig.~5 we show a snapshot from our simulation, at 300 Myr roughly
corresponding to the present day, accompanied by an HST image of M51. 
The figures show excellent agreement between the simulations and observations, 
in particular the overall shapes of the arms, and their pitch angles. 

We note several specific features seen in the simulation, and M51,
which are labelled on Fig.~5. Firstly there is a clear kink in the
spiral arm (marked A) in M51, and remarkably, the same feature appears
in our model. The kink is slightly further anticlockwise compared to
the actual M51, suggesting our model has evolved slightly too far,
presumably because the density structure, and hence the rotation curve we
have assumed for M51 is not quite right. This kink appears to arise
where the inner spiral arm, induced from the previous crossing of the
companion, meets a new spiral arm induced by the current passage of
NGC 5195. The feature marked B indicates a second spiral arm breaking
away from the main arm. This occurs in both model and the actual M51,
although again in our model the arm has rotated further round the
galaxy. We also see a large degree of gas between the spiral arms, in
particular indicated at C, and again this is typical for much of
M51.\footnote{At 180 Myr (Figure~4), there is also a stream of gas
  shearing away from the spiral arm. Thus the spiral arm appears to
  bifurcate moving radially \textit{inwards} along the spiral arm (at
  coordinates (-10,-7) kpc). Although not seen in M51, this is very
  similar to some of the structure seen in M81
  \citep{Kendall2008}. Identifications of bifurcations with pattern
  speed resonances (e.g. Elmegreen et al., 1989) generally require the
  bifurcation to occur as one proceeds radially \textit{outwards},
  along the arm. We see both in our simulations.}

A feature which is not seen in the Hubble image of M51, but is present
in our model, is a large stream of gas above the uppermost spiral arm
(i.e. above C). Interestingly though, we do see gas in front of the
companion galaxy (marked D). Possibly this gas extends further, and
would be visible in HI (see Fig.~6), or alternatively the gas disc in
our model is initially too extended.

In Fig.~6, we show a much larger scale image from our simulation at
the same time of 300 Myr, and below an HI image from
\citet{Rots1990}. We capture the HI tail of M51 very well. The width of the tail increases up
to the corner of the tail, before diminishing. There is also gas far
round the tail, i.e. at negative $x$ values in our coordinates, which
is apparent in the HI image. The entirety of the gas tail has a stellar counterpart. 
This is as expected since we base our models on the calculations of \citet{Theis2003}, 
who compared the stellar distribution of their results with \citet{Rots1990}.

The gas feature at (0,20) kpc in our
coordinates is not seen in the observations. We find that this gas has
been slung out of M51 by a close passage with the companion. Possibly
there is gas in this region which is not detected. Alternatively if we
properly modelled the companion, rather than using a point mass, the gas may
collide with gas internal to the companion and experience less acceleration, 
or even become accreted by the companion.

From Fig.~5, our simulated galaxy appears slightly stretched in the $y$ direction compared to the actual M51. Although we use the observed position and inclination angles of M51 in our initial conditions, these change over the course of the simulation, and the galaxy becomes less tilted on the plane of the sky. Thus at 300 Myr, the galaxy appears more face on than M51. This could suggest that we would need to start with a slightly different orientation to that currently observed for M51, or again reflect a simplification of our model (e.g. representing NGC 5195 by a point mass).

We show the detailed evolution of the disc from 280 to 320 Myr in
Fig.~7. The structure of the disc, and the spiral arms, are evidently
changing over timescales of $<$ 10 Myr. Interestingly we see that
\textit{the present apparent position of NGC 5195 at the end of one of
  the spiral arms is merely a coincidence} - 10 Myr earlier the
companion was nowhere near the spiral arm. In the direction along the
line of sight (right hand plots), the companion galaxy is passing
through the plane of the disc, from behind to in front of M51. We can
also see the evolution of the large branch in the spiral arm, marked B
in Fig.~5. The branch is due to material shearing away from the spiral
arm, and its presence and location change with time. Thus unlike
\citet{Elmegreen1989c}, who suggest branches in M51 lie at the 4:1
ultraharmonic resonance of some global pattern speed, we instead
suggest they are temporary features, due to the chaotic dynamics, and
evolve from material sheared from the spiral arms. We discuss the
formation of branches in more detail in Section~4.1.4.

Finally, it has been noted in some observations that there may be 
an oval distortion in the inner region of M51 \citep{Pierce1986, Tosaki1991,
Zaritsky1993}. An oval region is apparent in our simulations at $r\approx1$ 
kpc, with aspect ratio 2:1.

\subsubsection{The structure of the disc with a higher fraction of
  warm gas, and with cold gas}

We focus our analysis on model A, presented in Fig.~4 since this was
the most stable, with less fragmentation and few sink particles
forming. However before discussing this model in more detail, we first
describe the structure in models B and C.

In addition to our main model, we also ran calculations with 10\%
warm gas (model B) and 0.1\% cold gas (model C). The model with 10\%
warm gas represents a more realistic gas fraction for the
galaxy. However both models B and C were considerably less stable, and
formed many more sink particles. Consequently it was not possible (or
appropriate) to run these calculations for as long. We show the
structure of the disc at 180 Myr for these two simulations, as well as
the case with 1\% warm gas (model A) in Fig.~8. The structure for
both models with warm ($10^4$ K) gas is very similar (left and middle
panels). This is perhaps not that surprising as $Q_{\rm g}$ for both models
is fairly high. There are some small differences -- the spurs are more
compact, presumably because the gas is less gravitationally stable
(and in fact unlike model A, the spurs tend to contain sink
particles). There is generally more local collapse in the model with
10\% warm gas and consequently 32 sink particles have formed at this
point. The final panel in Fig.~8 (right) shows the disc with 0.1\%
100 K gas. As expected, there is much more fragmentation than in the
models with warm gas. Again widespread collapse occurs in the disc,
and after 180 Myr, 184 sink particles have formed.
\begin{figure*}
\centerline{
\includegraphics[bb=0 50 600 500, scale=0.28]{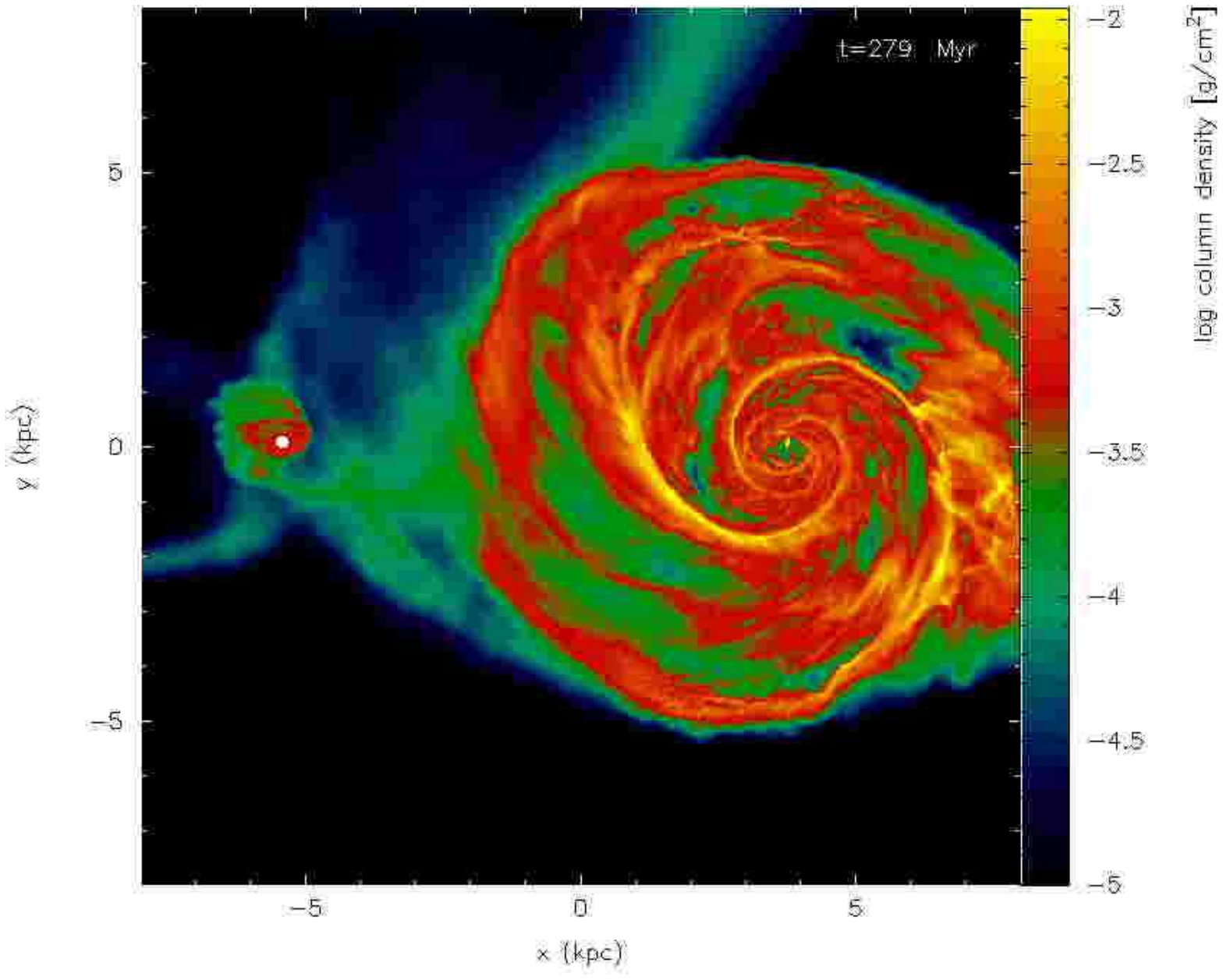}
\includegraphics[bb=0 50 600 500, scale=0.28]{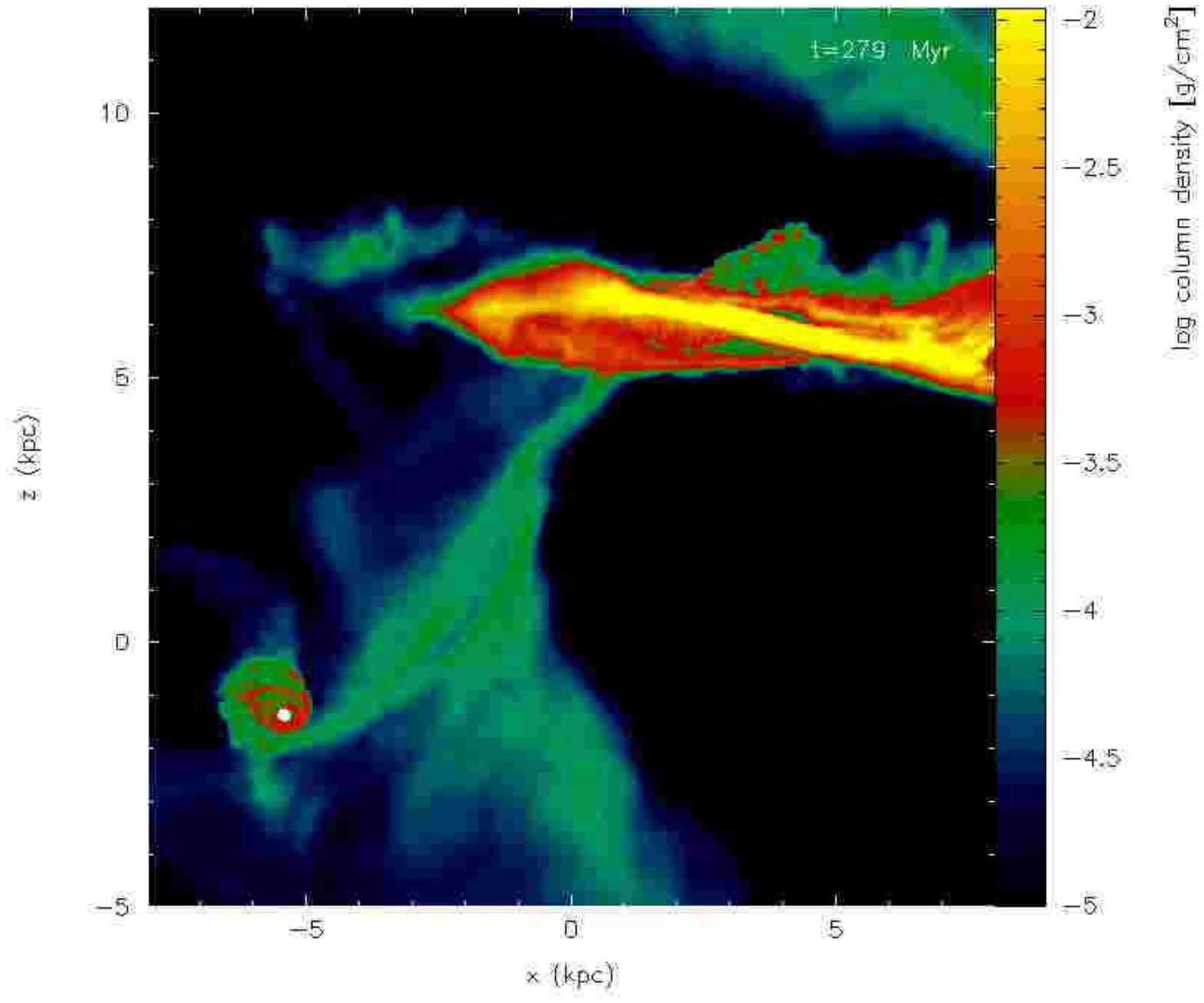}}
\centerline{
\includegraphics[bb=0 50 600 500, scale=0.28]{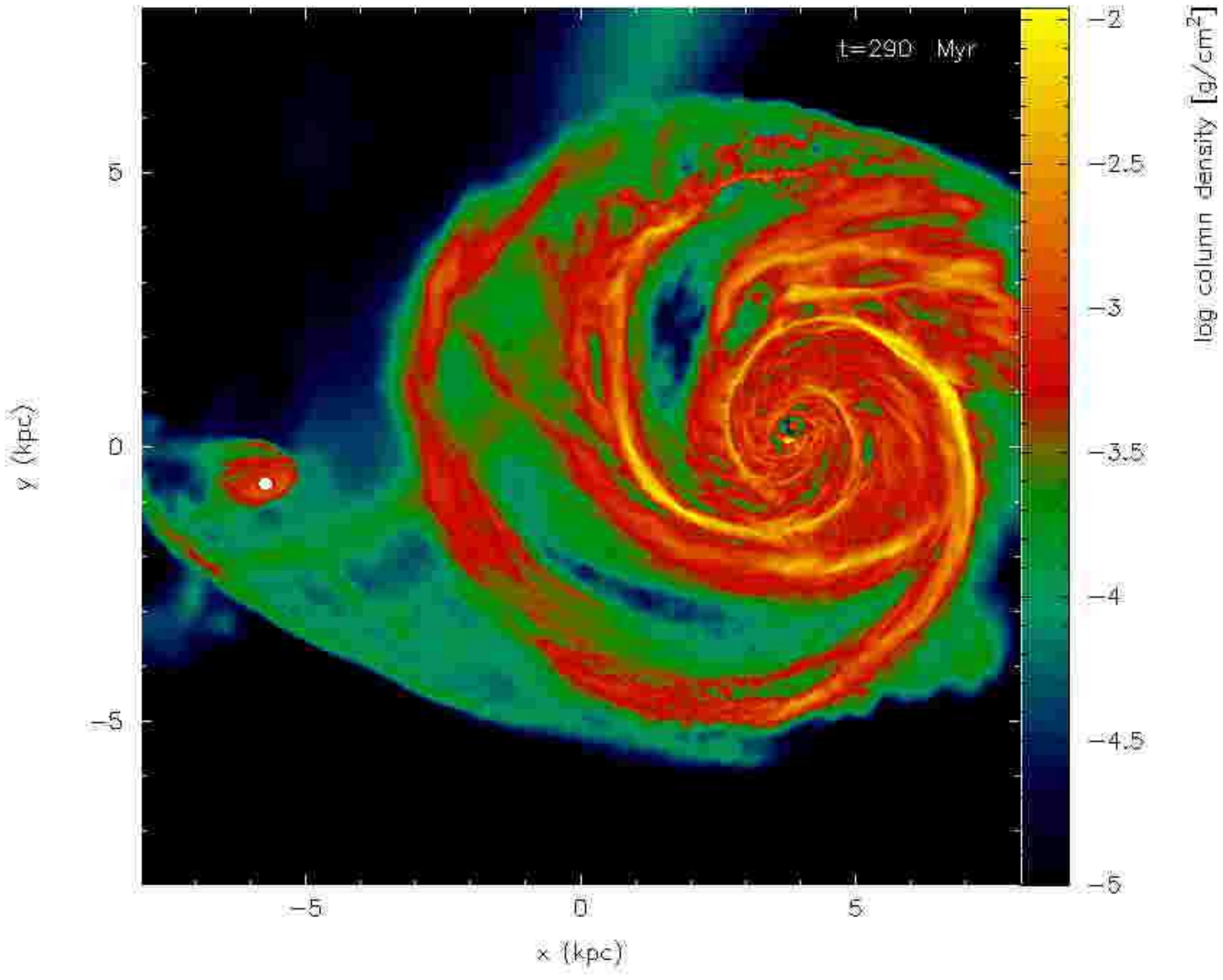}
\includegraphics[bb=0 50 600 500, scale=0.28]{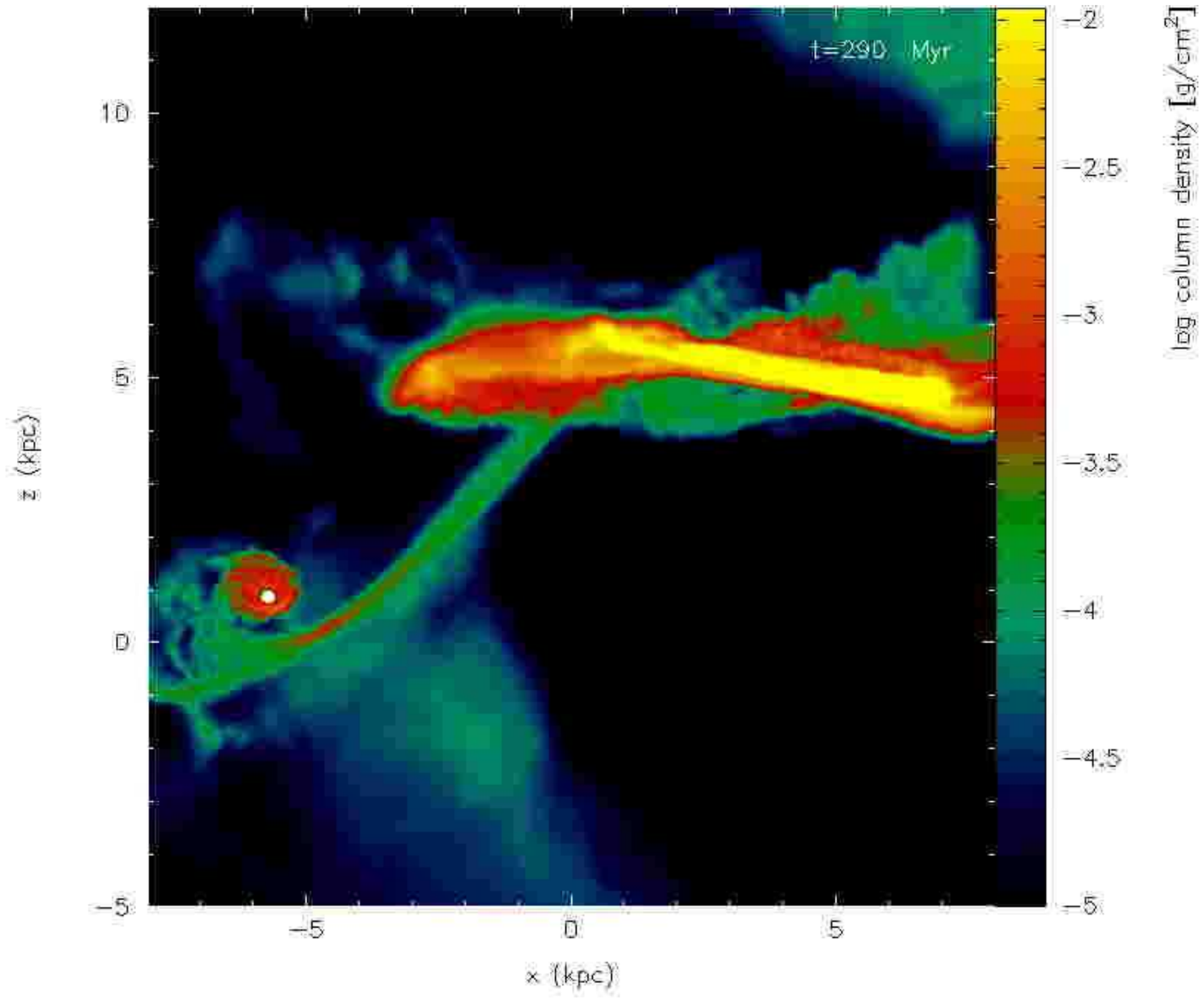}}
\centerline{
\includegraphics[bb=0 50 600 500, scale=0.28]{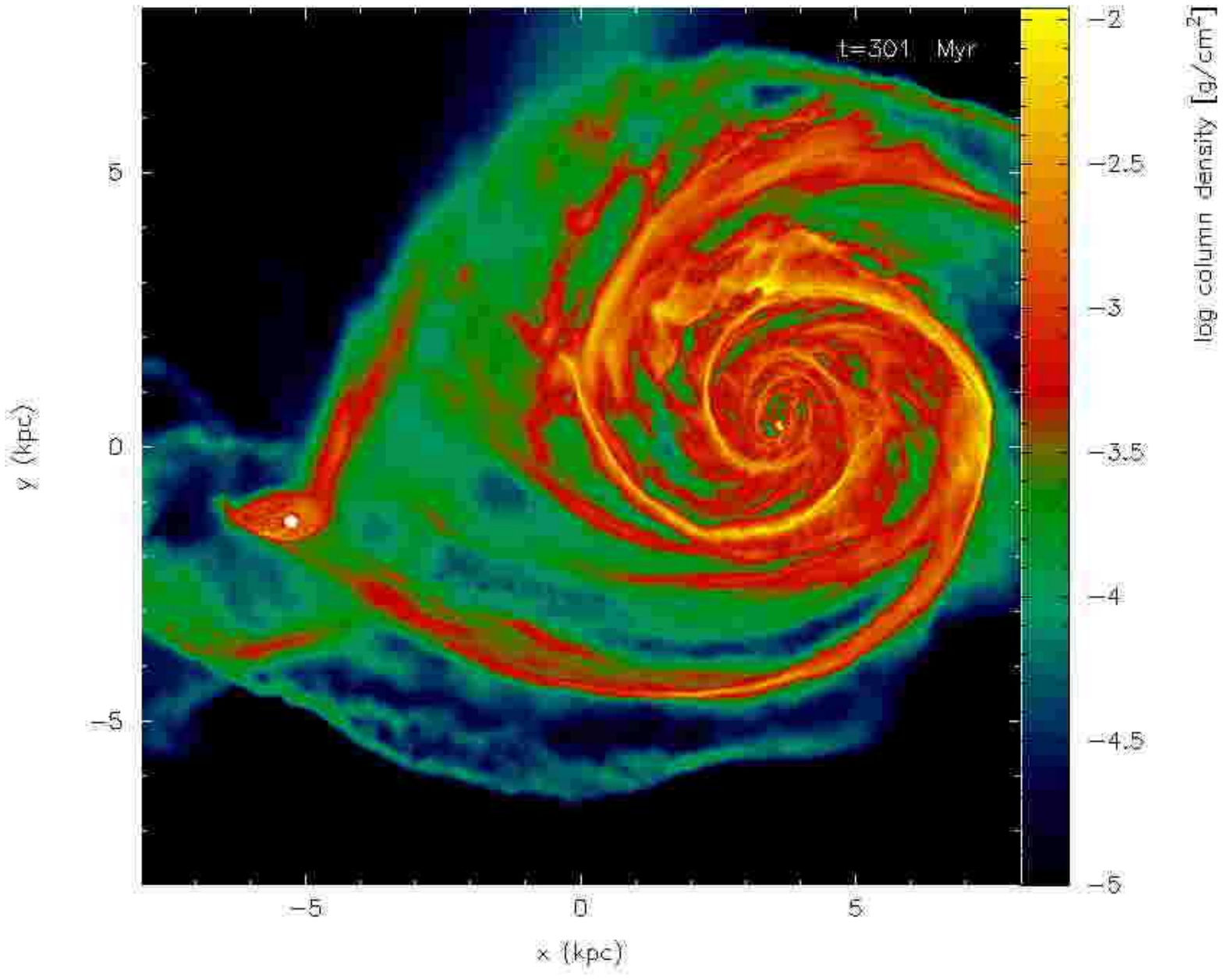}
\includegraphics[bb=0 50 600 500, scale=0.28]{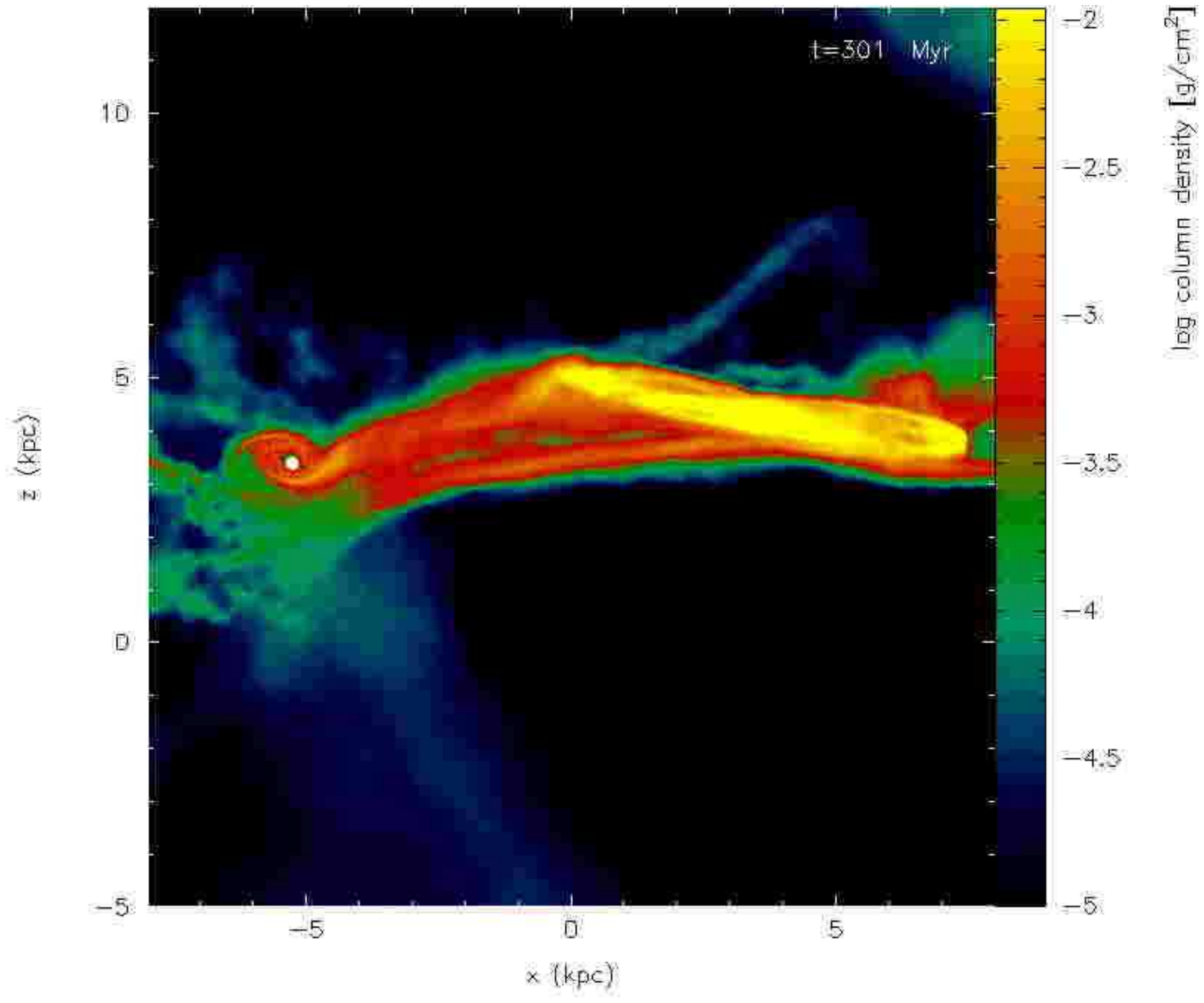}}
\centerline{
\includegraphics[bb=0 50 600 500, scale=0.28]{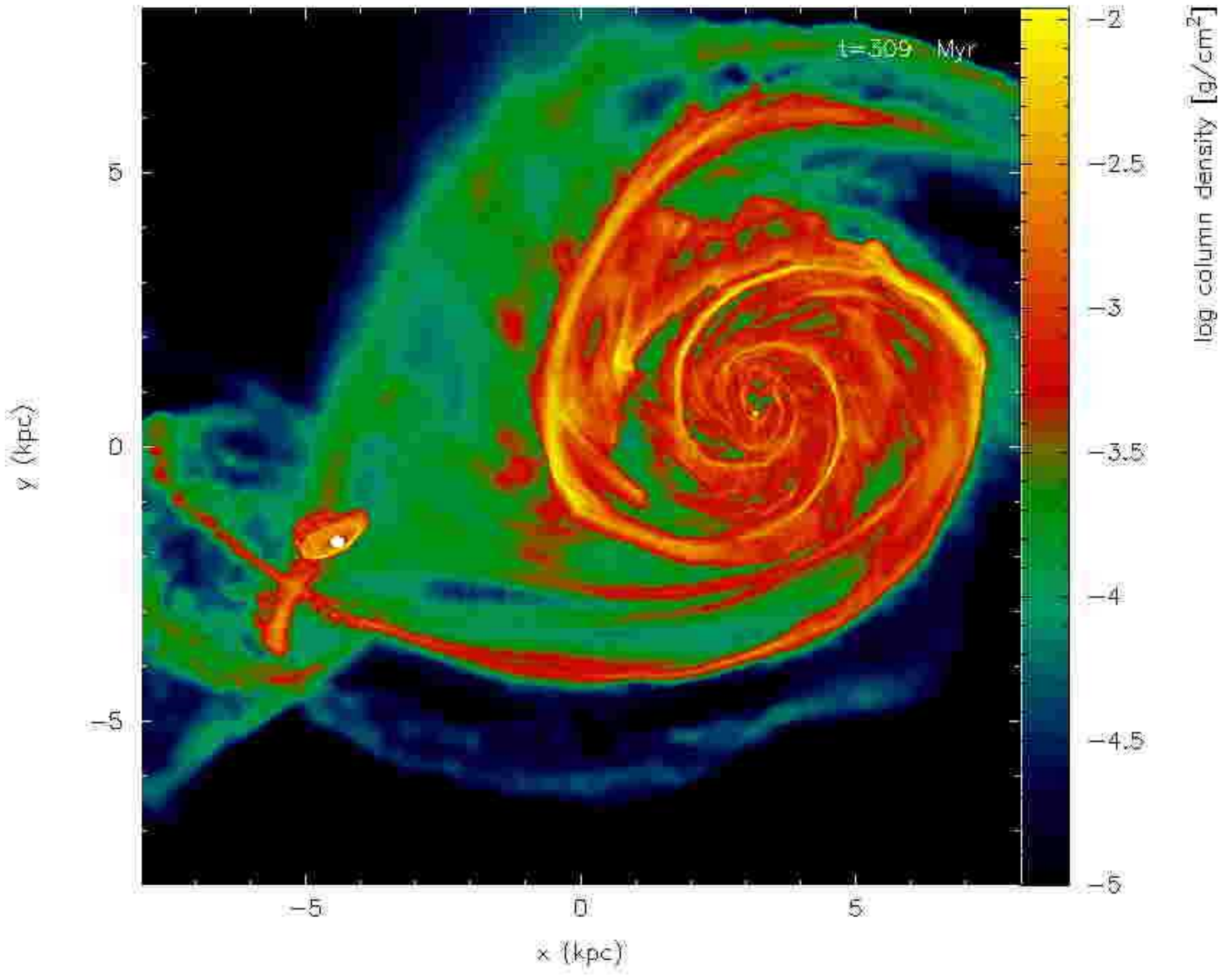}
\includegraphics[bb=0 50 600 500, scale=0.28]{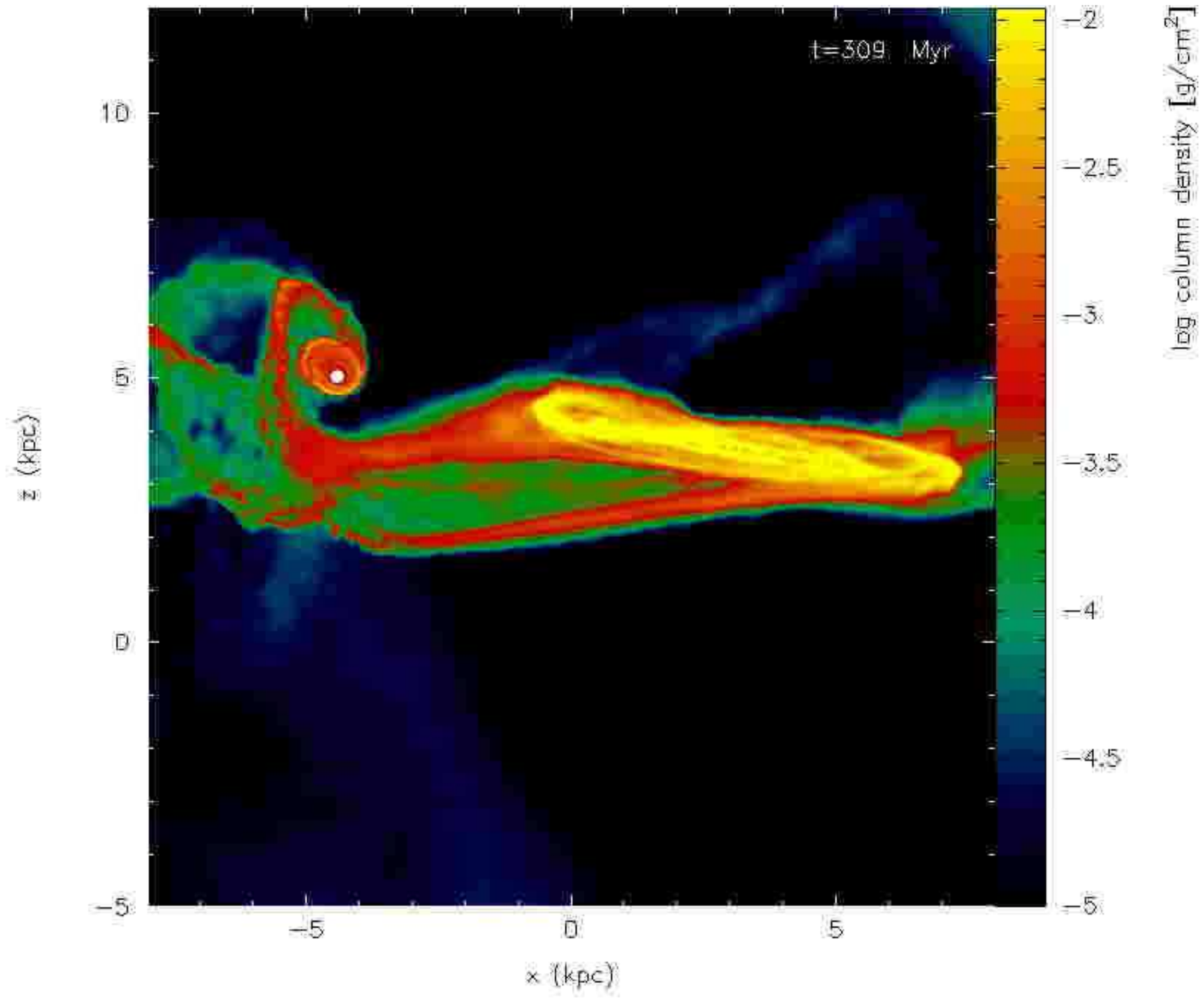}}
\centerline{
\includegraphics[bb=0 50 600 500, scale=0.28]{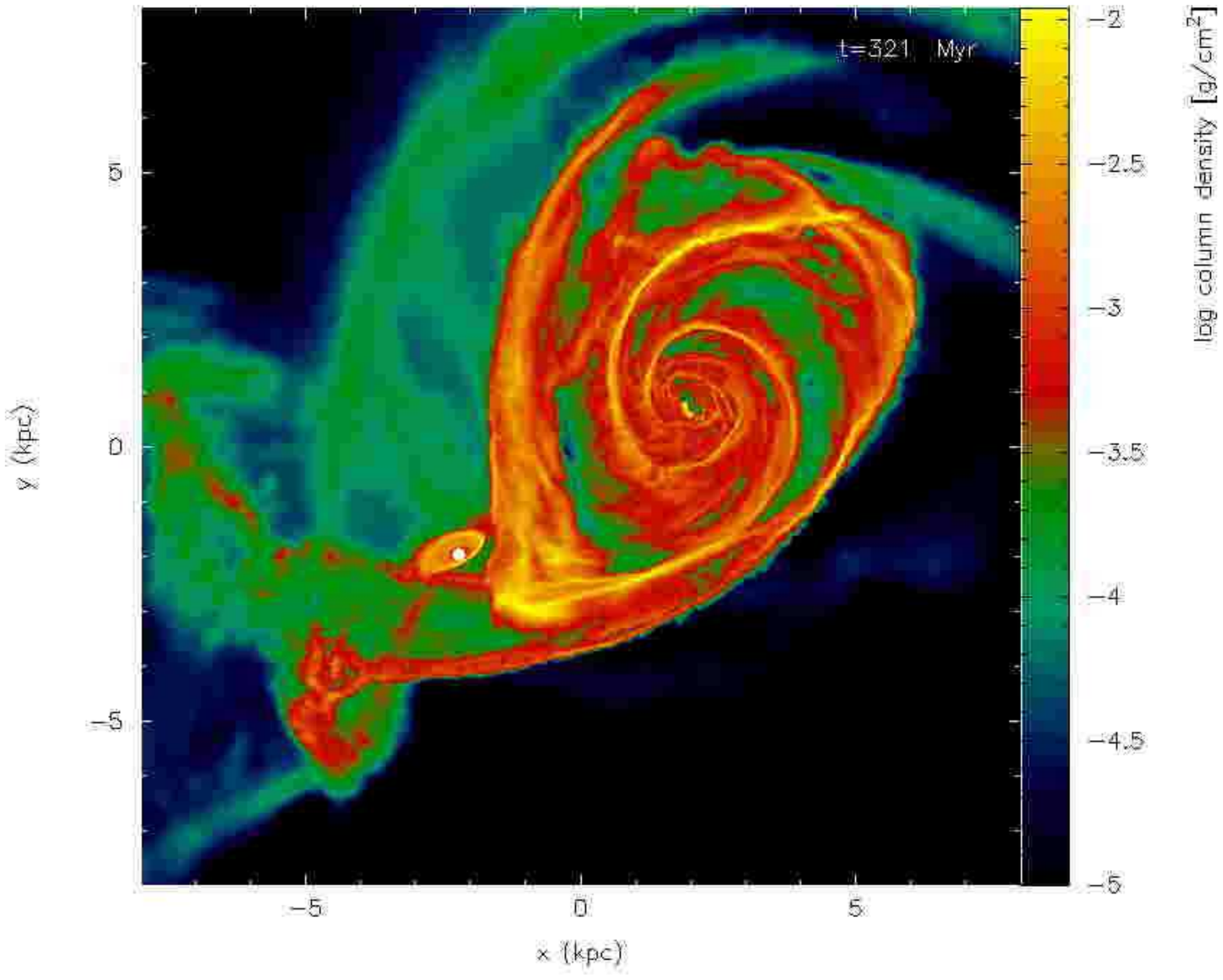}
\includegraphics[bb=0 50 600 500, scale=0.28]{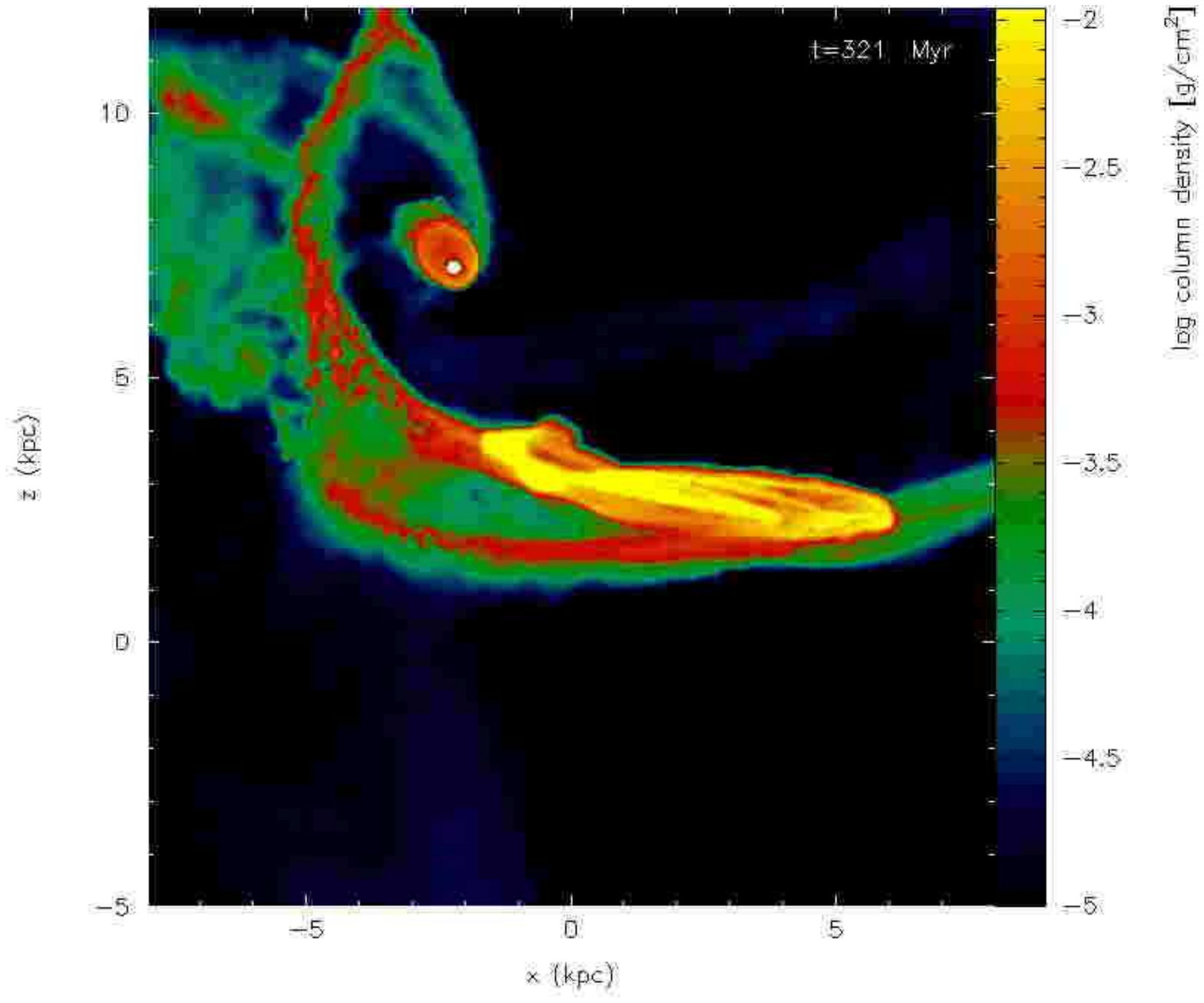}}
\caption{The column density of the gas disc is shown for model A (1\%
  warm gas) after 279, 290, 301, 309 and 320 Myr. The left hand panels
  show the galaxy face on (in the $xy$ plane in our coordinates)
  whilst the right hand panels show the galaxy along the line of sight
  (i.e. in the $xz$ plane) where the positive $z$ direction is towards
  us. The panels indicate that the current position of the companion
  apparently at the end of a spiral arm is a coincidence, whilst the
  left panels highlight the evolution of the spiral structure, and
  interarm structure over much shorter timescales compared to Fig.~4.}
\end{figure*}

\begin{figure*}
\centerline{
\includegraphics[scale=0.33, bb=0 -10 500 500, clip]{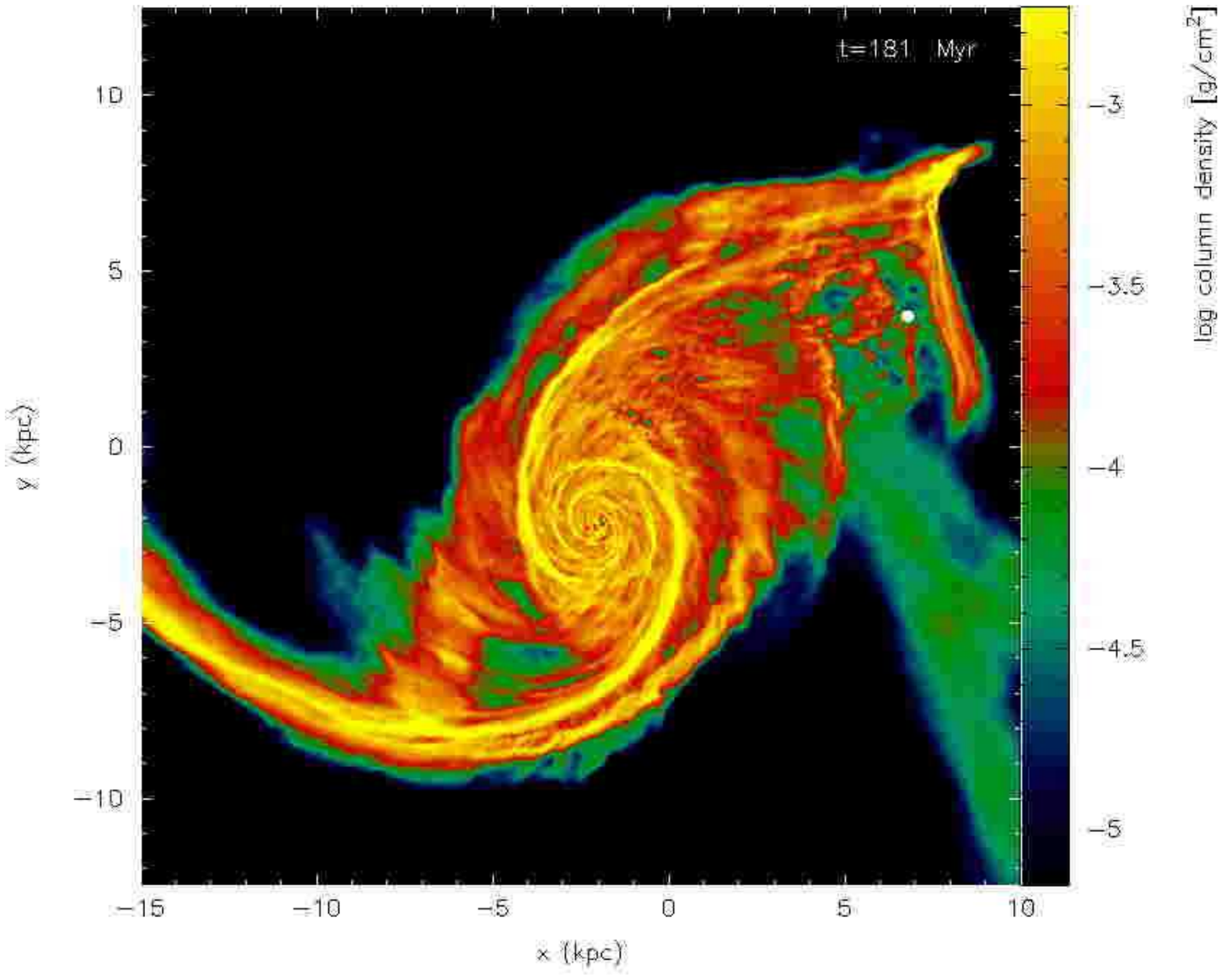}
\includegraphics[scale=0.34, bb=0 0 500 500, clip]{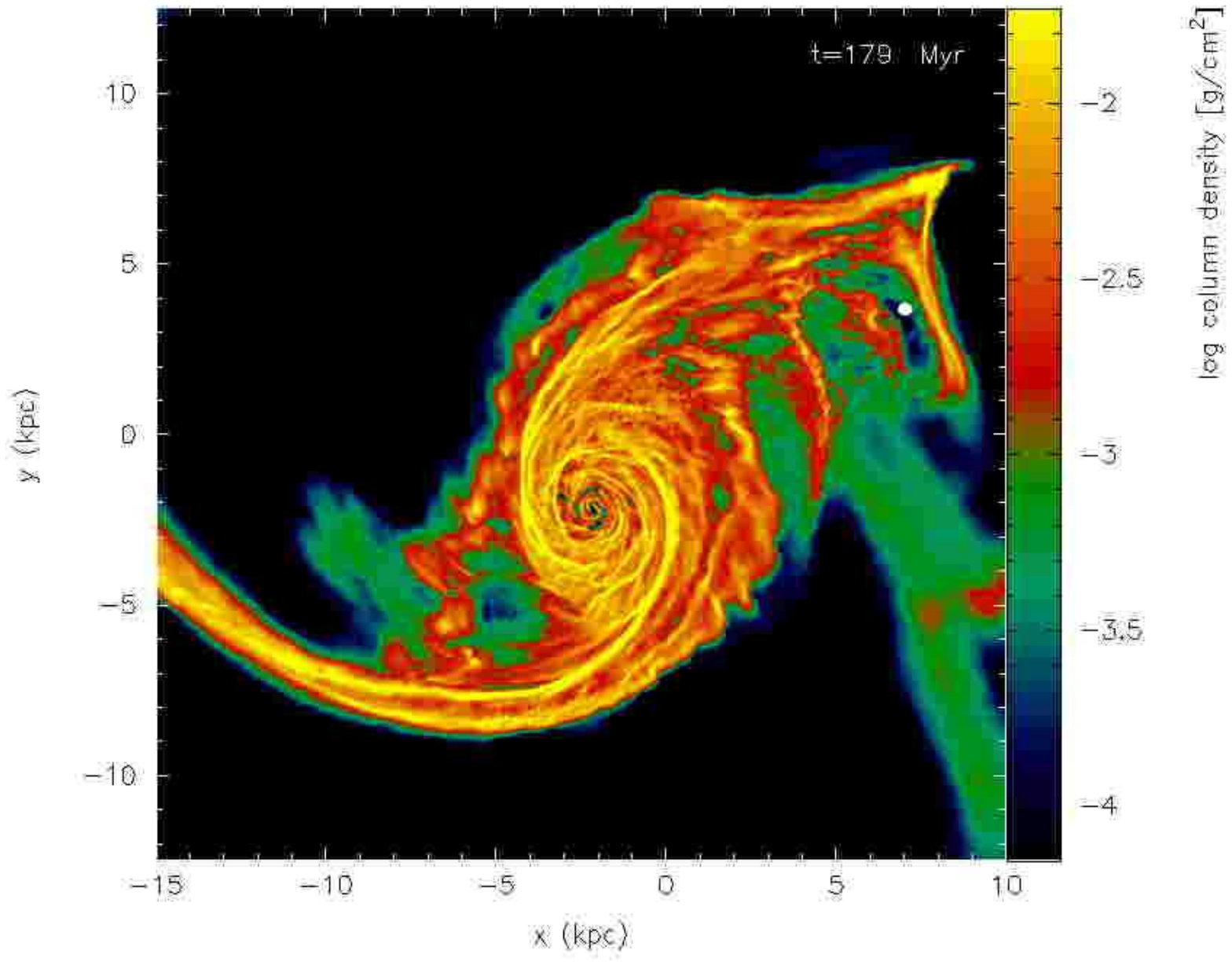}
\includegraphics[scale=0.327, bb=0 -10 500 500, clip]{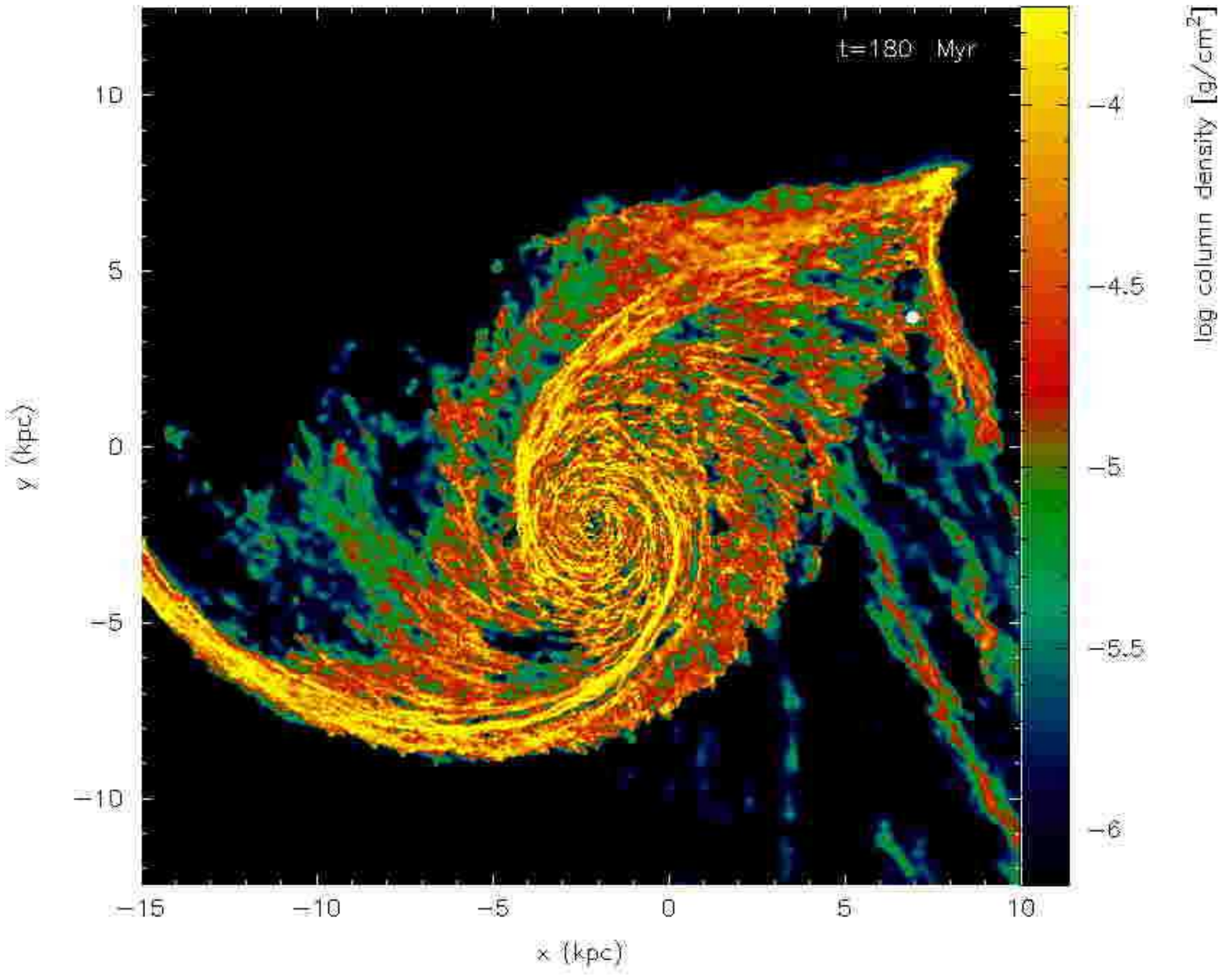}}
\caption{The column density of the gas disc is shown for the 3 models
  A, B and C. The figures show cases where 1 (left), 10 (middle) and
  0.1\% (right) of the disc is gas, by mass. The temperature of the
  gas is $10^4$ K in the left and middle panels, and 100 K in the
  right hand panel. The time of the simulation for each panel is 180
  Myr. Similar to Fig.~4 sink particles are not shown, only the point mass which represents 
  the galaxy NGC 5195. }
\end{figure*} 

\subsubsection{Evolution of spiral modes}
We consider specifically how the spiral structure of the disc changes
with time by computing the amplitude of the spiral modes.  We
calculate the Fourier amplitude of each spiral mode according to
\citet{Theis2004}:
\begin{equation}
  C_m=\frac{1}{M_{\rm disc}} \bigg|\int^{2 \pi}_{0} \int^{R_{\rm
      out}}_{R_{\rm in}} \Sigma(R,\theta) R \, dR \thinspace 
  e^{-im \theta} d\theta \bigg|,
\end{equation}
where $M_{\rm disc}$ is the mass of either the stellar or gas disc and
$\Sigma$ is the corresponding surface density. We calculate the
Fourier amplitudes over annuli of the disc, between $R_{\rm out}$ and
$R_{\rm in}$. Using cylindrical coordinates, $R$ is the distance from a point 
to the centre of mass of the galaxy, and $\theta$ is the angle which subtends to 
the centre of mass, measured anticlockwise round the galaxy. We did not perform 
any rotational transformations prior to
obtaining the Fourier transforms - the galaxy is still inclined with
respect to the line of sight.
\begin{figure}
\centerline{
\includegraphics[bb=0 380 650 770,scale=0.4]{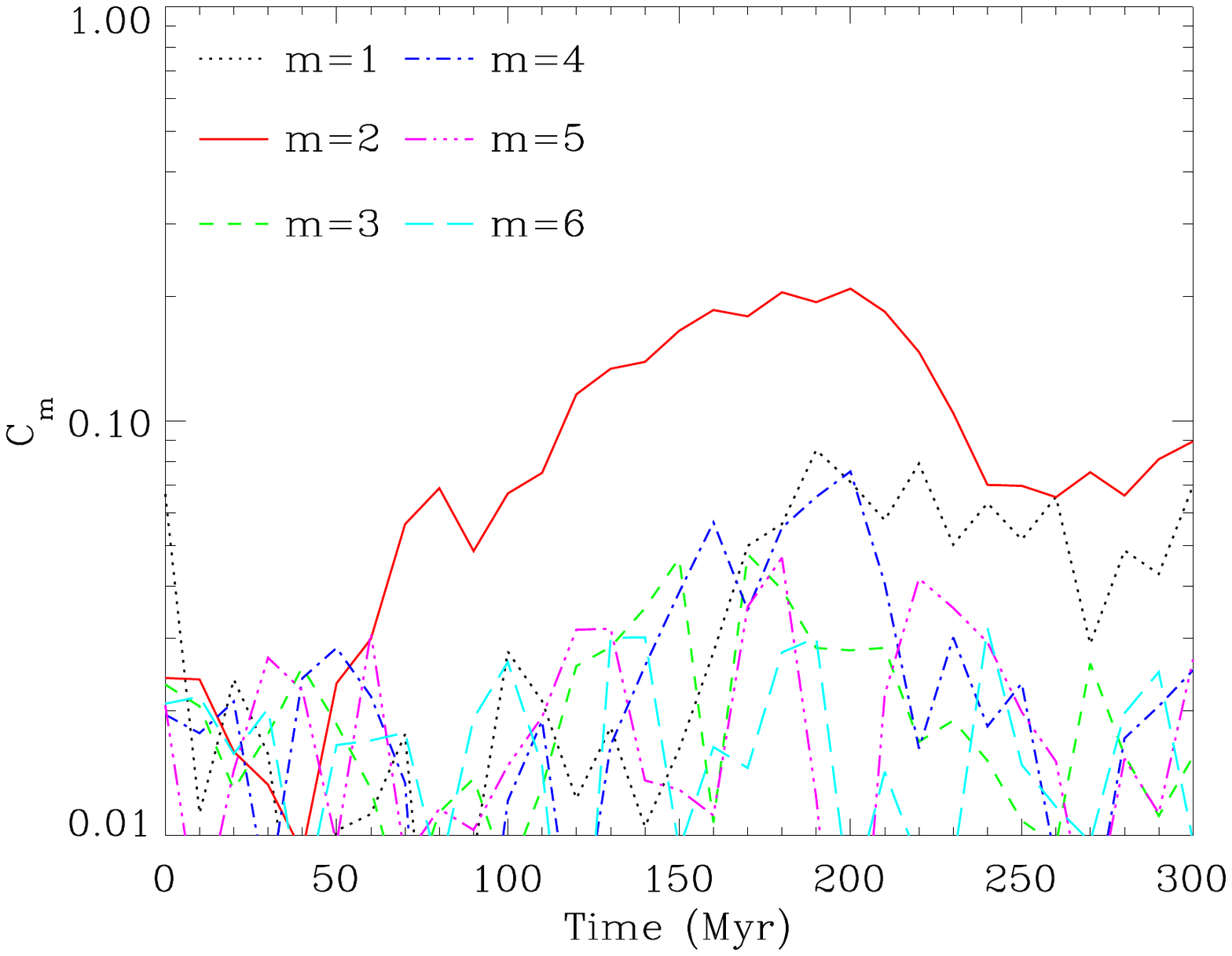}}
\centerline{
\includegraphics[bb=0 370 650 800,scale=0.4]{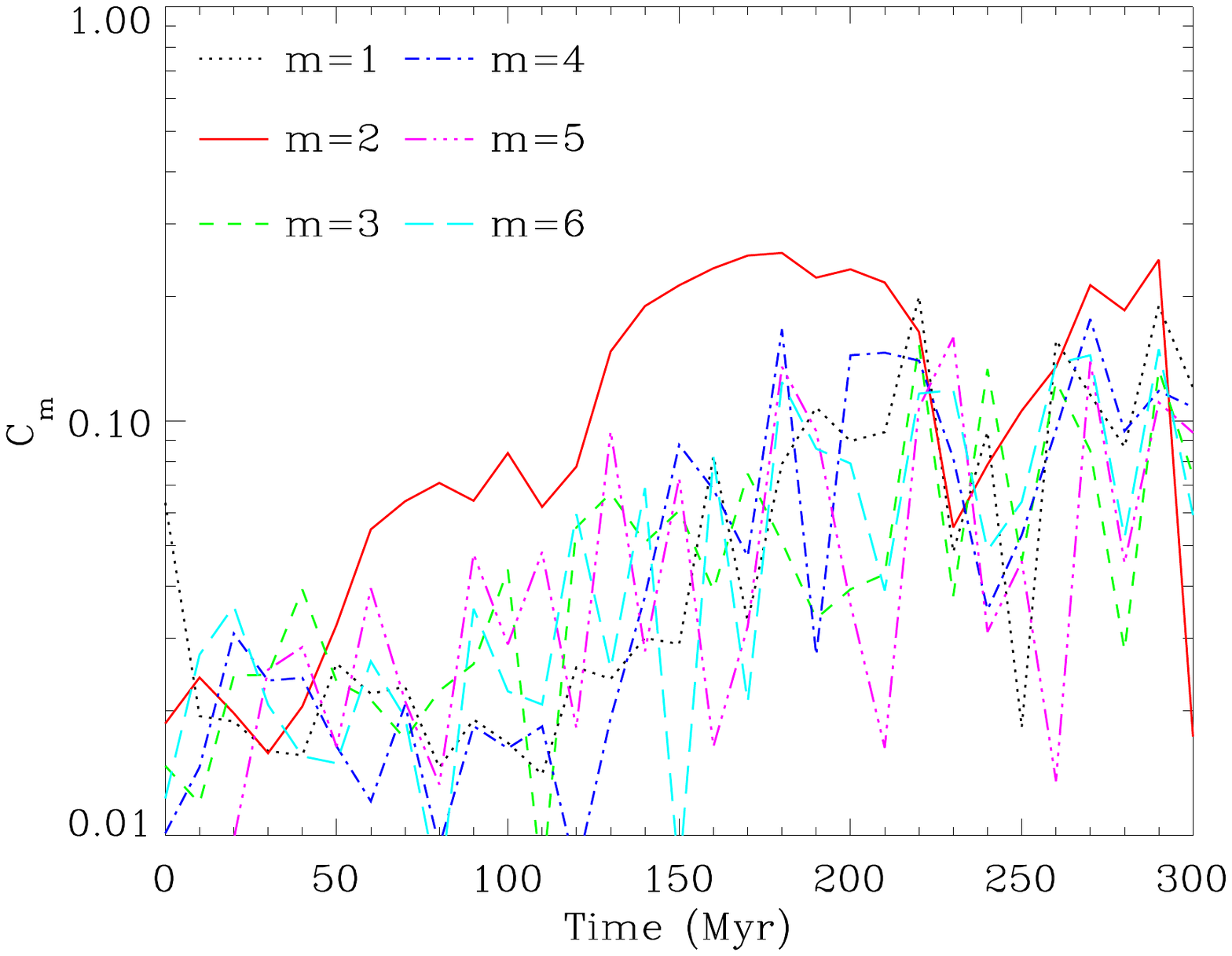}}
\caption{The spiral mode amplitudes are shown for the whole of the
  disc versus time. The top panel is for the stellar (disc) component
  and the lower for the gas component. The $m=2$ mode is generally
  highest, though surprisingly not substantially higher than the other
  modes, especially for the gas. The reason for this is evident from
  Figure~11, where it can be seen that although the azimuthal
  structure is predominantly double-peaked ($m=2$) it is sufficiently
  spiky, and the arms are sufficiently offset, a lot of power in
  higher values of $m$ is necessary to Fourier-decompose the azimuthal
  distributions.}
\end{figure} 

\begin{figure*}
\centerline{
\includegraphics[bb=200 370 500 770,scale=0.33]{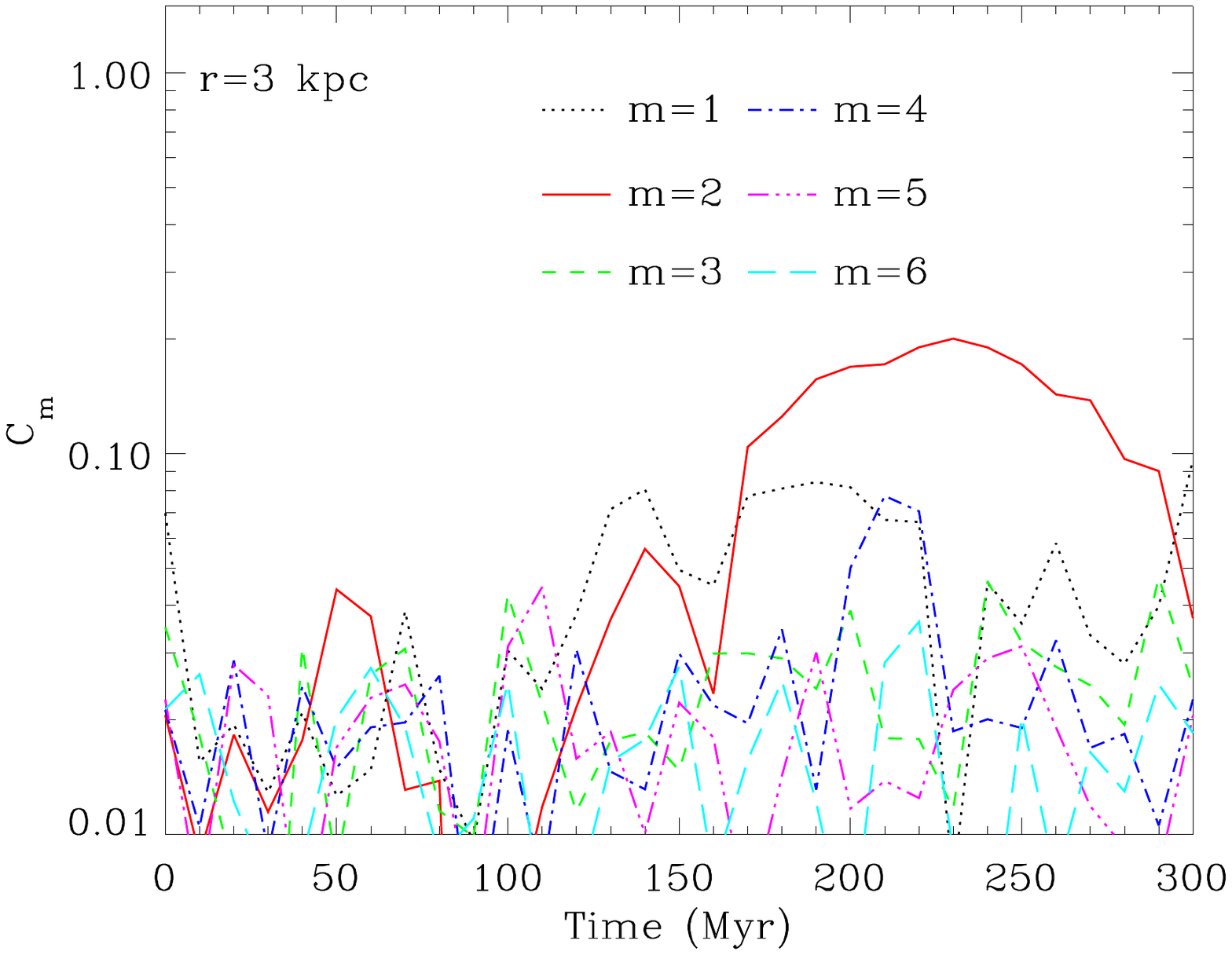}
\includegraphics[bb=0 370 500 800,scale=0.33]{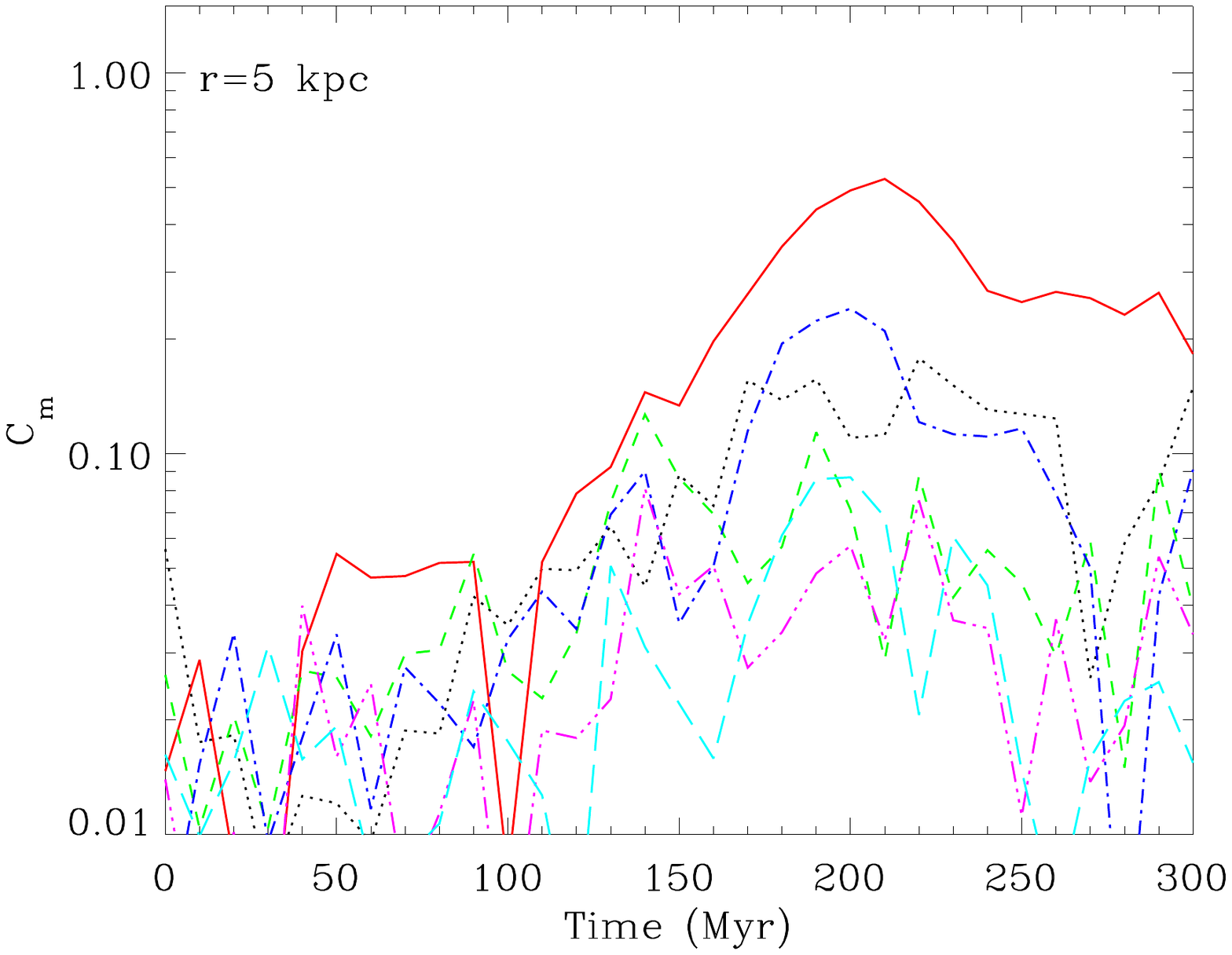}
\includegraphics[bb=0 370 500 800,scale=0.33]{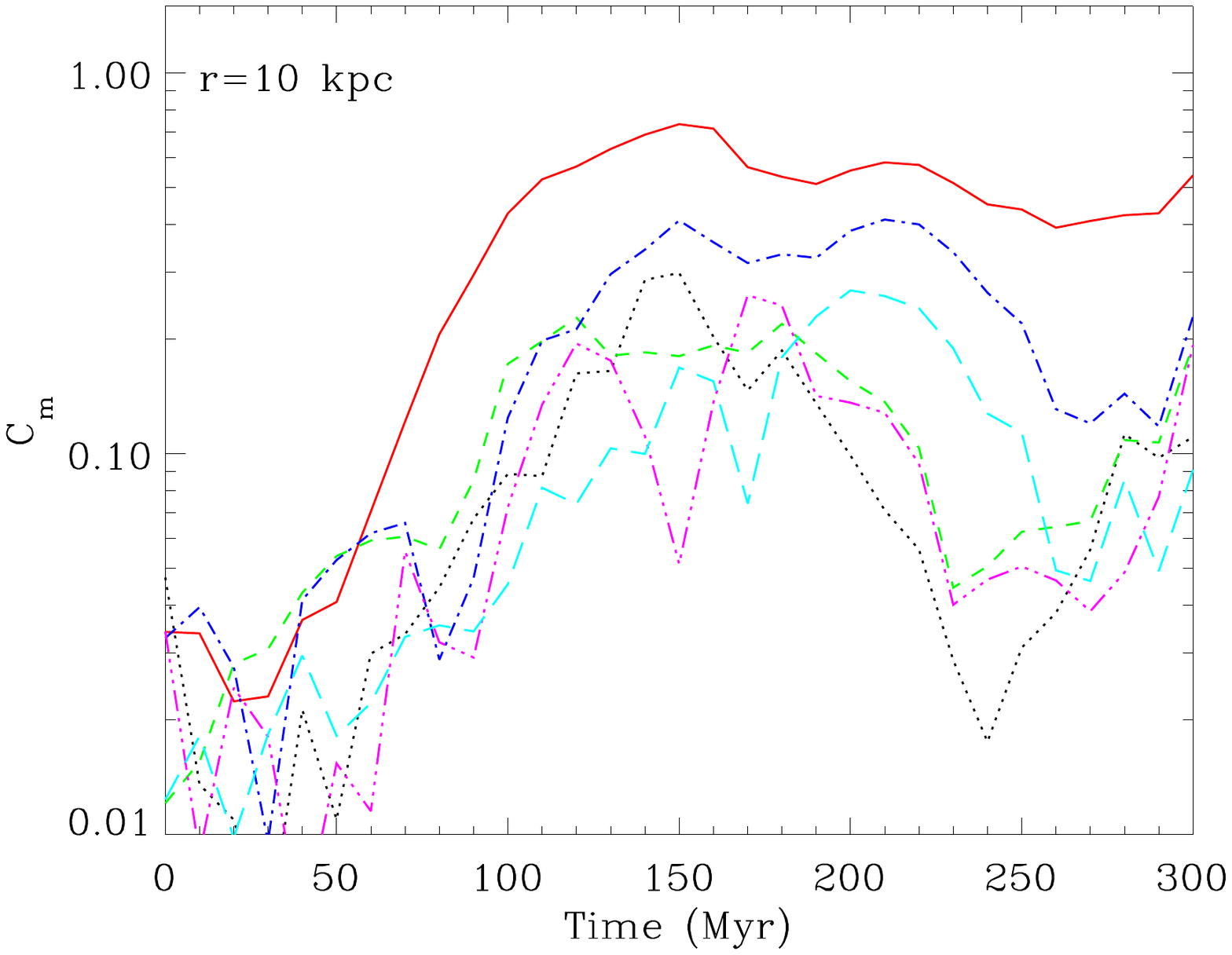}}
\caption{The spiral mode amplitudes are shown versus time at radii of
  3 (left), 5 (middle) and 10 (right) kpc for the stellar disc. There
  is an evident transition of the $m=2$ mode inwards through the
  disc At r=10 kpc, a two armed spiral emerges between 50 and 100
  Myr, but is not induced in the central 3 kpc until after 150 Myr. As
  well as the $m=2$ mode, all the Fourier amplitudes are seen to
  increase during the interaction. The spiral modes for the gas show
  similar behaviour, though as for Fig.~7, the $m=2$ mode is not quite
  as large relative to the other modes.}
\end{figure*} 

We show the amplitude of the modes versus time in Fig.~9 for the whole
of the disc (taking $R_{\rm in}=0$ kpc and $R_{\rm out}=10$ kpc). The
amplitudes of the modes increase during the interaction, and as
expected the $m=2$ mode is the dominant mode. However the $m=2$ mode
is not substantially higher than the other modes, particularly at
later times. For the gas, there is even less difference between the
$m=2$ and other modes. This comes about because the azimuthal
distribution of the gas density is very spiky (Figure~11).

In Fig.~10 we plot the amplitudes versus time at different radii in
the disc. Here we take annuli of width 2 kpc, about 3, 5 and 10
kpc. We only show the amplitudes for the stars, but the gas shows
similar behaviour. From this figure it is evident that the m=2 spiral 
structure takes longer to develop at smaller radii. This might be due 
to radial propagation, or it might be because the strength of the tidal 
forces decreases strongly with radius (note
that for the $R=10$ kpc plot, the Fourier amplitudes are not
particularly meaningful at later times since the companion is located
within 10 kpc). From viewing a movie of the stellar distribution,
there is no obvious break in the stellar arms with time. This suggests
that the tidal interaction is solely responsible for the inner
structure, in agreement with \citet{Toomre1977} and
\citet{Salo2000b}. For the gas however, the main spiral arms are not
always continuous, and it is possible (at earlier times) to have
shorter sections of spiral arms in the inner disc overlapping the main
tidally induced spiral arms.

The lack of a dominant $m=2$ mode is surprising, given that the column
density plots (Figs.~4 \& 5) show an obvious 2-armed spiral pattern,
both in the gas and stars. However this does appear to agree with
recent observations (Kendall et. al., in prep.), where the amplitude
of the $m=2$ mode is around 0.2-0.3, and the other modes $\lesssim
0.1$. \citet{Elmegreen1989c} found slightly higher amplitudes, between
0.1 and 1 for the $m=2$ mode, but still the amplitude of the $m=4$
mode is not significantly less than $m=2$ (and actually higher at
small radii).

In addition to the Fourier amplitudes, we show azimuthal profiles of
the density in Fig.~11. We determine the mass averaged 
volume density ($\rho$) over an annulus
of width 1 kpc, divided into 64 sections azimuthally.
The figure shows the density at times of
120, 180, 240 and 300 Myr. At 120 Myr, the interaction is still at an
early stage, so as expected there is no obvious 2-armed pattern,
though Fig.~11 and the column density plot (Fig.~4, top middle)
indicate that there is one prominent spiral arm in the gas, due to the
interaction. The density profiles generally show that the arms are
considerably more peaked than sinusoidal curves. In some cases
(e.g. $r=5$ kpc, at 180 Myr or $r=3$ kpc at 300 Myr) there is also a
considerable degree of substructure in the gas. Furthermore the main
arms may be of unequal densities (e.g. $r=5$ kpc, 240 Myr) and not
necessarily symmetric. As we remarked above, this explains the
relatively low amplitude of the $m=2$ mode in relation to the other
modes.

\begin{figure*}
\centerline{
\includegraphics[bb=200 370 500 770,scale=0.33]{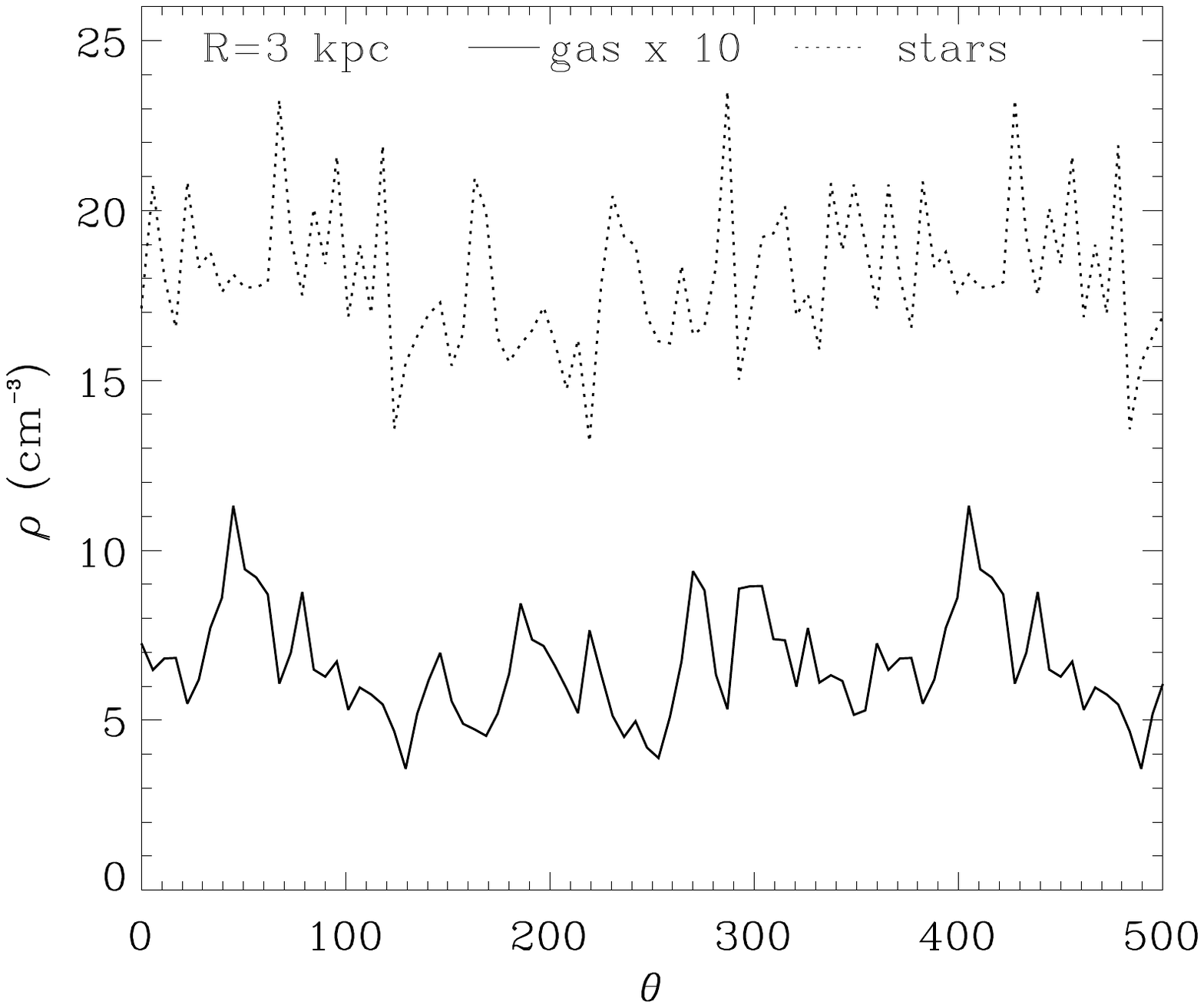}
\includegraphics[bb=0 370 500 800,scale=0.33]{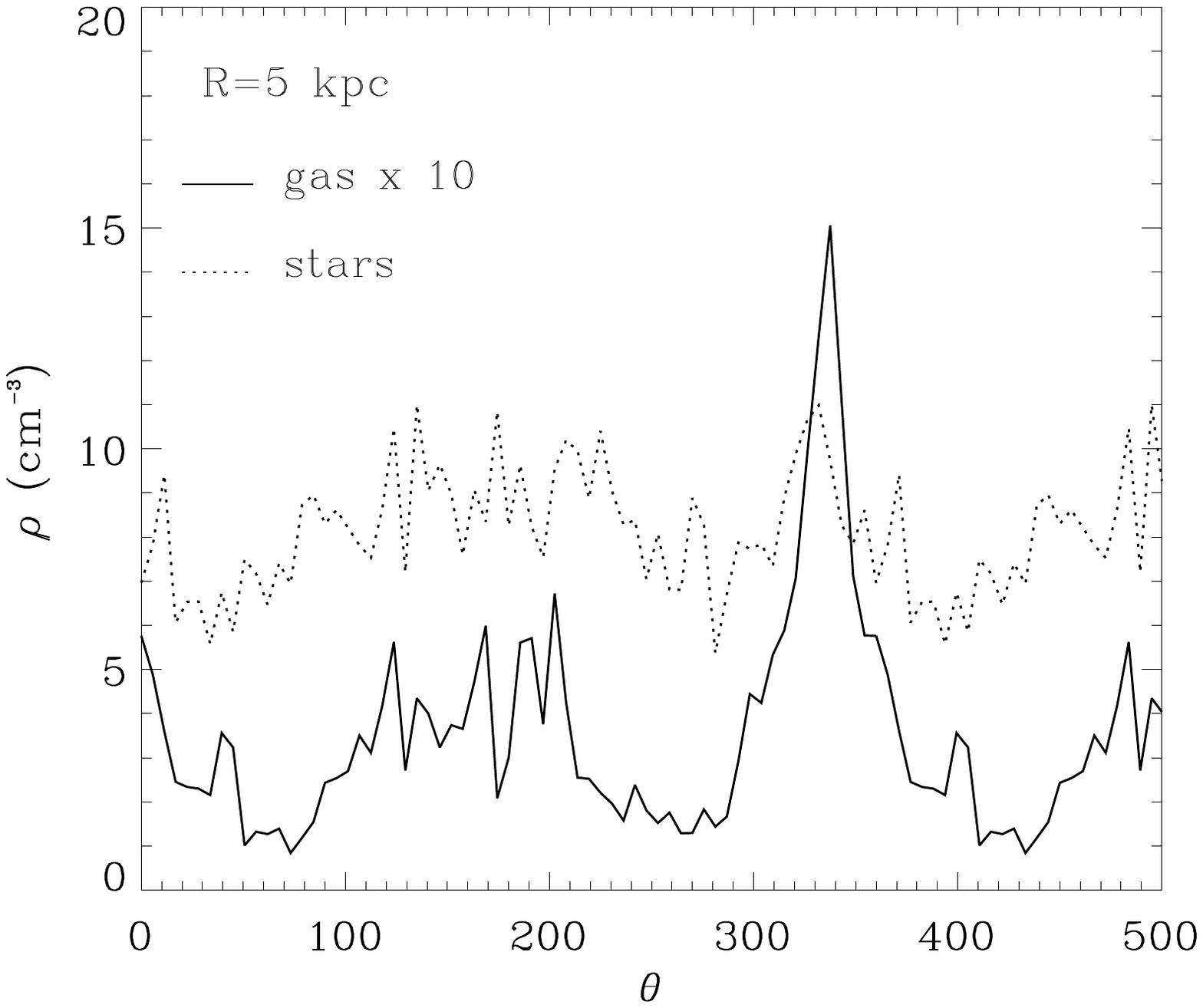}
\includegraphics[bb=0 370 500 800,scale=0.33]{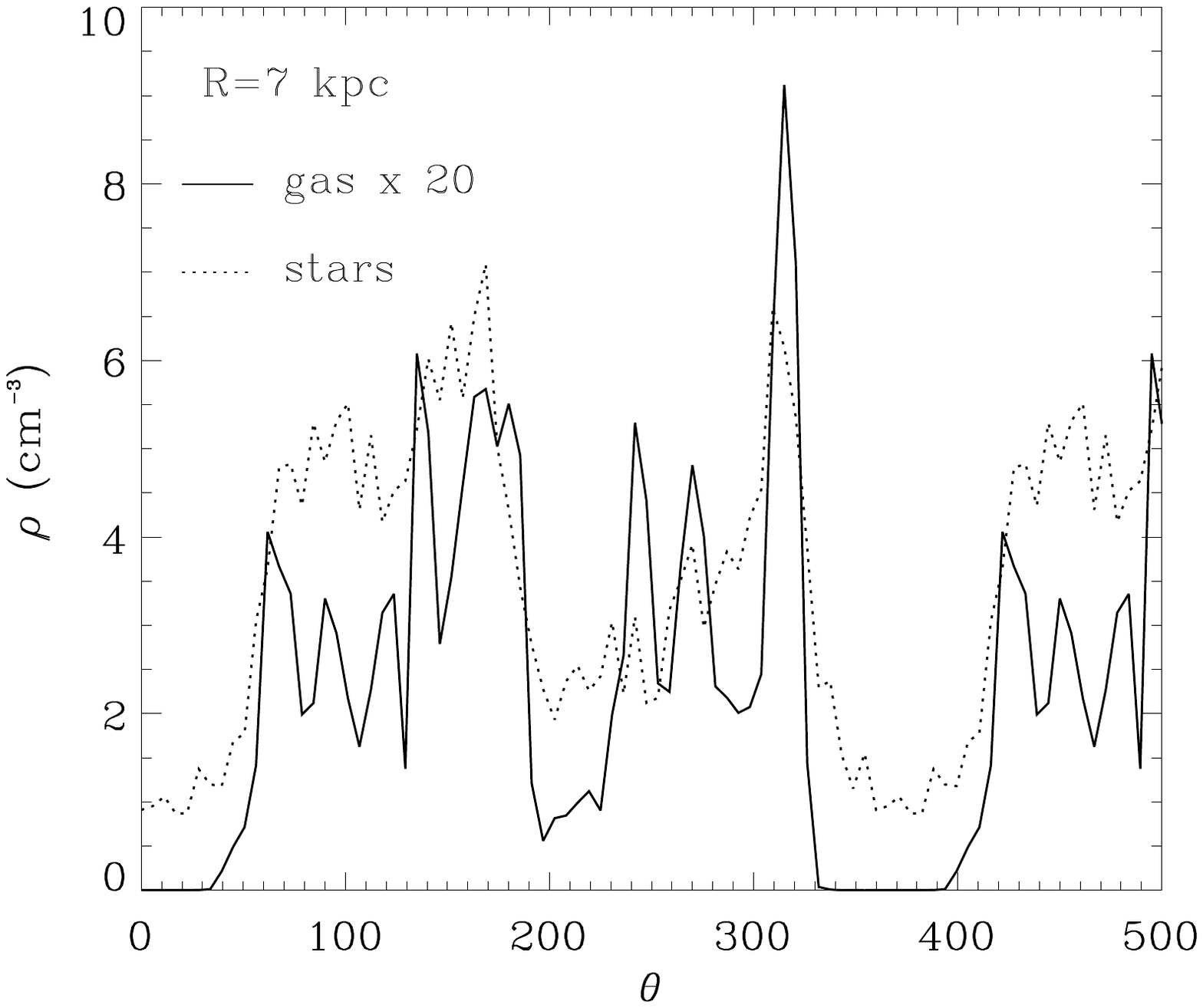}}
\centerline{
\includegraphics[bb=200 370 500 770,scale=0.33]{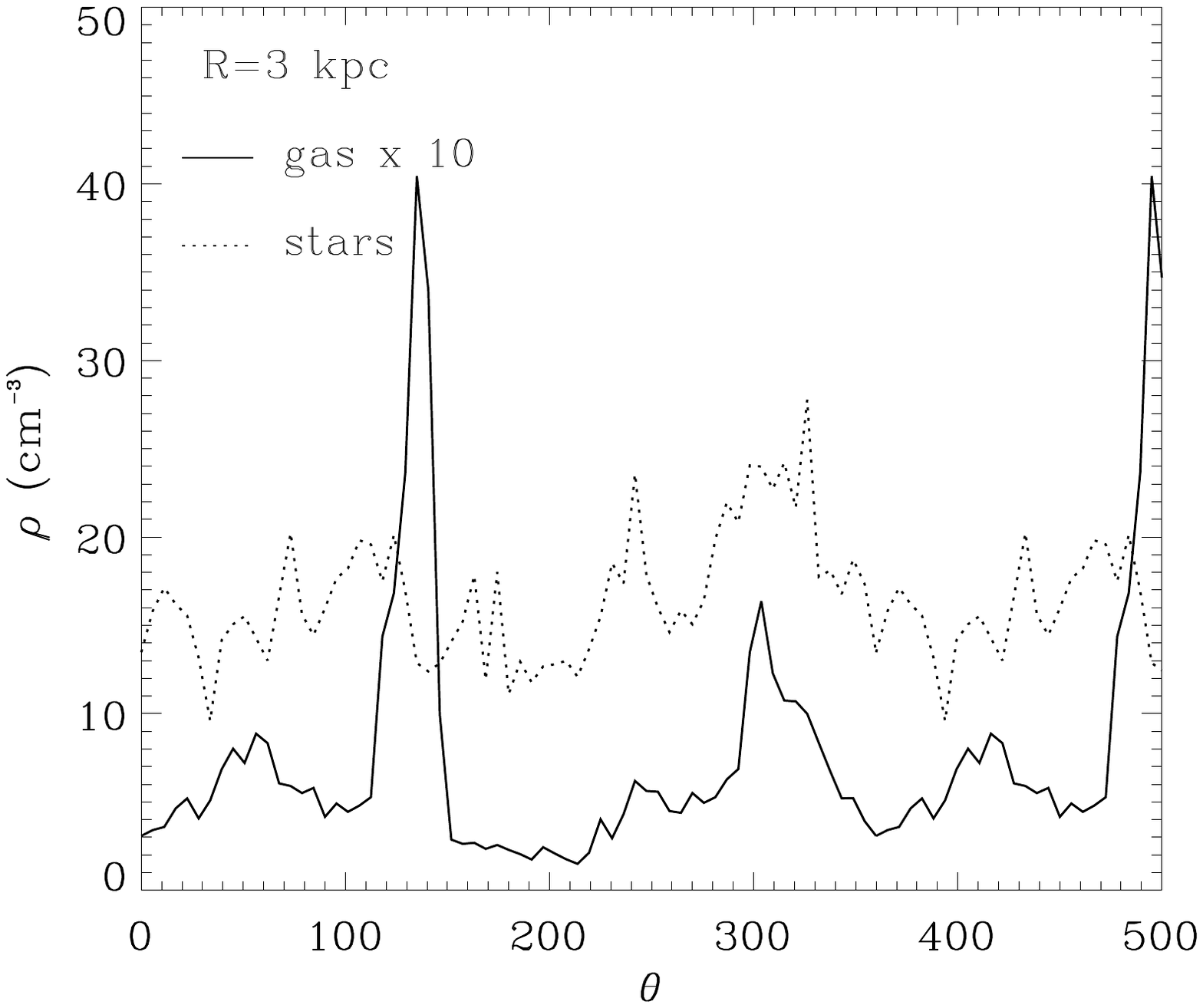}
\includegraphics[bb=0 370 500 800,scale=0.33]{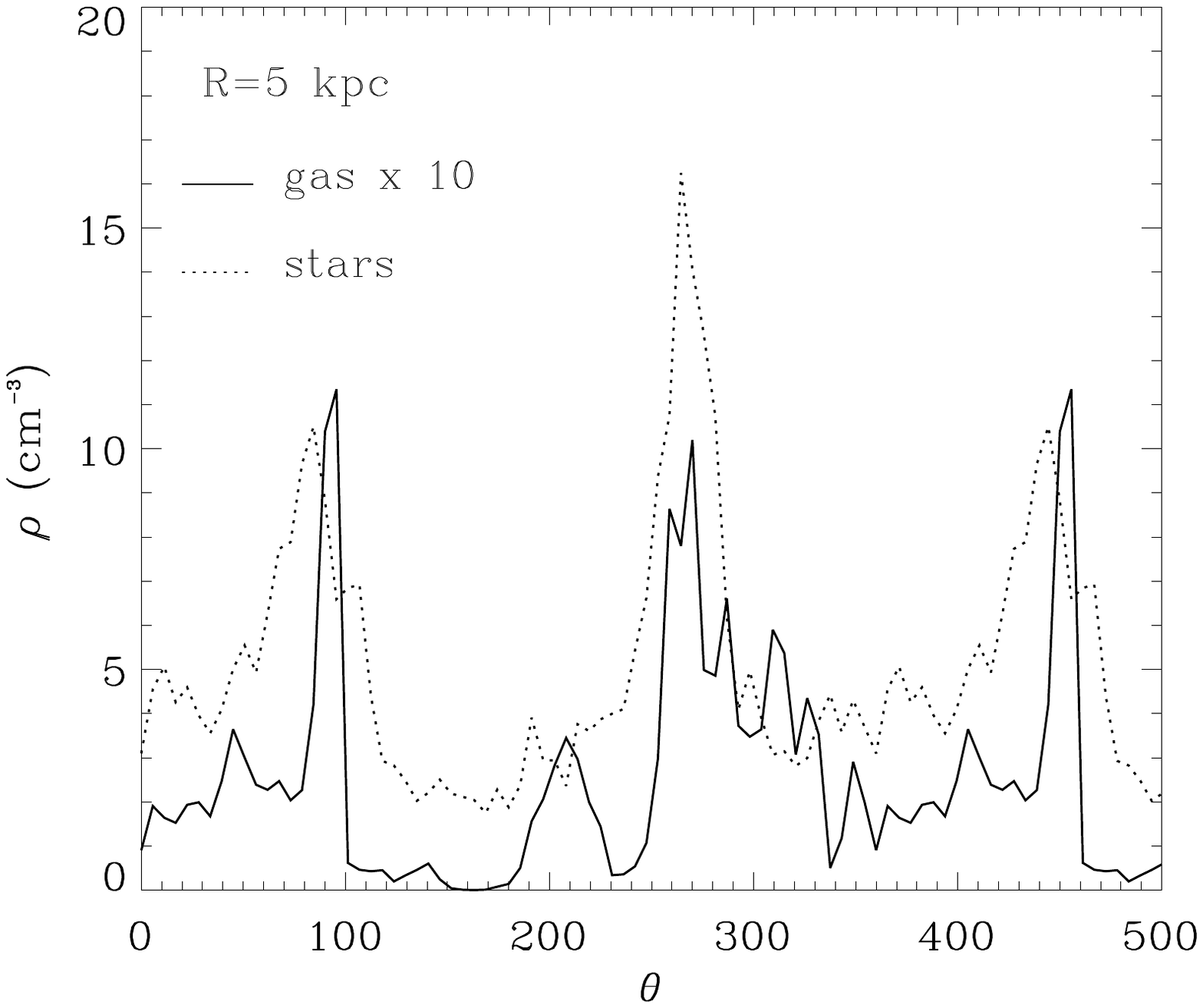}
\includegraphics[bb=0 370 500 800,scale=0.33]{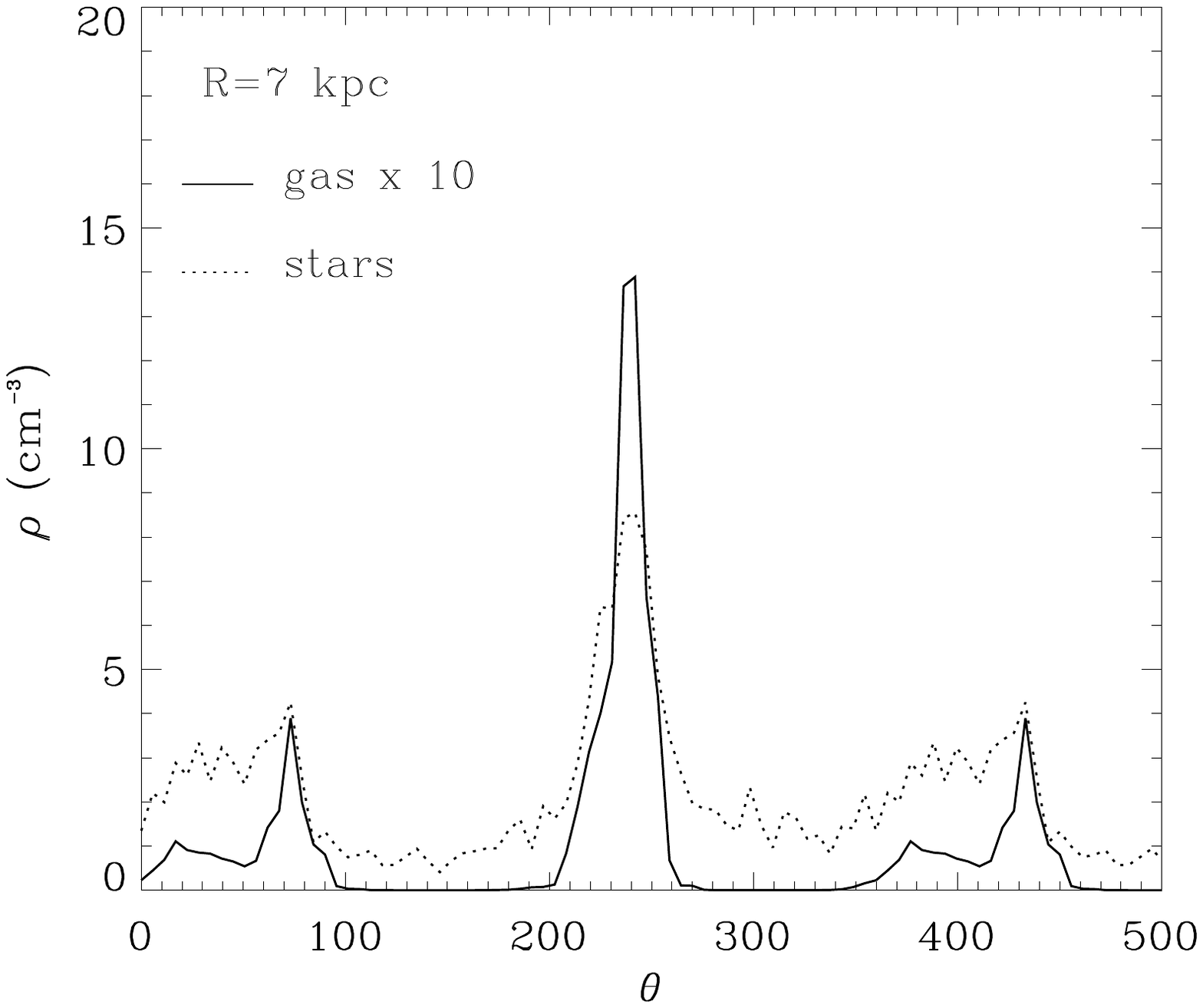}}
\centerline{
\includegraphics[bb=200 370 500 770,scale=0.33]{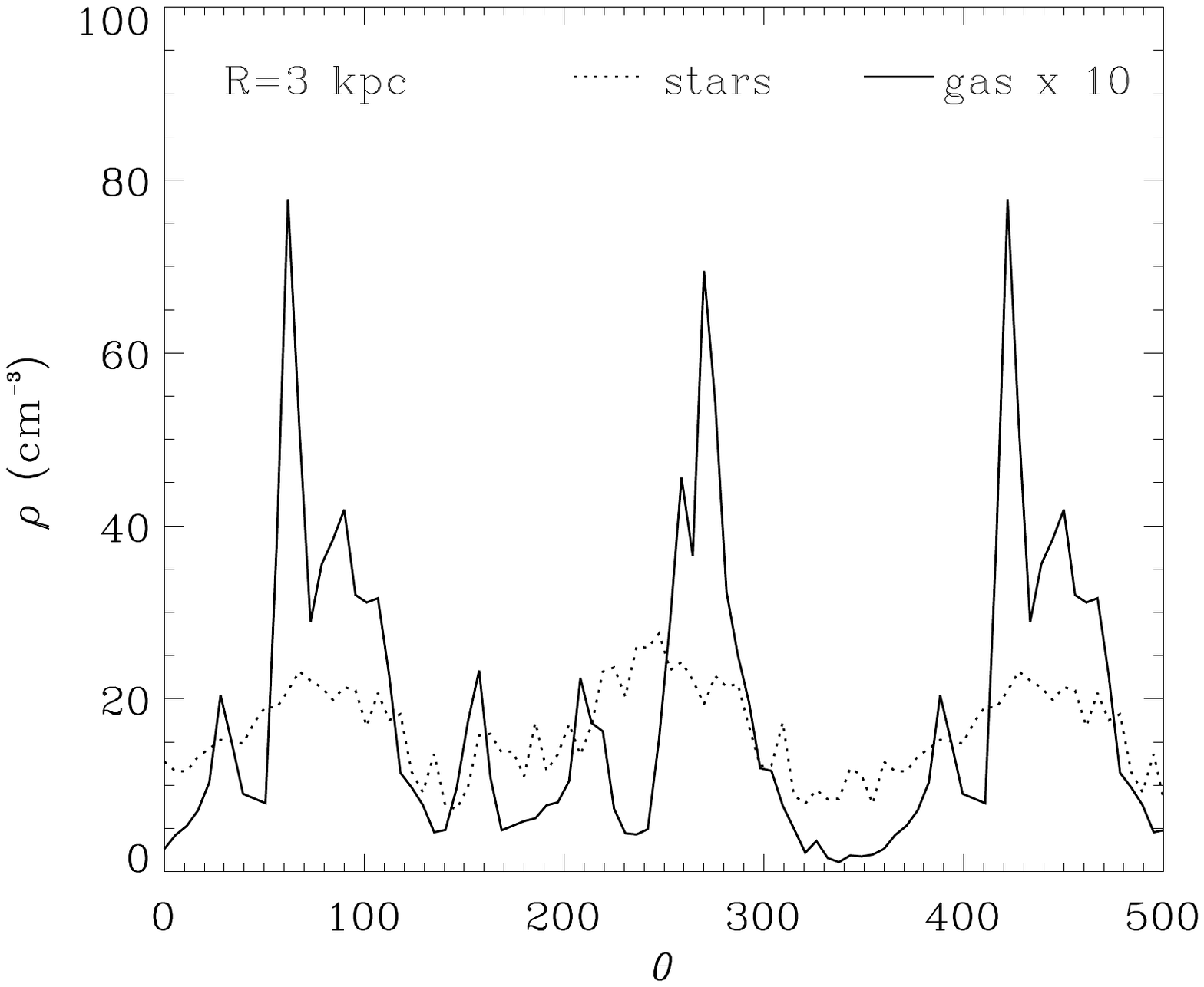}
\includegraphics[bb=0 370 500 800,scale=0.33]{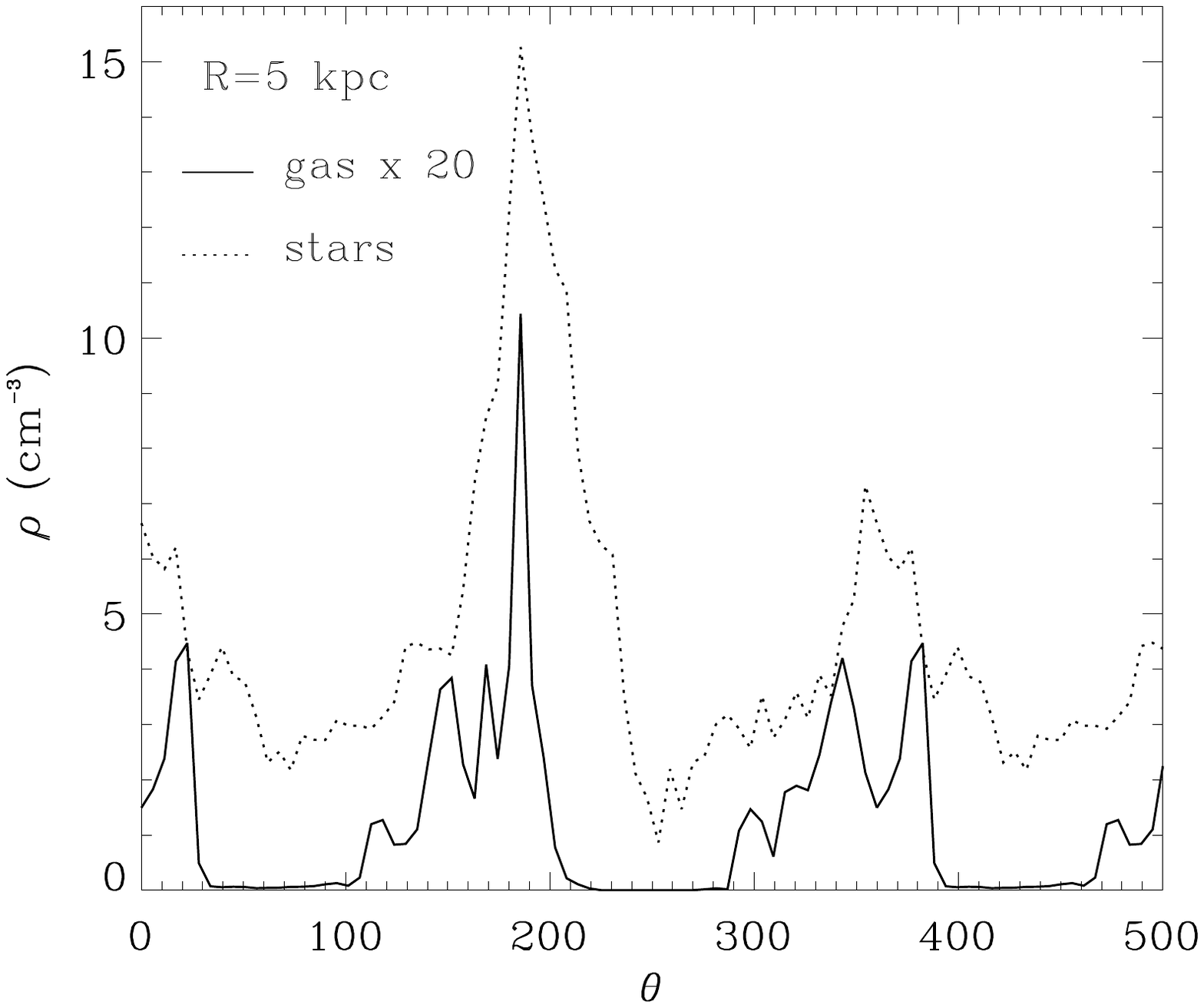}
\includegraphics[bb=0 370 500 800,scale=0.33]{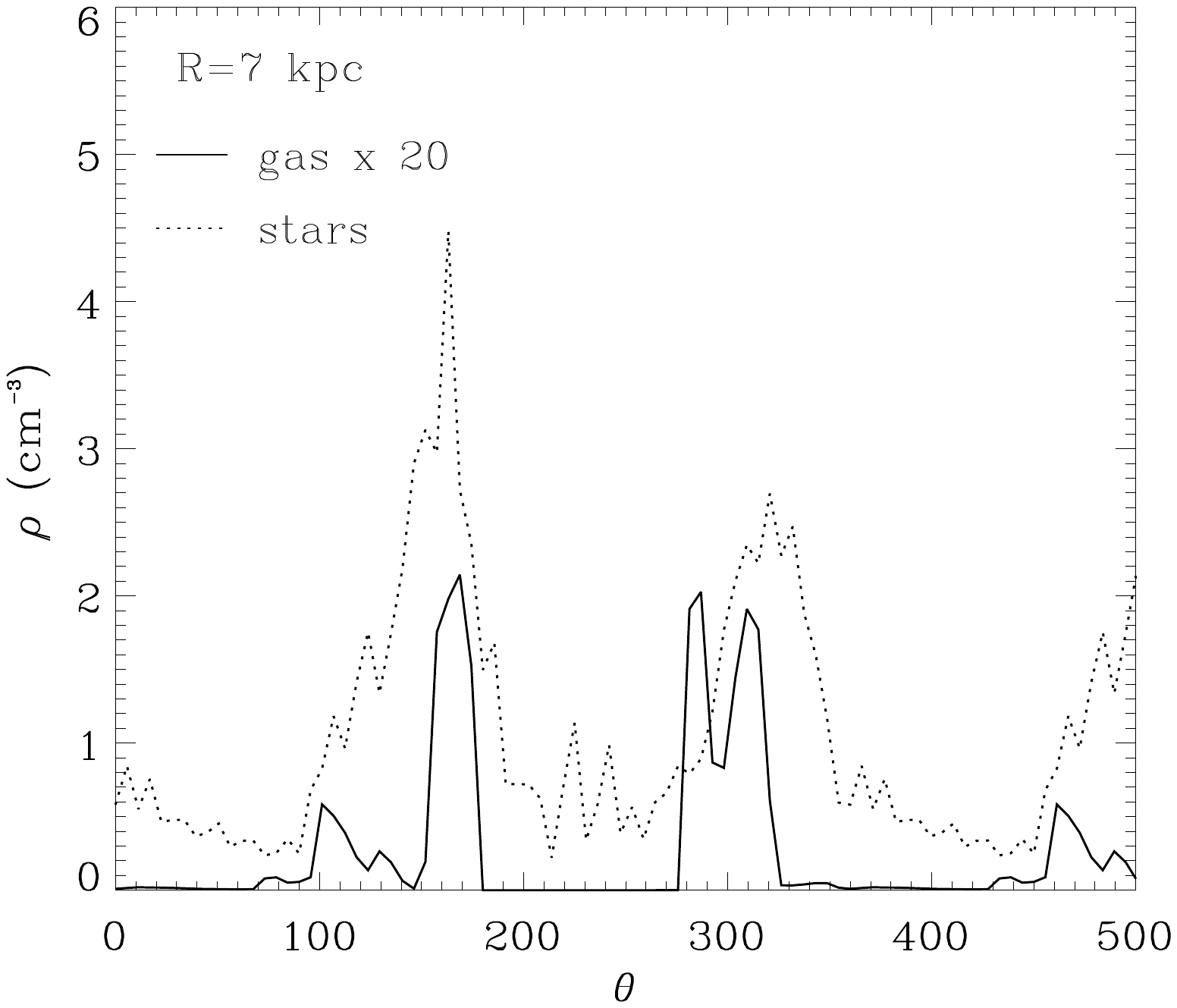}}
\centerline{
\includegraphics[bb=200 370 500 770,scale=0.33]{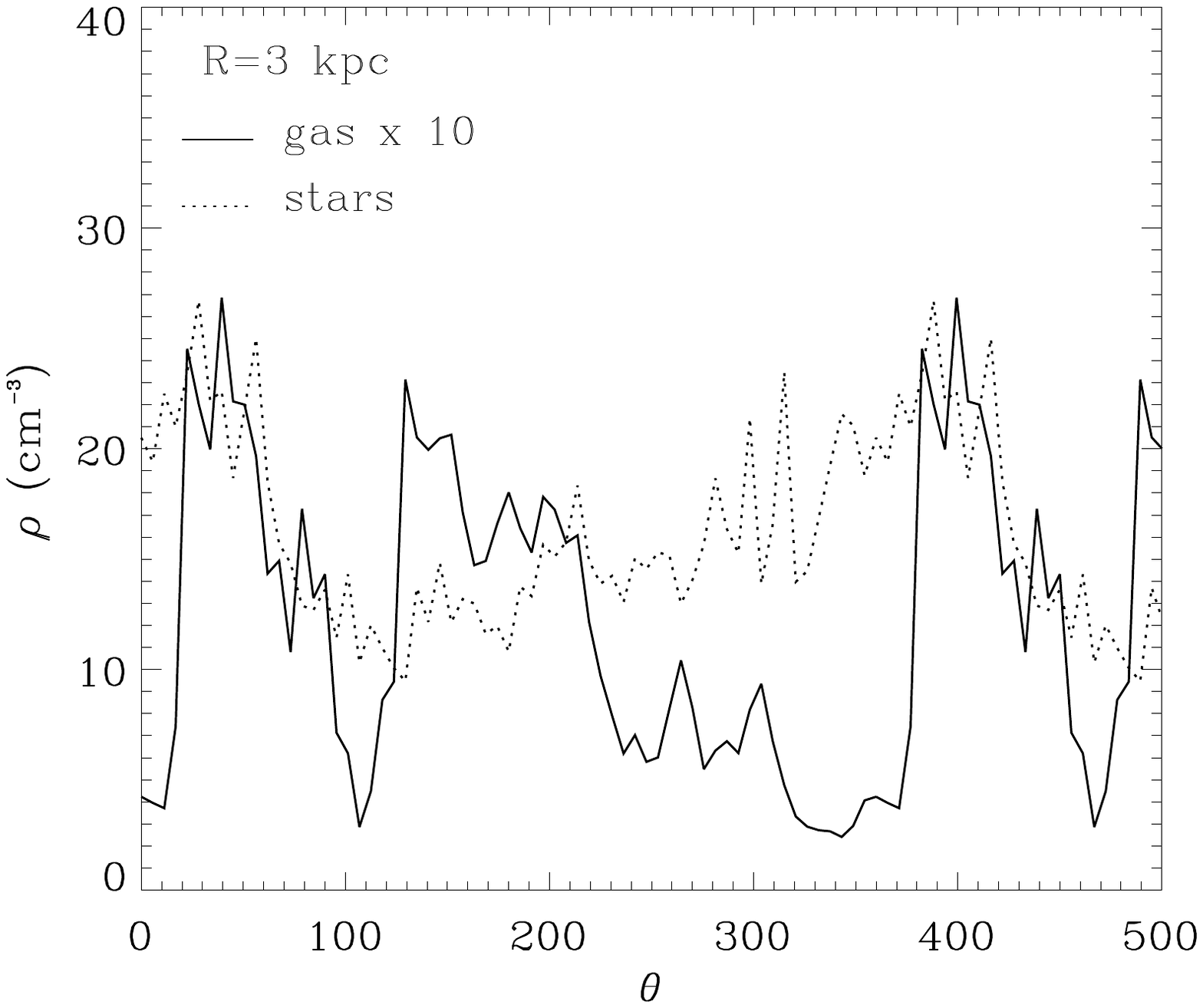}
\includegraphics[bb=0 370 500 800,scale=0.33]{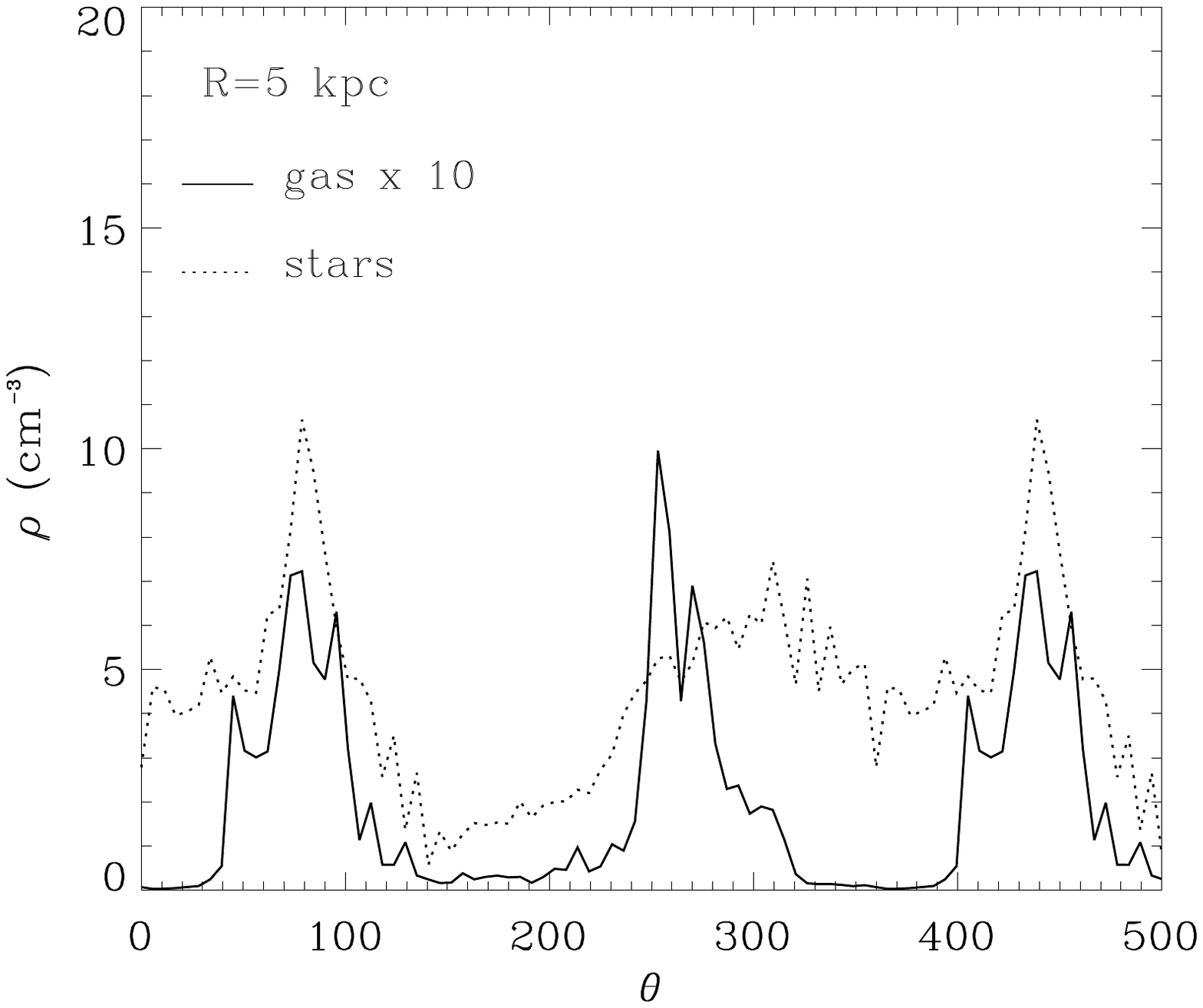}
\includegraphics[bb=0 370 500 800,scale=0.33]{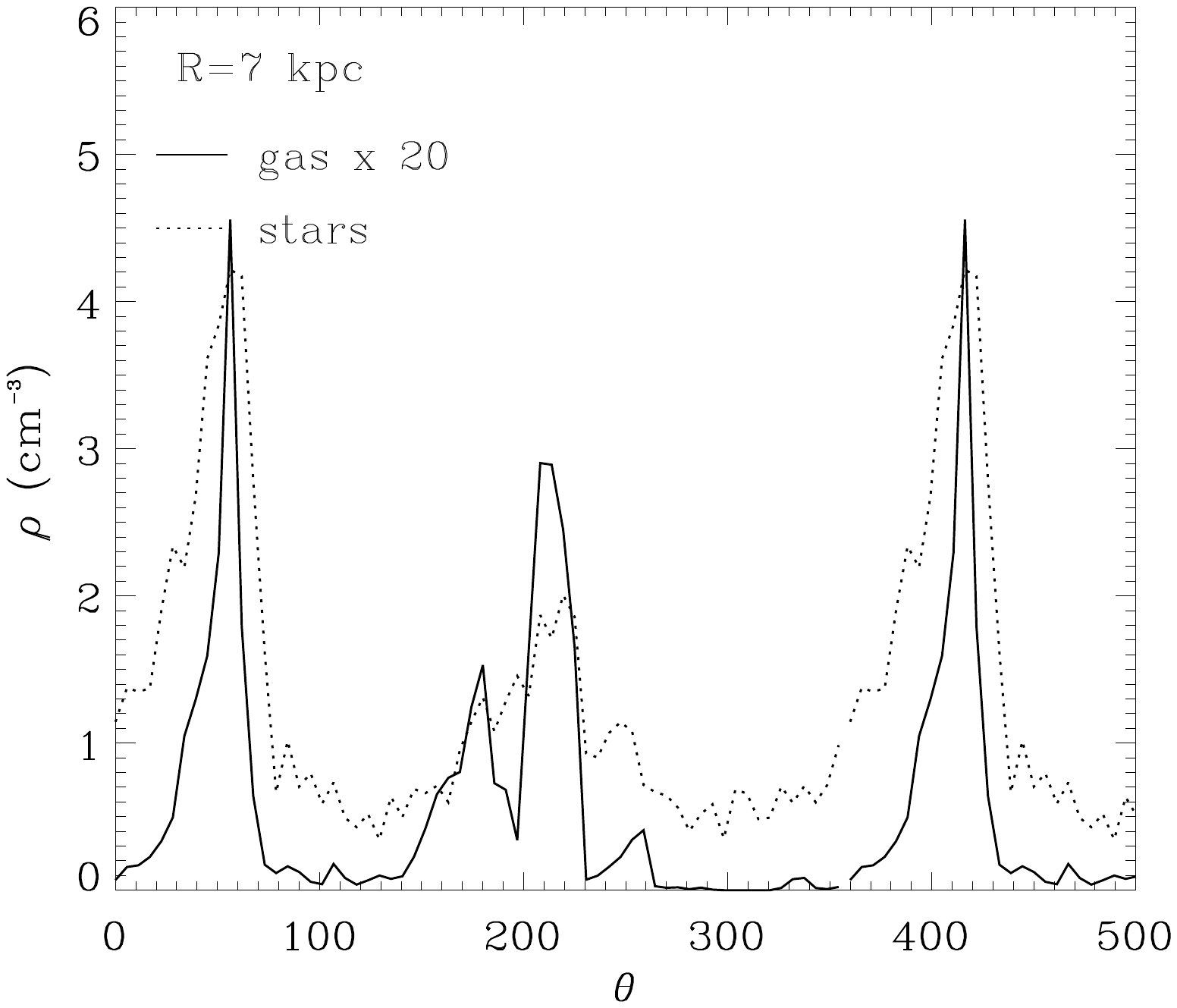}}
\caption{These panels show azimuthal plots of the stellar and gas
  densities at times of 120 (top), 180 (second), 240 (third) and 300
  (lower) Myr. The mass averaged 
  density is calculated from annuli of width 1 kpc,
  which are positioned at 3 (left), 5 (centre) and 7 (right) kpc, and
  both the gaseous (solid) and stellar (dotted) densities are shown.
  The azimuthal angle is calculated anti-clockwise round the disc (in the direction of flow).
  At the earliest time, the interaction is only commencing, so there
  is no dominant two armed structure. Though the gas and stellar peaks
  are generally correlated, which is not surprising if they are
  generated by gravitational instabilities simultaneously. After 180
  Myr, both the gas and stars now have two peaks. However at the later
  times there tends to be substructure in the gas which leads to
  multiple peaks as well as the main spiral arms.}
\end{figure*}

\subsubsection{Origins of substructure and spurs}

The later part of this paper focuses predominantly on comparisons
with observations, but beforehand we briefly consider the origins of
spurs and branches which lie between the main spiral arms.  As
mentioned in the introduction, gaseous spurs could be due to
perturbations in the underlying stellar disc, or occur as GMCs
experience shear when they leave the spiral arms. Numerical
simulations of grand design spirals have largely neglected the first
hypothesis, as they have assumed an underlying stellar potential,
rather than incorporating a live stellar component. Alternatively for
a quasi-steady structure driven by some fixed pattern speed, spurs can
occur at specific resonances in the disc.

We investigate the origins of substructure in the disc further by
tracing back the gas which constitutes a large branch at the time of
300 Myr. Fig.~12 shows the gas at earlier times of 256 and 271 Myr. At
271 Myr, the gas lies in a clump along the spiral arm, which then
becomes sheared into the branch we see at 300 Myr.

The origin of the clump in the spiral arm at 271 Myr could be due to
gravitational instabilities, which may potentially lead to
fragmentation along the spiral arms, and in turn to spurs. However
this does not appear particularly likely, firstly since as mentioned
in Section~2.2.2, the gas has a relatively high $Q$. Moreover, as
Fig.~8 indicates, there is not a significant difference in structure
between the models with 1 and 10\% gas (A and C), implying that
self-gravity of the gas is not regulating the structure. A second
possibility is cloud collisions. However the formation of large clumps
in this manner also seems unlikely when the temperature of the gas
($10^4$ K) is high \citep{Dobbs2006}, since the gas is largely
smooth. Nevertheless the gas in these models is clumpy, not because
the gas is cold, but rather due to the flocculent spiral arms induced
at earlier times by gravitational instabilities in the stellar and
gaseous components. Thus flocculent spiral arms merge with the tidally
induced spiral arms to form clumps, which are then sheared into
branches or spurs, which in turn seed clumpy structure in the next
spiral arm.

A third option is that clumpy structure along the spiral arms is
associated with compressive tidal forces \citep{Renaud2008}, which
could account for structure at predominantly larger radii, a
possibility that we will address in a later paper.
\begin{figure}
\centerline{
\includegraphics[bb=-200 105 500 240,scale=0.85]{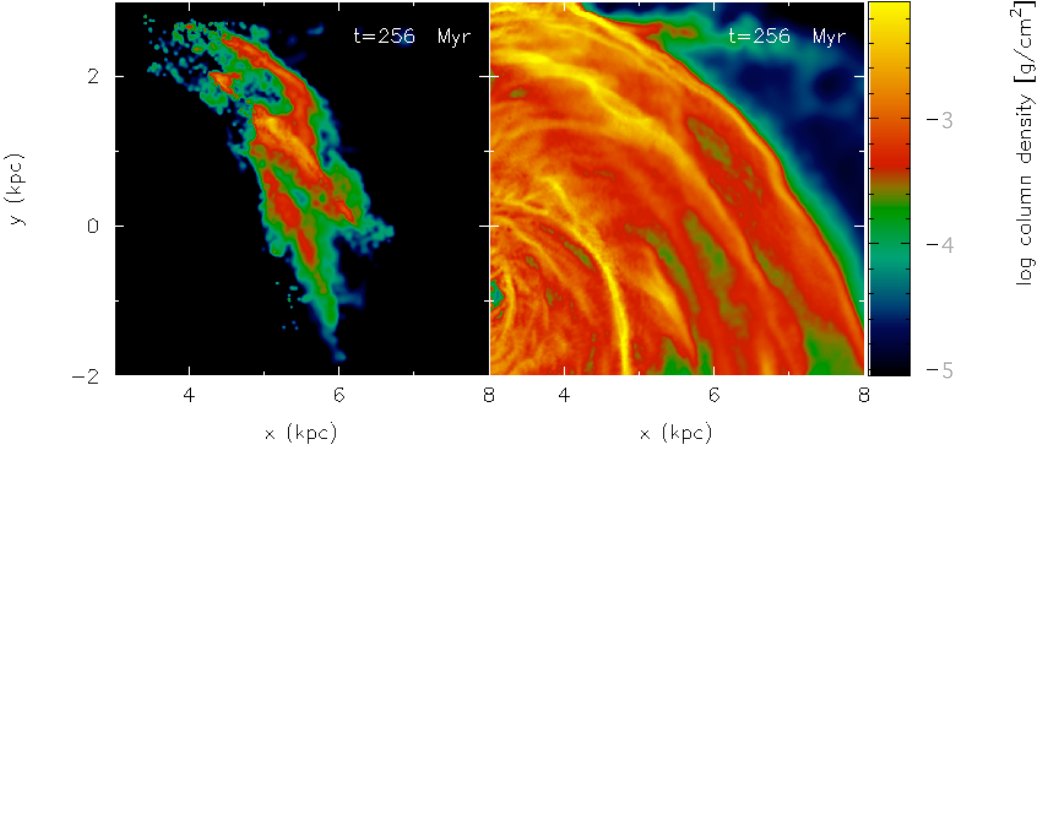}}
\centerline{
\includegraphics[bb=-200 105 500 240,scale=0.85]{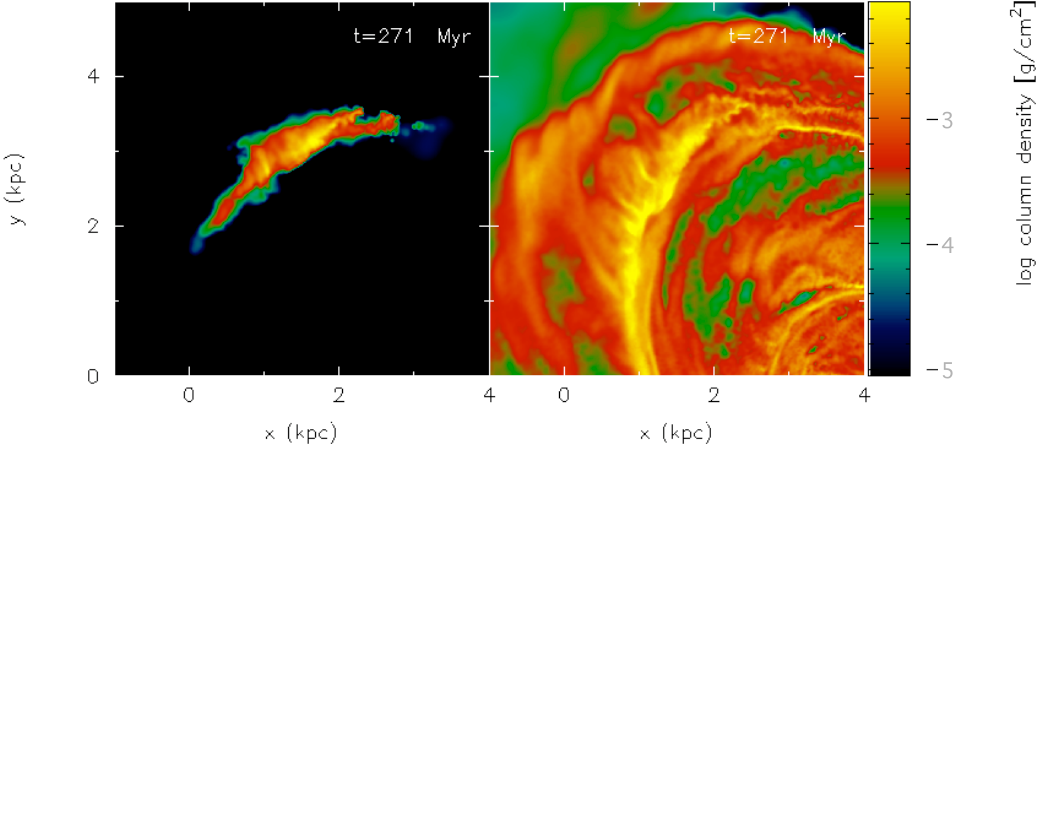}}
\centerline{
\includegraphics[bb=-200 105 500 240,scale=0.85]{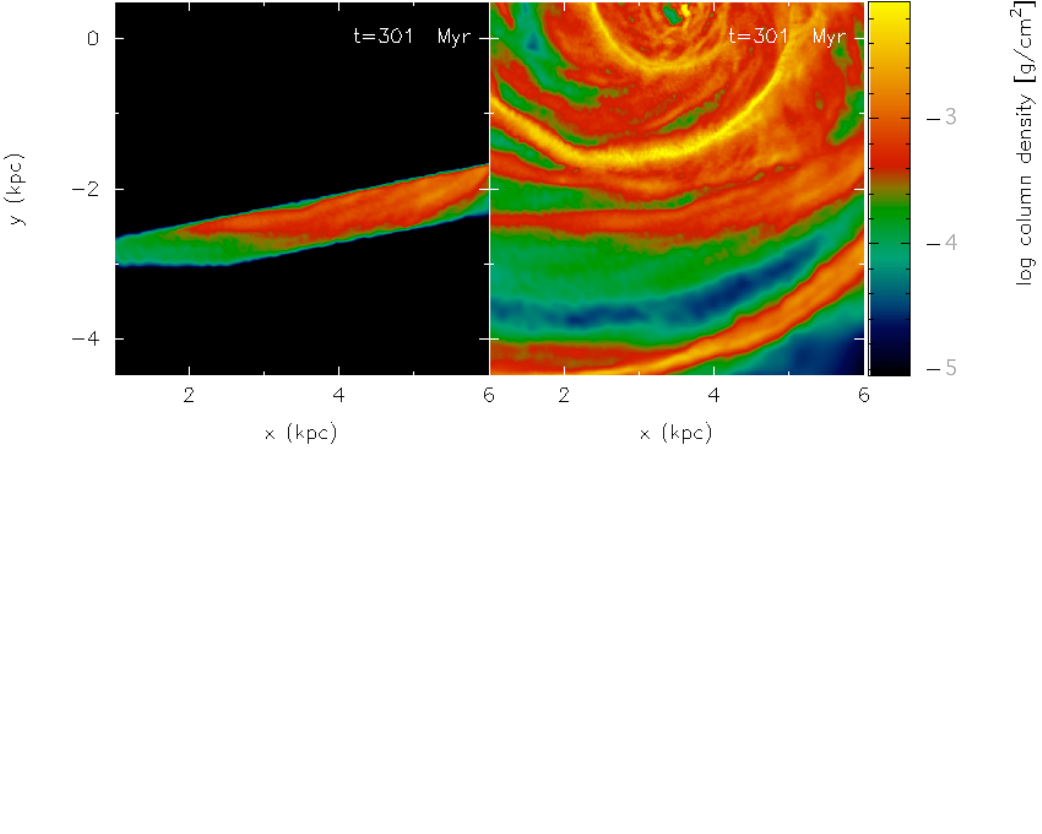}}
\caption{In this figure we focus on the formation of a large secondary
  spiral arm, or branch which appears in the simulation. Gas particles
  are selected from the branch, and then shown at two earlier time
  frames (left) whilst all the gas in the locality is shown on the
  right hand column density plots. The branch originates from a clump
  of dense gas in the spiral arm which gets sheared in the interarm
  region. This in turn originates from substructure in the disc.}
\end{figure}

\begin{figure}
\centerline{
\includegraphics[bb=80 20 480 480,scale=0.34]{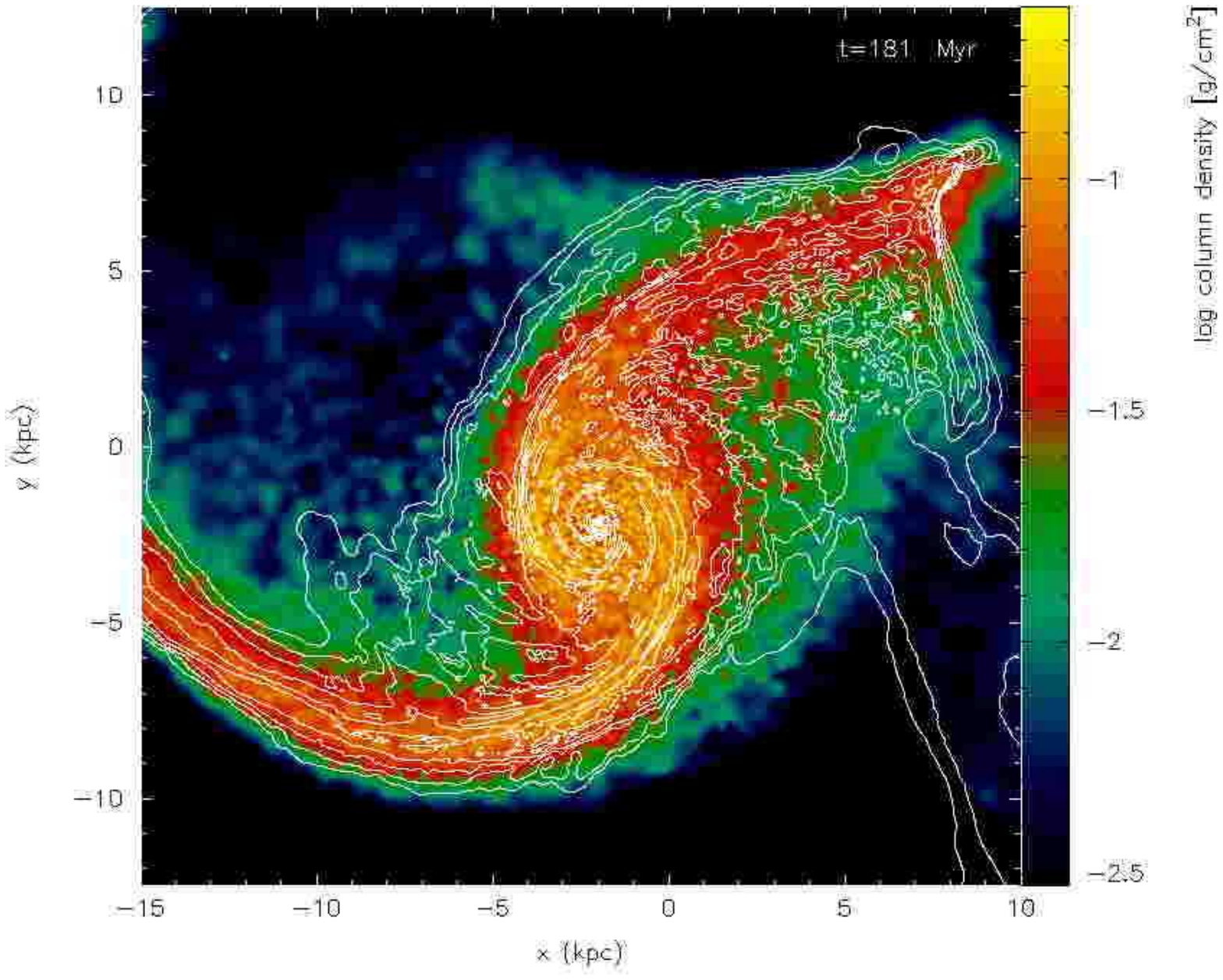}} 
\centerline{
\includegraphics[bb=80 20 480 480,scale=0.34]{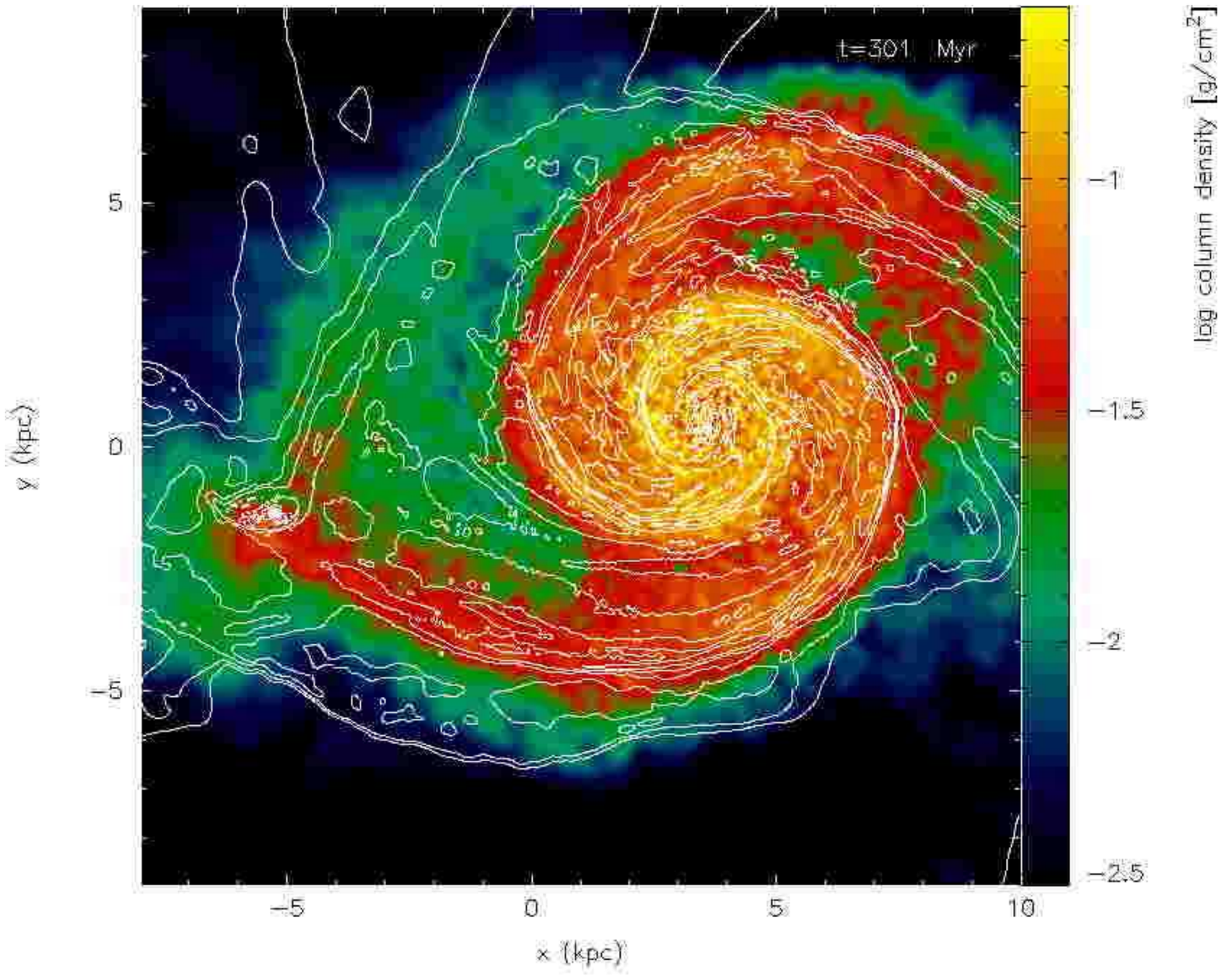}}
\centerline{
\includegraphics[scale=0.6,angle=90]{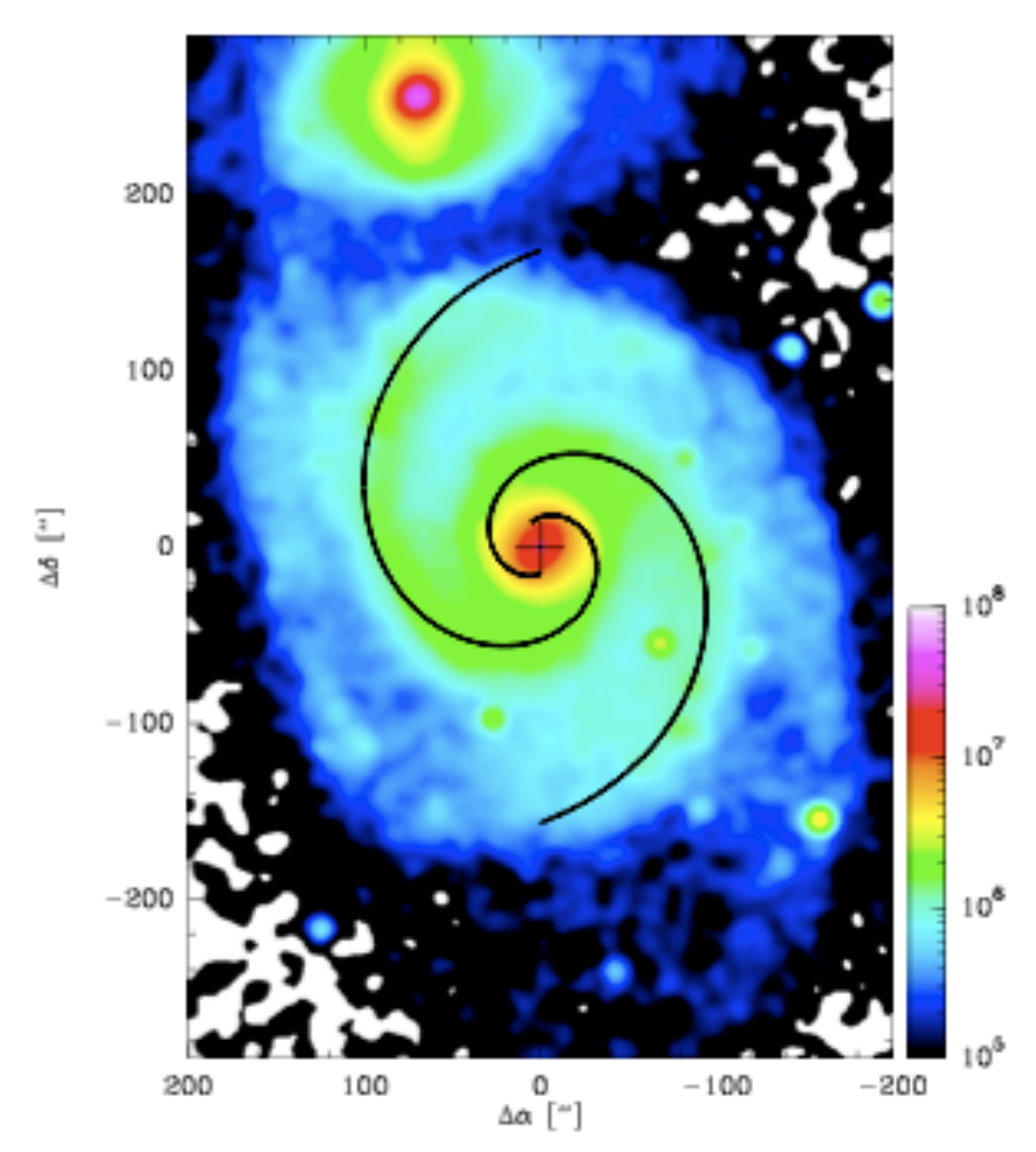}}
\caption{The column density for the stellar disc is shown for the model with 1\%
  warm gas at times of 180 (top) and 300 (middle) Myr. Contours of gas
  column density are overlaid. The lower panel shows a K-band image
  of M51, indicating the old stellar population, from
  \citet{Hitsch2009}. The black lines represent 
  symmetric logarithmic spirals fitted to $^{12}$CO data by
  \citet{Shetty2007}.} 
\end{figure} 

\subsubsection{Comparison of spurs with M51}

The surface densities of spurs in the 1 and 10\% warm gas models are
$10^{-3}$ and $10^{-2}$ g cm$^{-2}$ respectively, or 4.5 and 45
M$_{\odot}$ pc$^{-2}$. The surface densities of spurs seen in M51 are
$\sim$ 50 M$_{\odot}$ pc$^{-2}$ \citep{Corder2008}. Thus the features
found in our 10\% gas model are approaching the surface densities of
spurs in M51, although the surface densities are deliberately low in
our models to avoid gravitational collapse.

\subsection{Stellar spiral arms and the offset between stars and gas}

There have been numerous studies of the offset between different
tracers in M51,
e.g. \citealt{Tilanus1989,Rand1990,Petit1996,Patrikeev2006}. The CO is
typically seen upstream of H$\alpha$, and the offset interpreted as
the time for stars to form \citep{Tamburro2008}. Density wave theory,
i.e. the assumption that the stellar arms rotate more slowly than the
gas, predicts that the CO should also be upstream of the underlying
old stellar population \citep{Roberts1969,Gittins2004}, although this
is dependent on the properties of the model, e.g. sound speed of the
gas, strength of the potential \citep{Slyz2003,Thesis}. Observations
of CO overlaid on optical images of M51 indicate such an offset for
one spiral arm, though oddly for the other they are coincident
e.g. \citealt{Schinnerer2004}.

In Fig.~13, we show the column density of the stars (in the disc
only), and overlay contours of the gas column density.  As expected
the stellar distribution is much smoother, and the spiral arms much
broader compared to the gas. Also shown on Fig.~13 is a K-band image 
(which traces the old stellar population)
of M51 from \citet{Hitsch2009}, which again shows good agreement
with the stellar distribution from our simulations. Similarly to the gaseous arms, 
the stellar arms in the simulations are wound slightly further than in the observations due to
adopting a too high rotation curve initially (Section~2.2).
\begin{figure*}
\centerline{
\includegraphics[bb=100 320 570 770,scale=0.33]{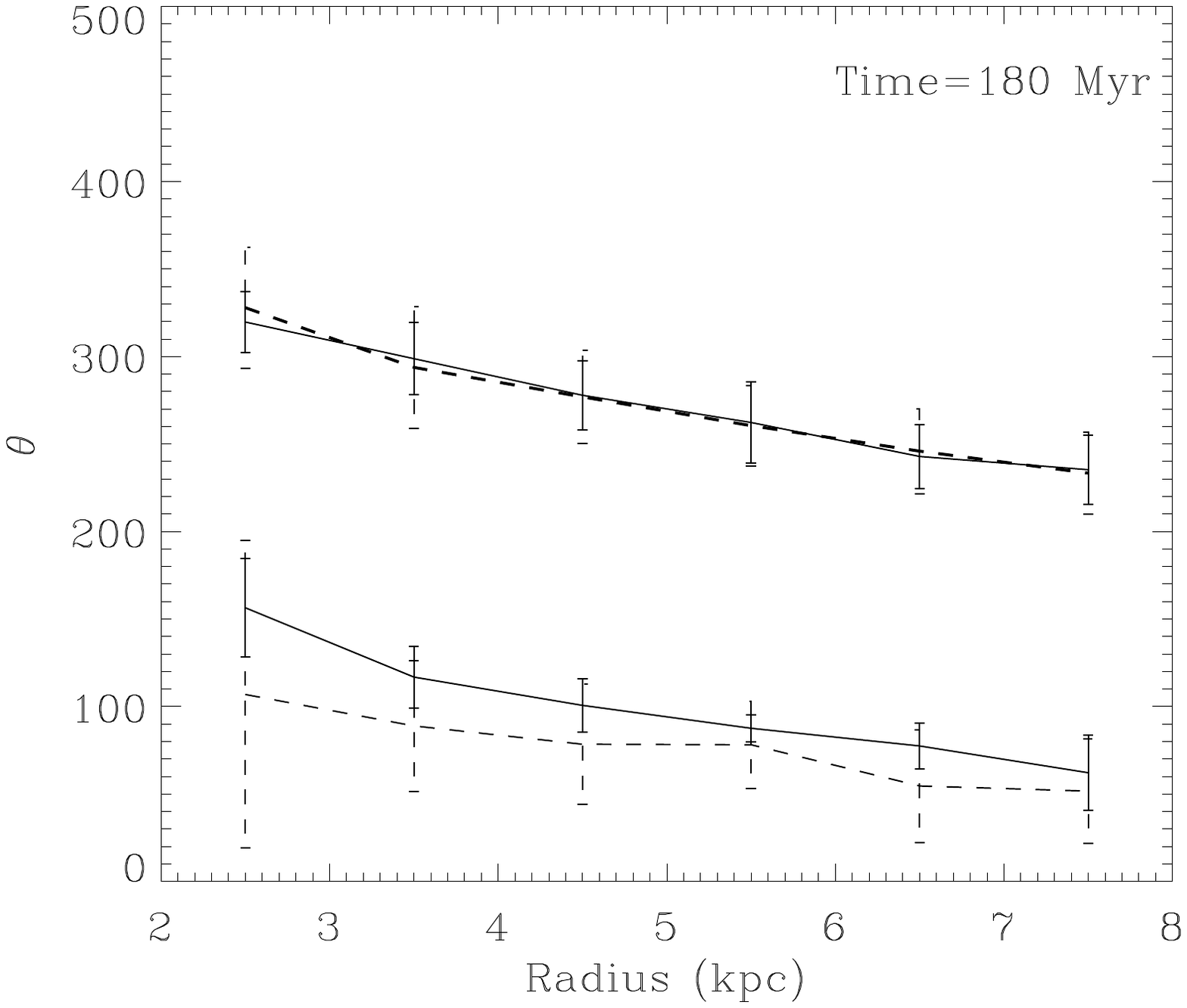}
\includegraphics[bb=100 320 570 770,scale=0.33]{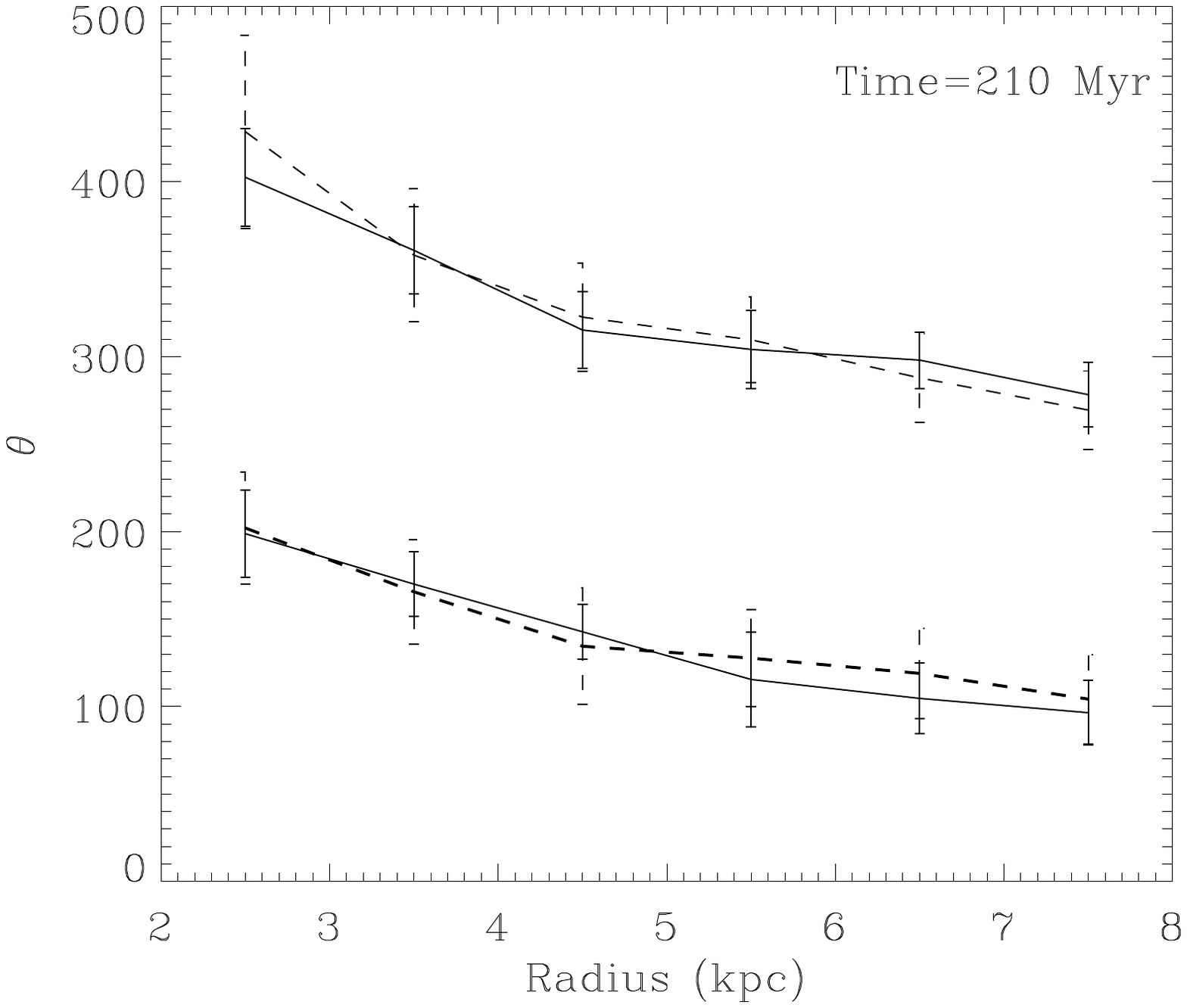}
\includegraphics[bb=100 320 570 770,scale=0.33]{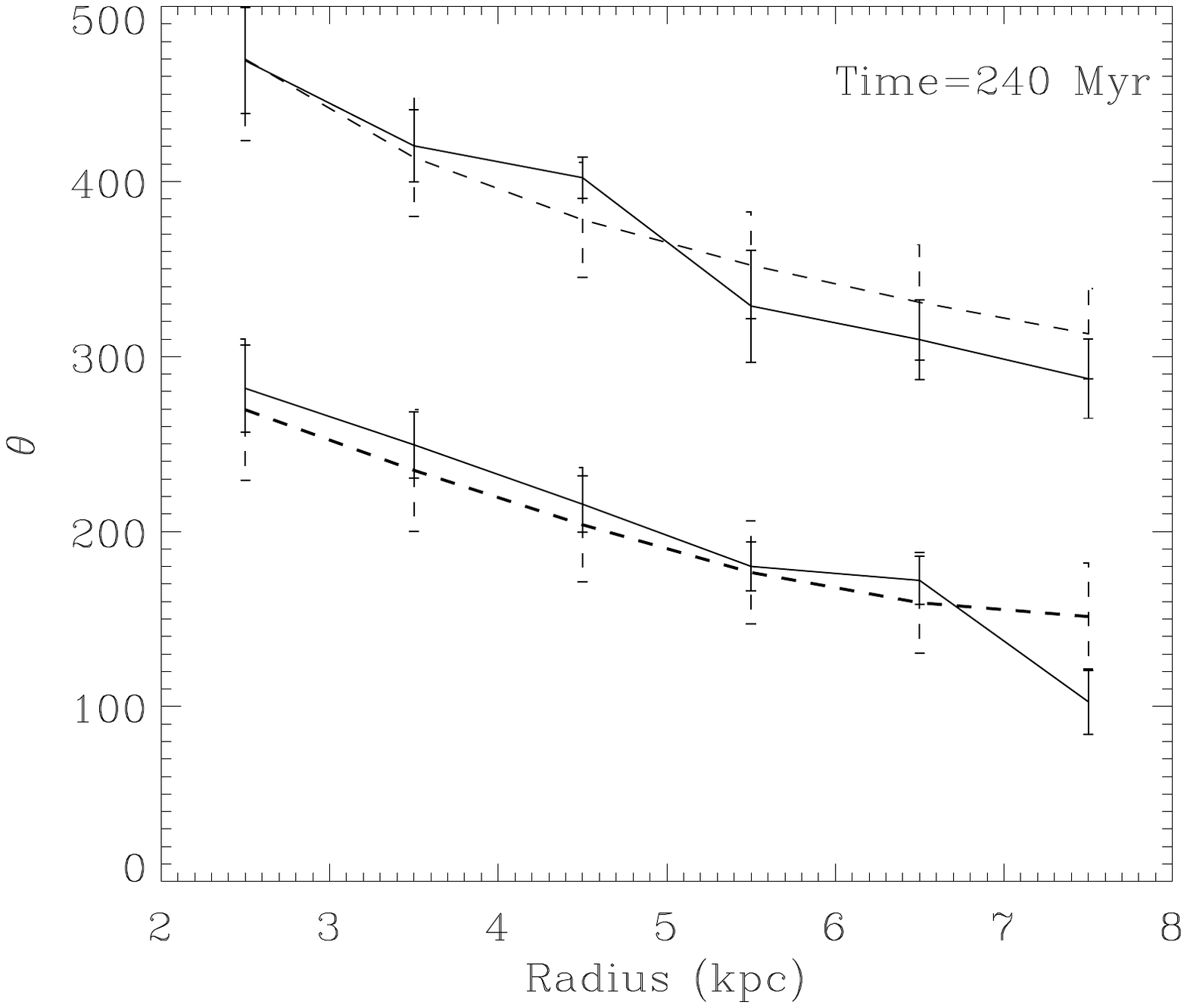}}
\centerline{
\includegraphics[bb=100 370 570 770,scale=0.33]{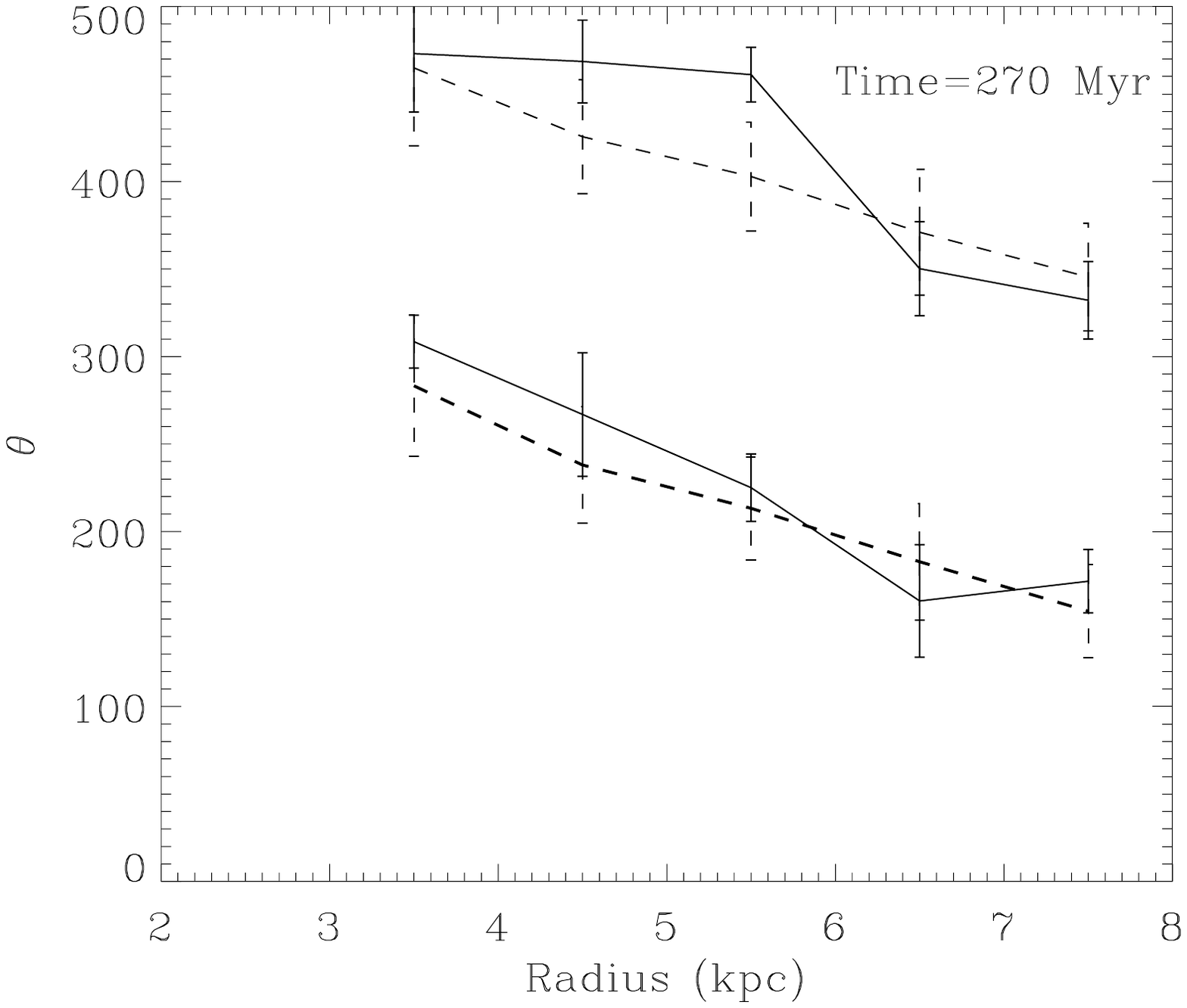}
\includegraphics[bb=100 370 570 770,scale=0.33]{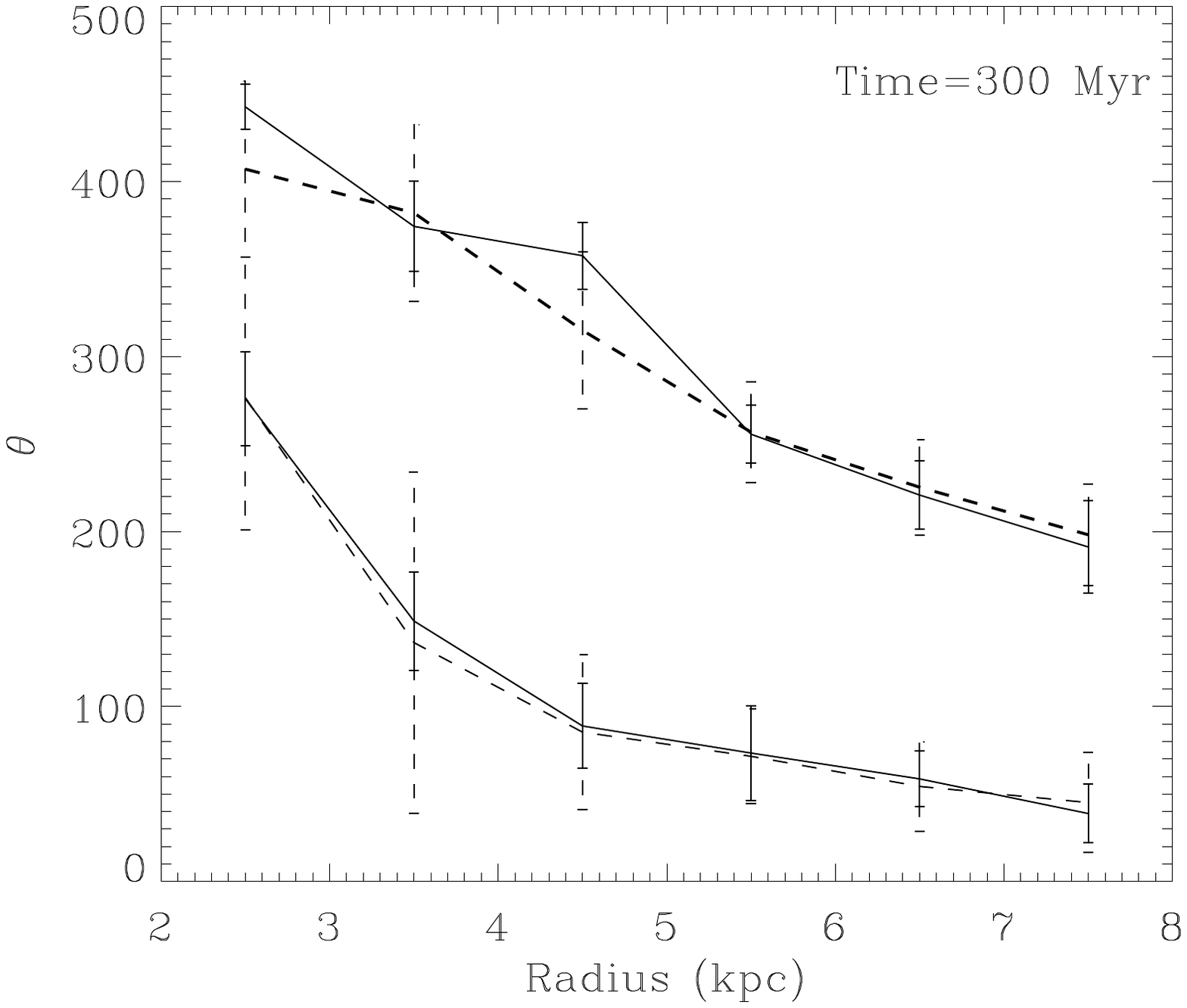}}
\caption{In this figure we plot the peaks of the gaseous (solid) and
  stellar (dashed) spiral arms as determined by fitting Gaussian
  profiles to azimuthal density plots. The azimuthal angle ($\theta$) is calculated anti-clockwise round the disc (in the direction of flow) so a solid line lying above a dashed line indicates the shock lies upstream of the stellar. Generally, there is little
  offset between the stars and gas, and since the stellar arms are
  very broad, the error bars tend to encompass those for the gaseous
  spiral arms. The main exception is at a time of 270 Myr, which is
  likely due to the complicated structure at this time (see Fig.~19)
  and the subsequent difficulty of picking out the main spiral arms.}
\end{figure*}

Fig.~13 does not indicate any obvious offset between the gas and
stars. The main departures of the gas from the stellar component are
interarm features. A similar conclusion can be drawn from Fig.~11,
which shows that the stellar and gaseous peaks tend to be
coincident. Any offset between peaks of stellar and gas density in
azimuth are temporary, and tend to be localised, e.g. at $R=5$ kpc, one
gas arm lies downstream of the stars (higher $\theta$) whilst the
other is essentially coincident. At 240 Myr, one arm is still largely
coincident (if anything, upstream) whilst the other has split into two
separate peaks.

We determine the offset more systematically by locating the peak of
the gas and stellar spiral arms. However rather than use the $m=2$
Fourier amplitude, which implicitly assumes symmetric spiral arms and
the dominance of the $m=2$ harmonic, we fit Gaussians to our azimuthal
density profiles. As can be seen from the column density plots and
Fig.~11, the spiral arms may often be asymmetric. We took density
profiles over annuli of width 1 kpc, similar to those shown in
Fig.~11. We used a routine in Numerical Recipes \citep{Press1992} to fit a function of
the form
\begin{equation}
  \rho(\theta)=A_1 \exp \bigg[-\Big(\frac{\theta-B_1}{C_1}\Big)^2 \bigg]+ A_2 \exp \bigg[-\Big(\frac{\theta-B_2}{C_2}\Big)^2 \bigg]+A_3
\end{equation}
with the amplitudes of the spiral arms given by $A_1$ and $A_2$, the
offsets by $B_1$ and $B_2$ and the dispersions by $C_1/\sqrt{2}$ and
$C_2/\sqrt{2}$.

The position of the spiral arms is shown versus radius in Fig.~14, for
the gas and stars at five different times during the interaction. We
took annuli from 1-8 kpc, but at the innermost radius it was often
impossible to fit the distribution to two peaks, indicating that in
our model the spiral structure was not strongly induced at such small
radii. The error bars in Fig~14 correspond to the width of the Gaussian peak for each arm.
In nearly all cases the stellar and gaseous spiral arms are
coincident within the error bars, and in particular at the present day
time of 300 Myr. Had we used the errors of the fit instead, 
the error bars would have been smaller, typically about one third the size 
shown in Fig.~14. At most time frames however, the two types of arm would still coincide 
within the smaller errors.

Kinks in the positions of the gaseous spiral arms are
either due to a sharp dip in the spiral amplitude between the inner
and outer parts of the annulus, which results in an overestimate of
the azimuthal angle of the peak, or occur where the spiral arm
bifurcates. There is a slight tendency for the shock to be upstream of the stelar arms at smaller radii and downstream at larger radii. However if anything the location of the shock oscillates upstream and downstream (as typically found in the grand design galaxies studied by \citet{Kendall2009}, including M51) and moreover, the offsets are typically too small to be very meaningful.
\begin{figure}
\centerline{
\includegraphics[bb=210 370 500 770,scale=0.4]{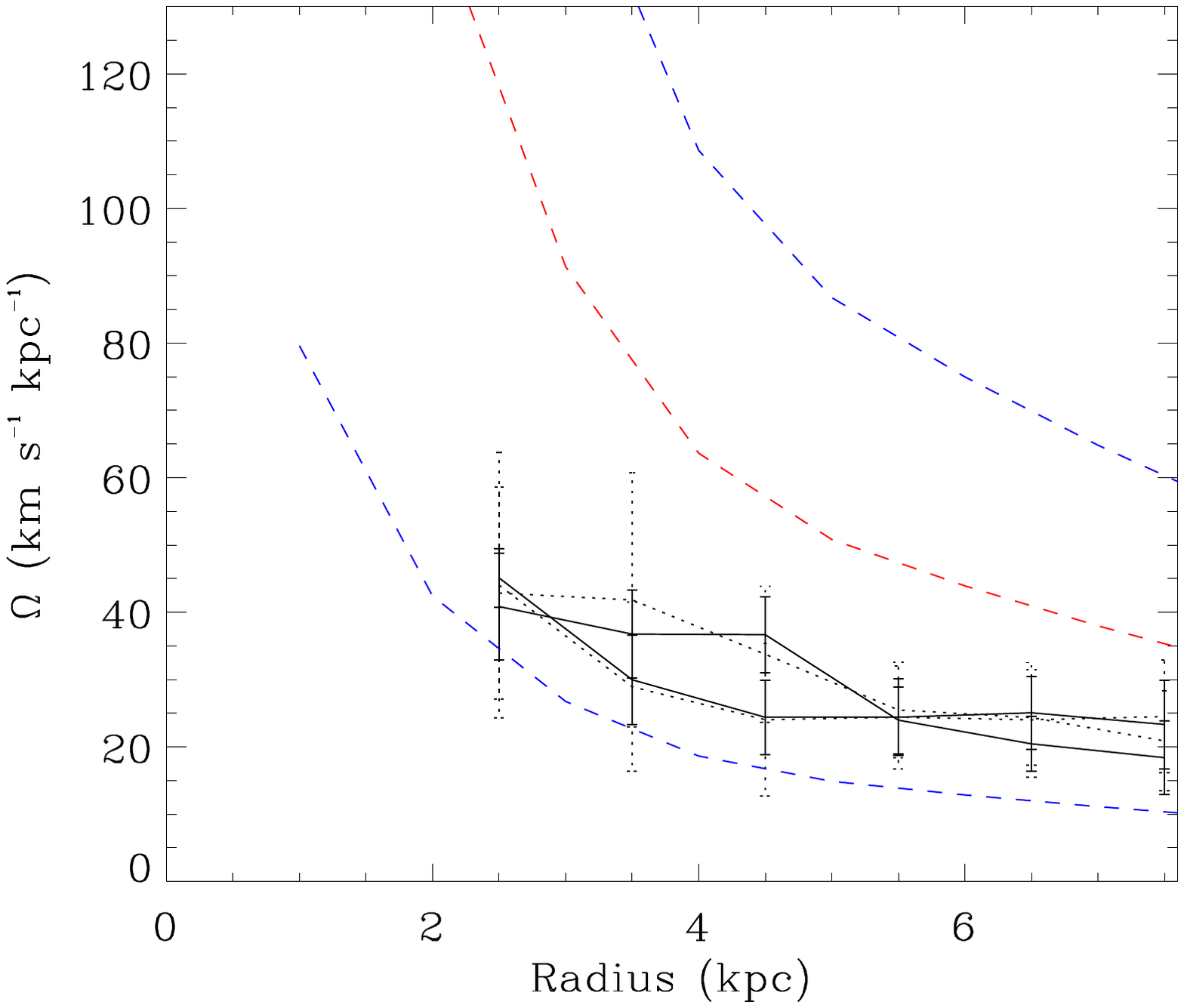}}
\centerline{
\includegraphics[bb=0 00 300 200,scale=0.74]{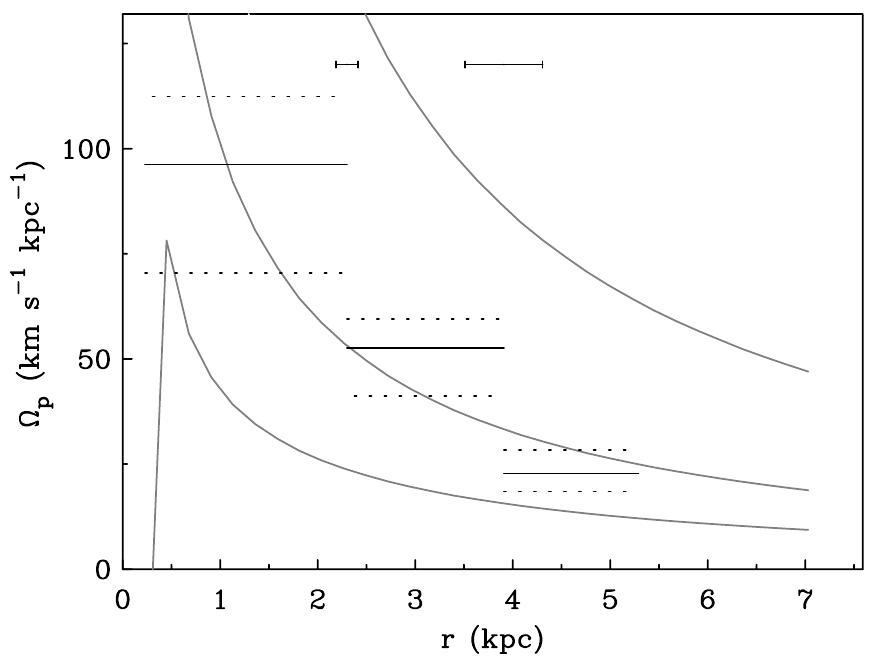}}
\caption{Here we show the pattern speeds of the gaseous (solid) and
  stellar (dotted) spiral arms (top panel) with error bars. We plot
  both arms (i.e. two gaseous arms and two stellar arms) separately,
  as we don't assume they have the same pattern speed. The pattern
  speed is calculated using the locations of the density peaks (shown
  in Fig.~14) at 180 and 300 Myr. Also shown is the angular velocity
  of the stars (red dashed line) and $\Omega \pm \kappa/2$ (blue
  dashed lines). The lower panel shows a corresponding plot form
  \citet{Meidt2008}, who used the Tremaine Weinberg method to
  determine the pattern speed in the inner regions of M51. }
\end{figure}

\begin{figure}
\centerline{
\includegraphics[bb=200 370 500 770,scale=0.33]{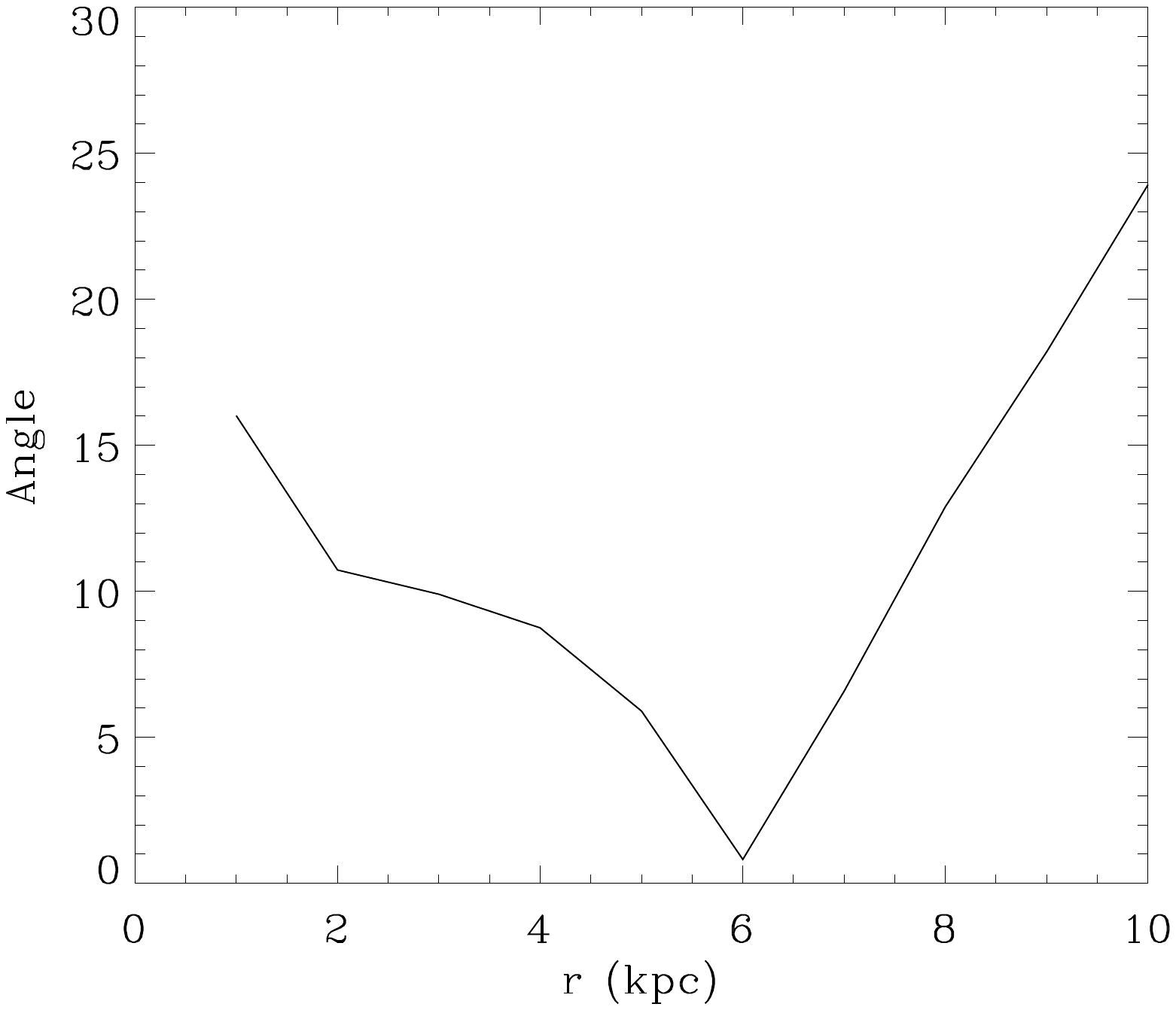}}
\centerline{
\includegraphics[bb=200 370 500 770,scale=0.33]{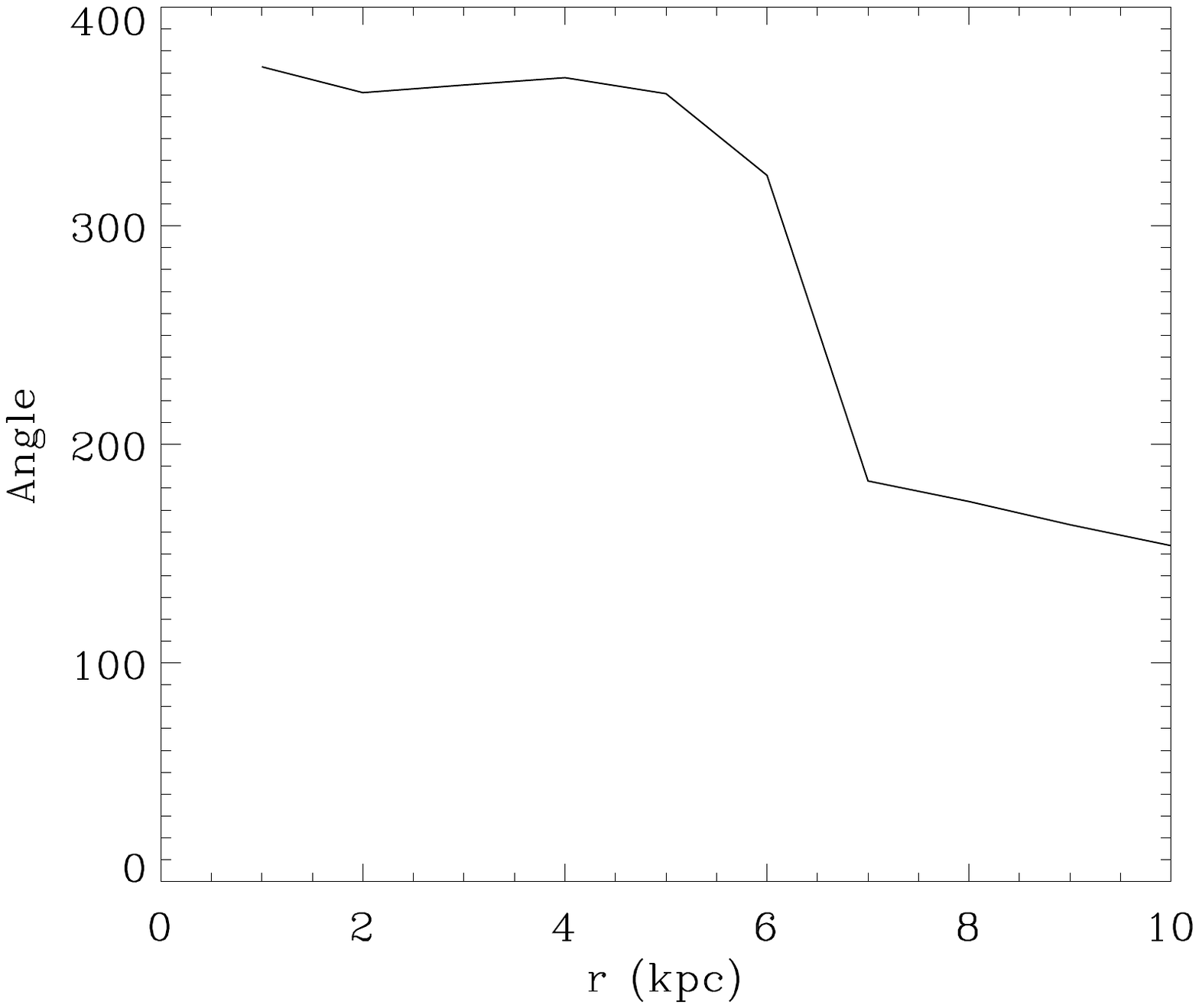}}
\caption{This figure shows the inclination angle (top) and position
  angle (lower) against radius. The disc is slightly warped, but by
  less than 10.3$^o$ for radii within 5 kpc, where observations are
  focused.}
\end{figure}

\subsection{Pattern speeds of the spiral arms}

One of the implicit assumptions in Lin-Shu density wave theory is that
the spiral arms have a coherent pattern which rotates at a fixed
pattern speed. However recent observations by \citet{Meidt2008} have
been unable to find a single pattern speed; rather they assign
different pattern speeds to different radii. In an attempt to retain
some contact with density wave theory, they suggest that there may be
multiple spiral arm modes in the disc, but also agree that their
results can be explained by a winding pattern. The most obvious
interpretation of their results is the latter, and even more simply,
that there is no global pattern speed. This can be seen from our
simulations, e.g. by comparing the galaxy at times of 180 and 300 Myr
(Fig.~17), where it is evident that the spiral arms are visibly
winding up.

We use the results of our Gaussian fitting to find the pattern speed
of the arms from our simulations, simply taking 
$\Omega_p=d\theta/dt$ at each radius $R$. In Fig.~15 we show the pattern speed for the
stars and gas versus radius, accompanied by a similar figure from
\citet{Meidt2008}. We show the pattern speeds for each arm, as we do
not assume they are necessarily the same. Fig.~15 shows that the
pattern speed is not constant with radius. Rather the pattern speed
decreases with radius. For Fig.~15, we calculate the pattern speed
between times of 180 and 300 Myr. Taking 240 Myr to 300 Myr produces a
similar result but with larger error bars. We note that our pattern
speeds agree roughly with those determined by \citet{Meidt2008}, lying
between 25 and 50 kpc km s$^{-1}$ compared to 50 kpc km s$^{-1}$ at
radii between 2 and 4 kpc and 25 kpc km s$^{-1}$ between 4 and 5 kpc. 
\citet{Egusa2009} also measure a pattern speed of 30 kpc km s$^{-1}$. 
In the simulations, the pattern speed of one arm starts
relatively flat, then drops at 4.5 kpc, corresponding again to the
kink observed in Fig.~5.
\begin{figure*}
\centerline{
\includegraphics[bb=0 200 600 350,scale=0.38]{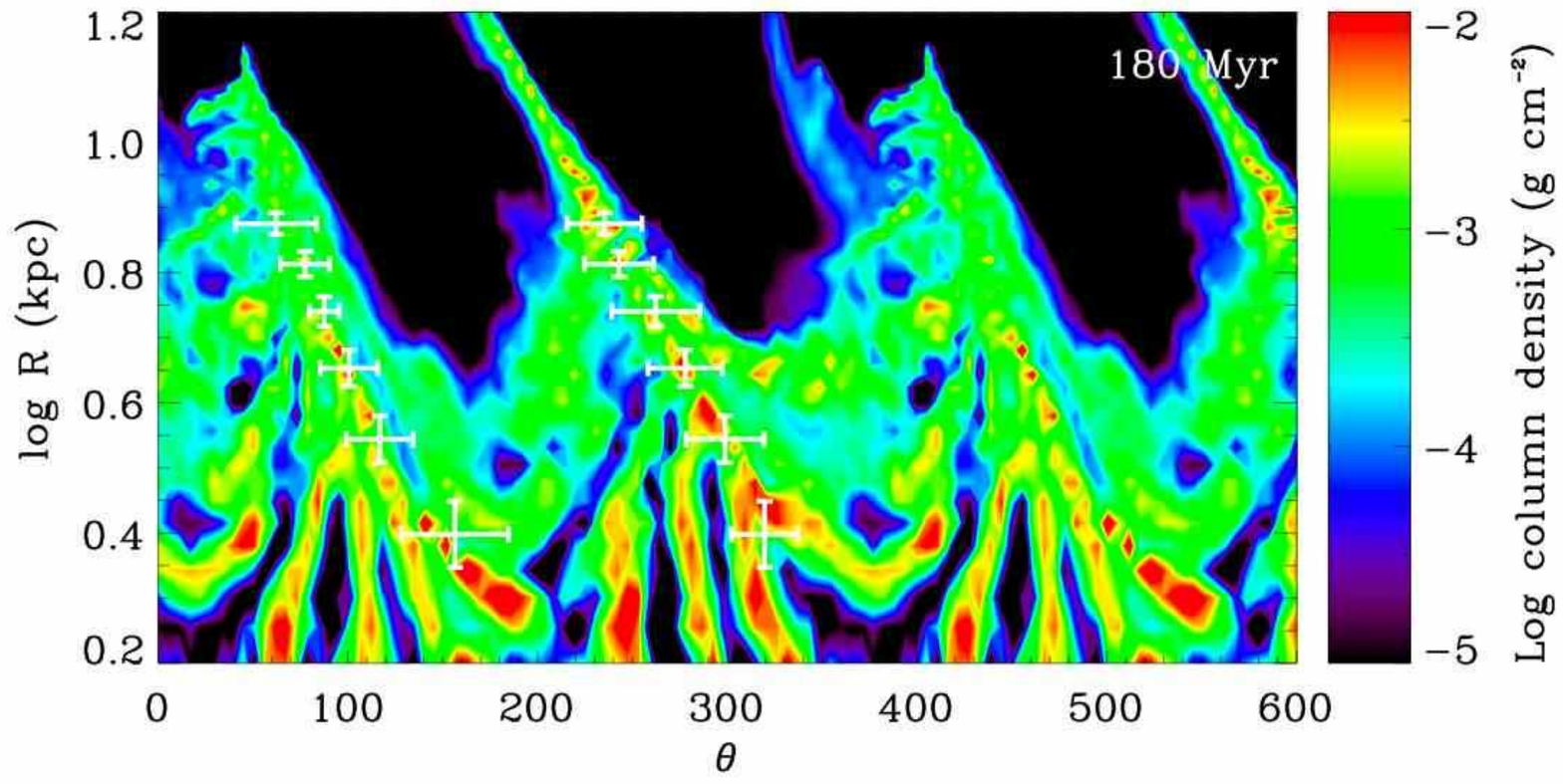}
\includegraphics[bb=0 200 600 350,scale=0.38]{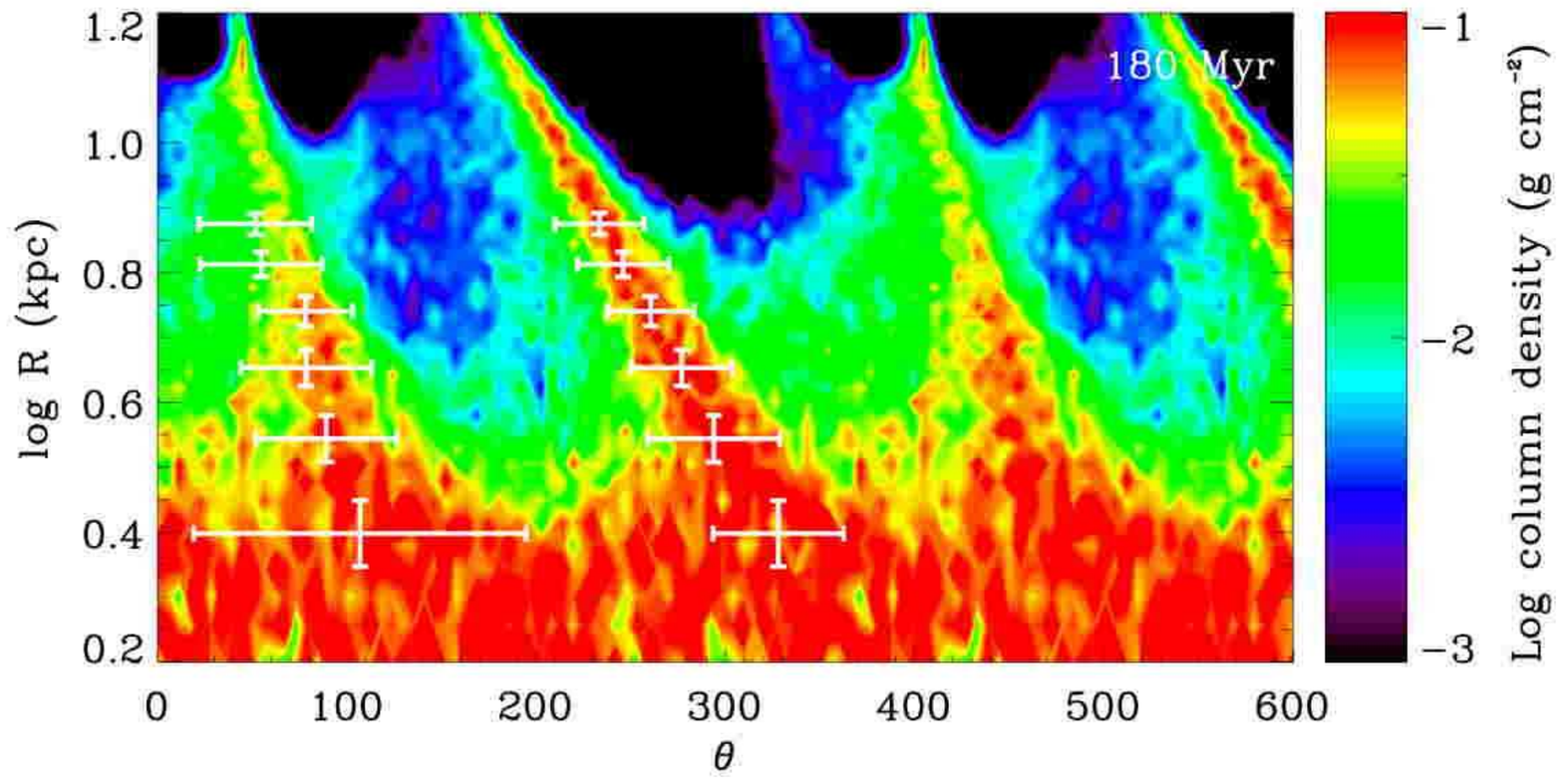}}
\centerline{
\includegraphics[bb=0 200 600 470,scale=0.38]{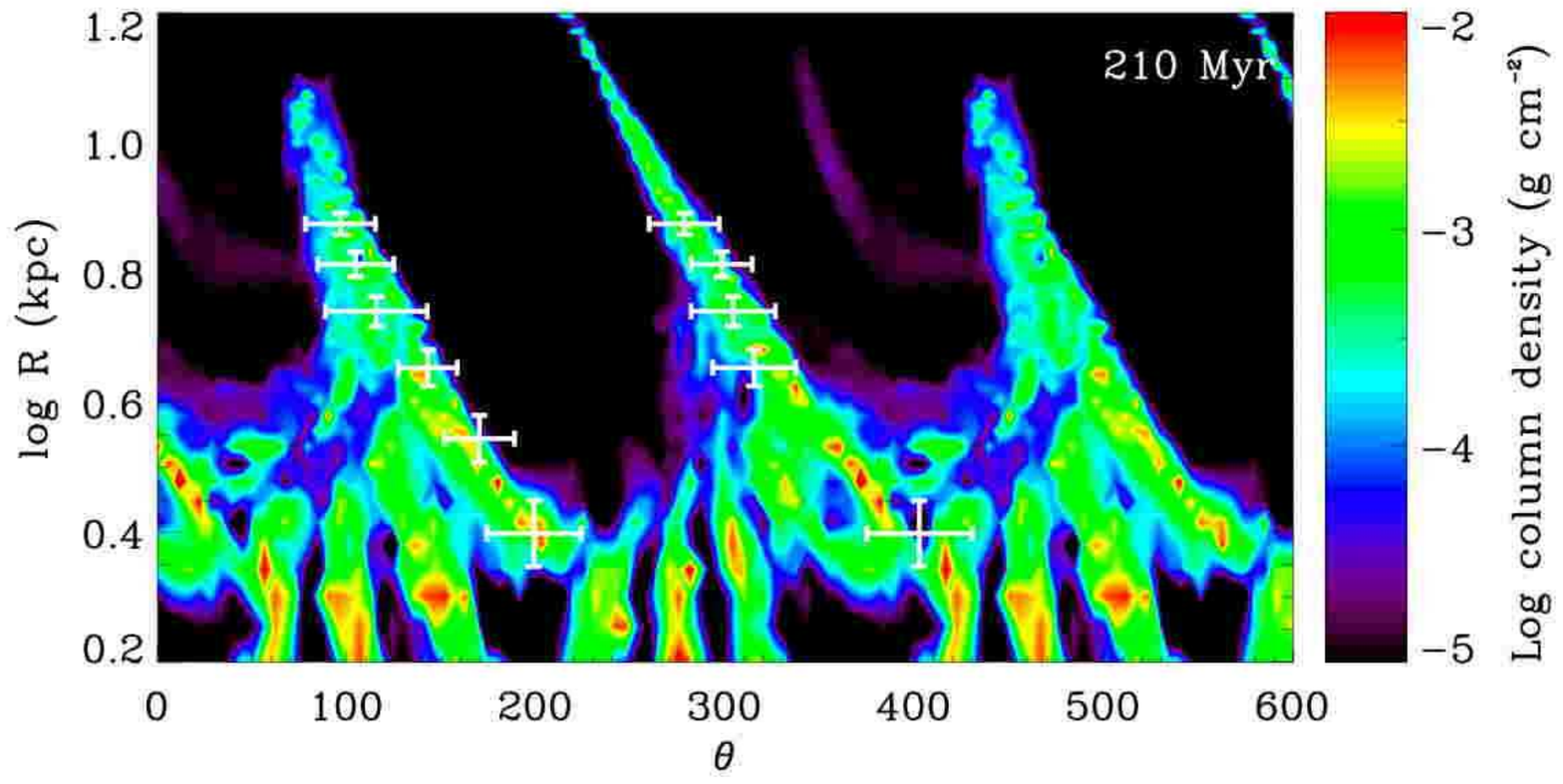}
\includegraphics[bb=0 200 600 470,scale=0.38]{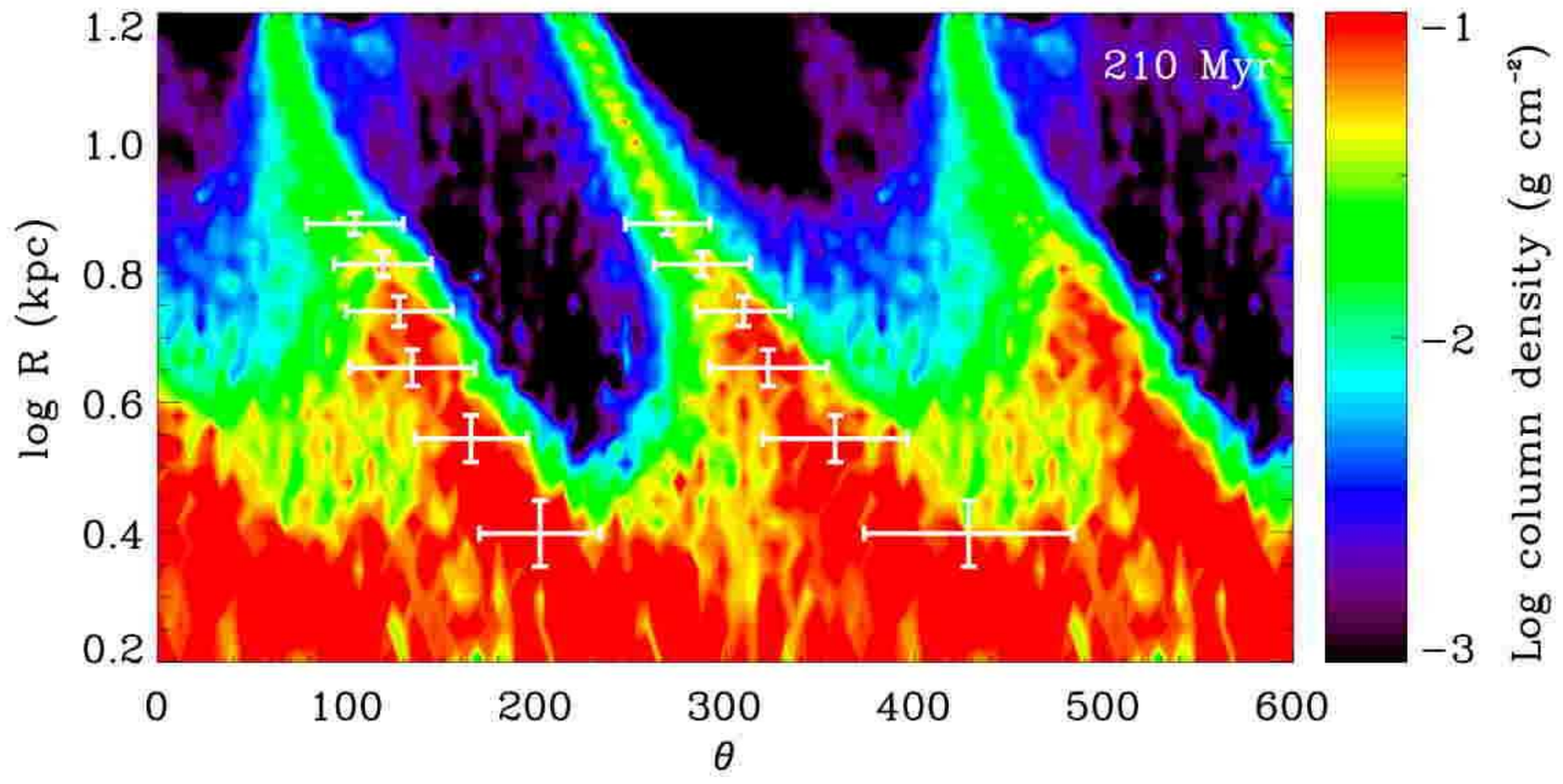}}
\centerline{
\includegraphics[bb=0 200 600 470,scale=0.38]{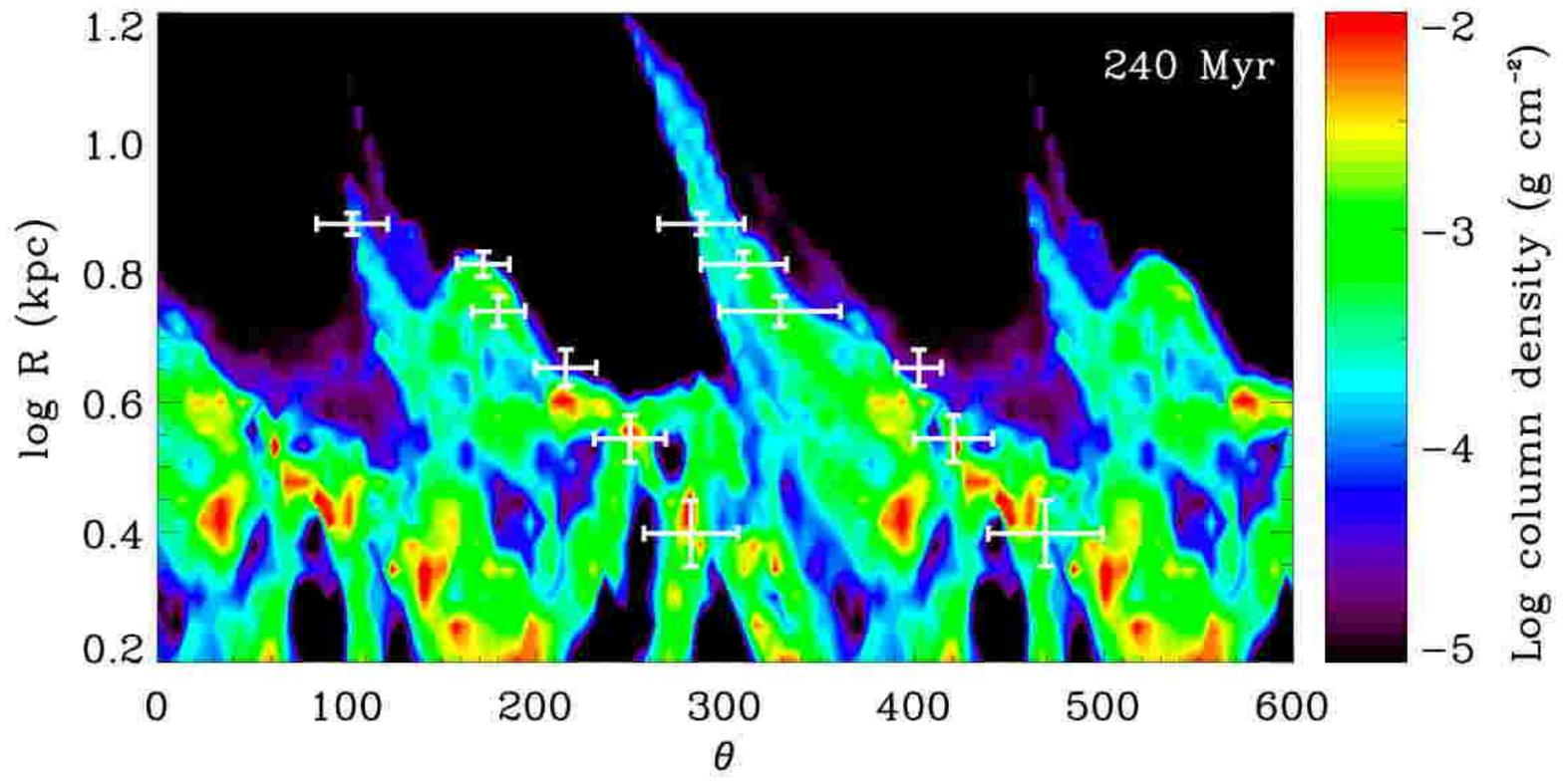}
\includegraphics[bb=0 200 600 470,scale=0.38]{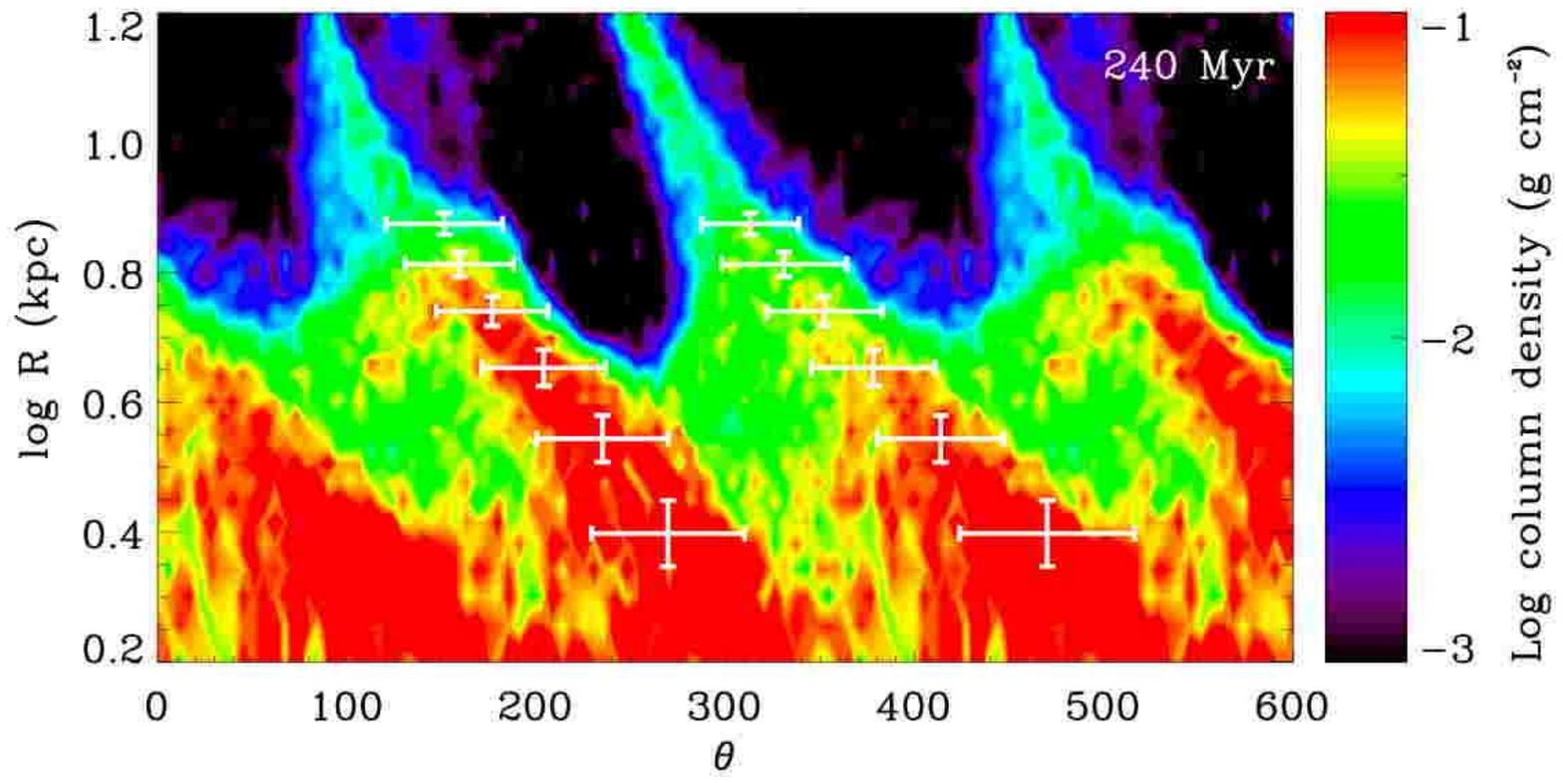}}
\centerline{
\includegraphics[bb=0 200 600 470,scale=0.38]{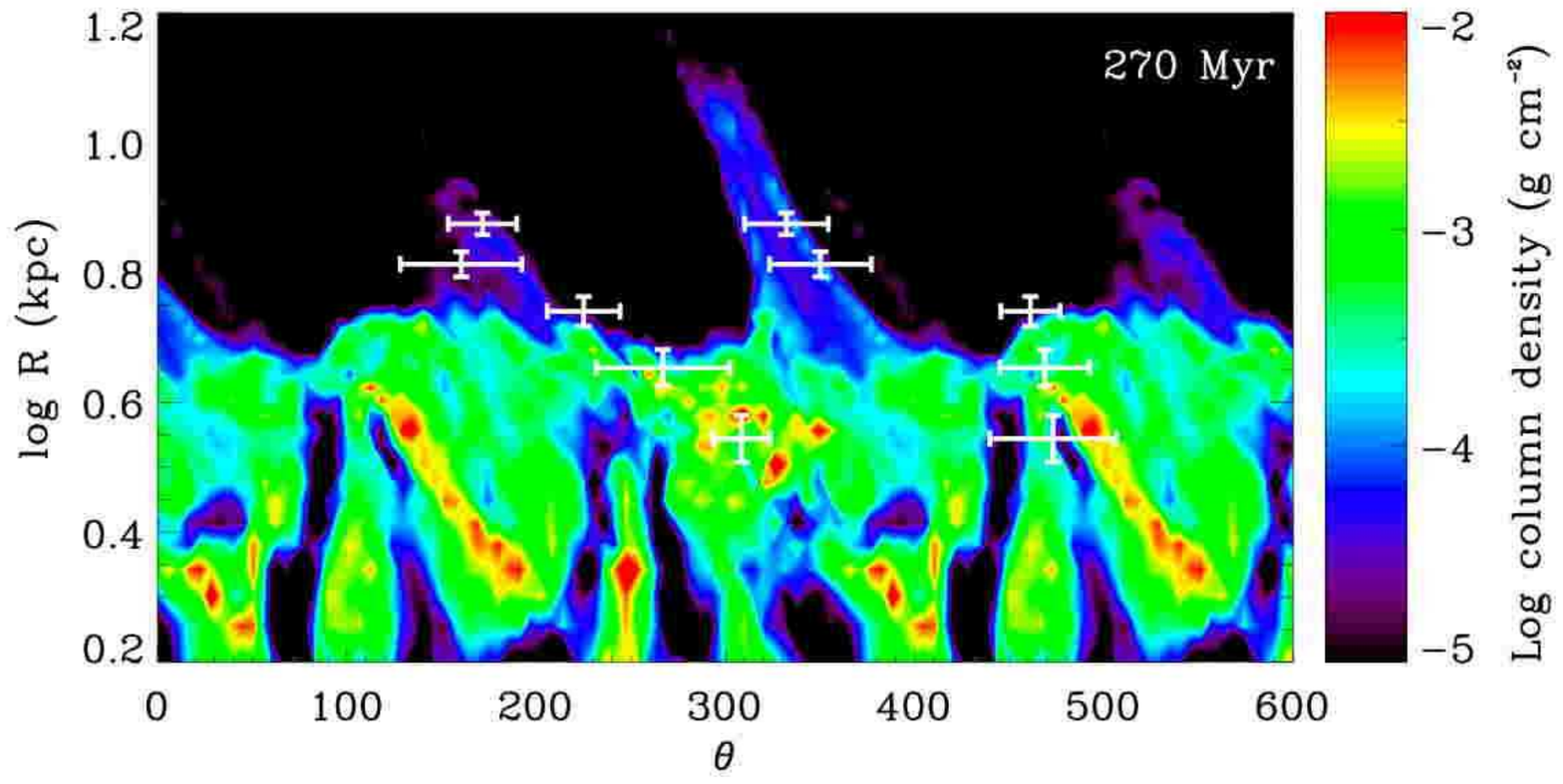}
\includegraphics[bb=0 200 600 470,scale=0.38]{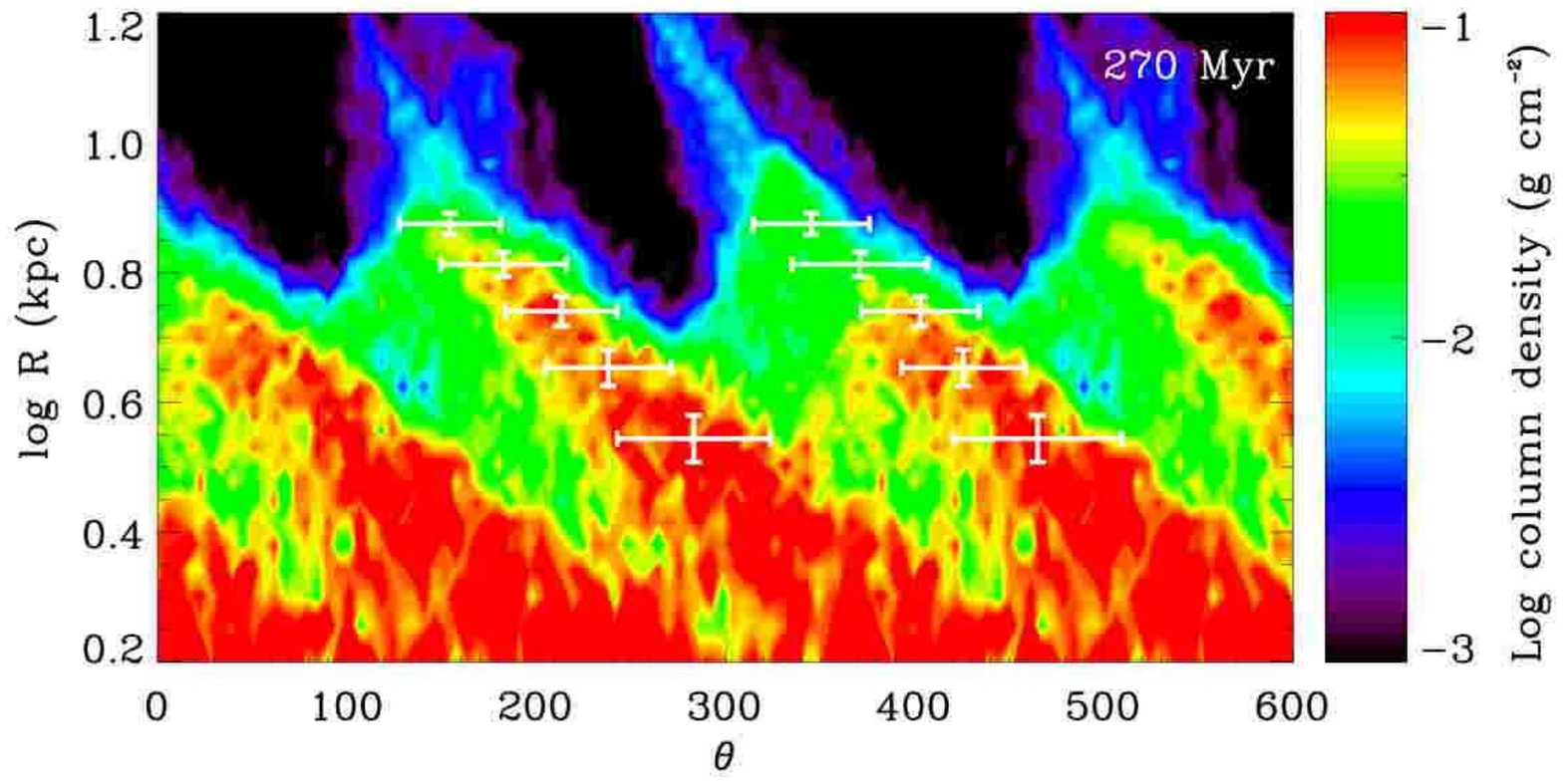}}
\centerline{
\includegraphics[bb=0 0 600 470,scale=0.38]{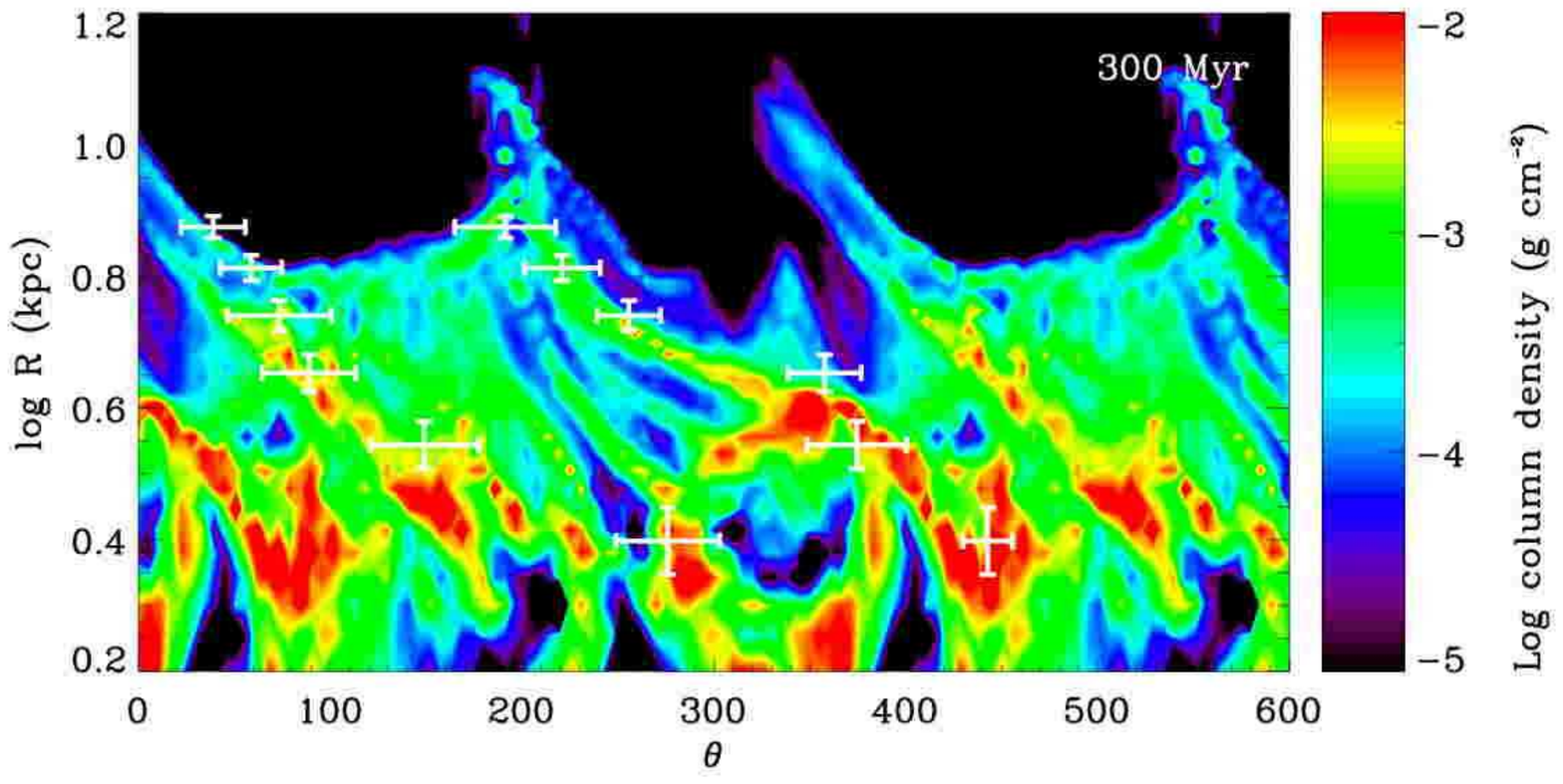}
\includegraphics[bb=0 0 600 470,scale=0.38]{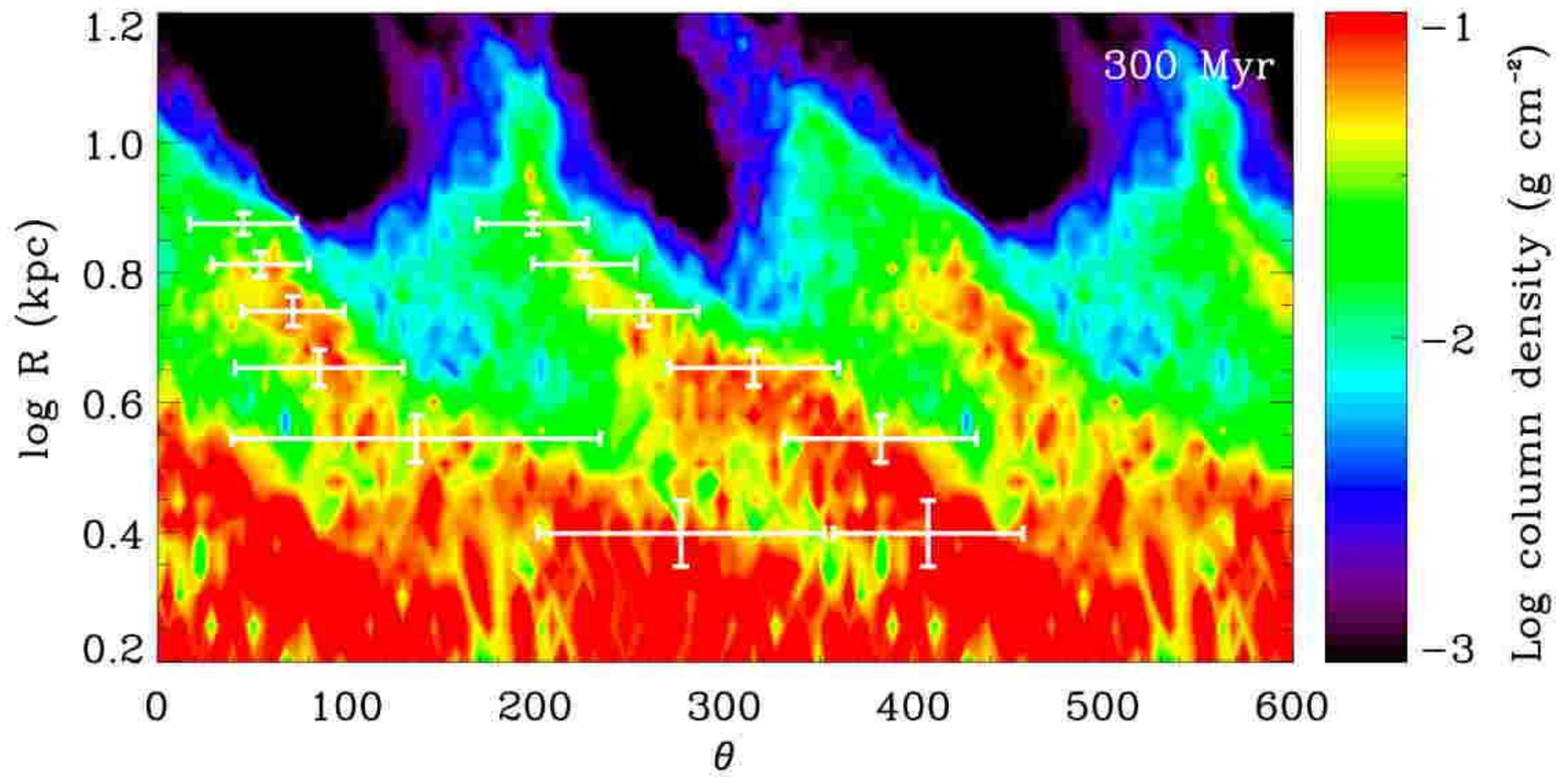}}
\caption{These plots show the column density of the gas (left) and
  stars (right) in $\log R$ versus $\theta$ space. For logarithmic
  spirals, we would expect to see straight lines indicating the spiral
  arms.  This is often not the case, particularly at later times
  (e.g. 210 Myr, 300 Myr). Instead the slope changes as the arms
  become more or less tightly wound. Even when the arms can be fitted
  approximately by straight lines (e.g. 180 Myr), the gradient of each
  appears to differ. Also shown on these plots are the density peaks
  found using out Gaussian fits, with 1$\sigma$ error bars.}
\end{figure*}

\section{Comparison with observations of density and velocity}

In this Section we make a comparison between our results and some of
the analysis presented in \citet{Shetty2007}. \citet{Shetty2007} note
some discrepancies between their observations and density wave theory,
in particular large circular and radial velocities.  We follow
the procedures outlined in \citet{Shetty2007} for analysing M51. They
show plots of column density on a $\log R$ versus azimuth space, and fit
straight lines to the spiral arms, which assumes an underlying shape
of the arms to be that of a equiangular spiral. They then plot
velocity against an angle $\psi$, where $\psi$ is the angle extending
from a spiral arm.

We first assessed whether there was a warp in the disc, as
\citet{Shetty2007} include the position angle in their
analysis. However although our disc shows a slight warp (see Fig.~16),
we considered it would make little difference to our results, at least
in the central parts of the disc.
\begin{figure*}
\centerline{
\includegraphics[bb=0 -5 550 200,scale=0.44]{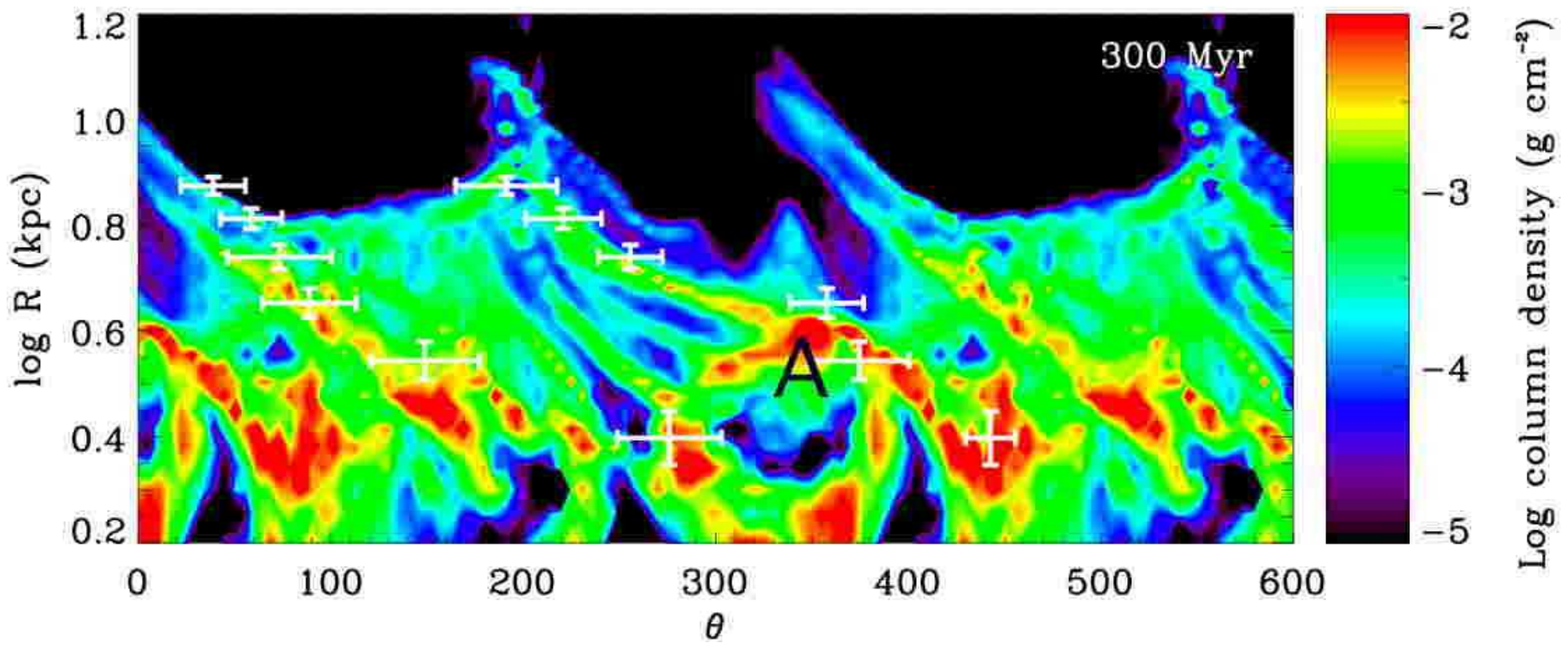}
\includegraphics[bb=10 -25 550 200,scale=0.5]{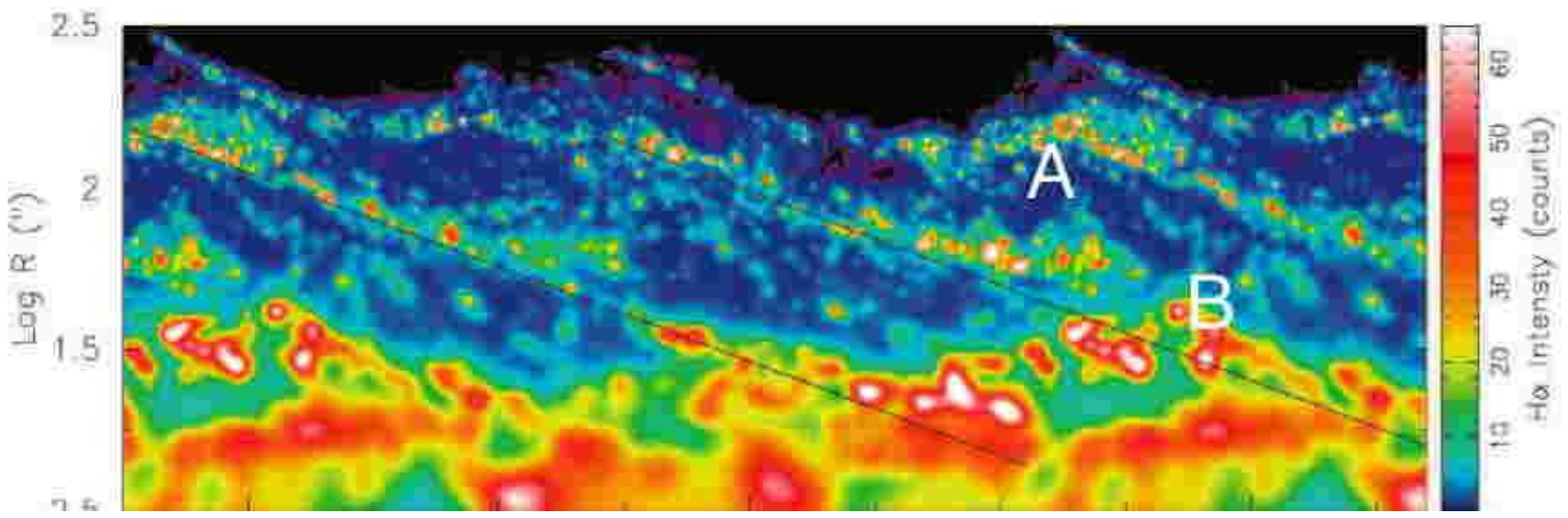}}
\caption{We select the $\log (R)$ versus $\theta$ plot at a time of 300
  Myr, corresponding approximately with the present day (left) and
  show the H$\alpha$ plot from \citet{Shetty2007} on the right 
  (where the observations and data reduction are taken from \citet{Vogel1993} and 
  \citet{Gruendl1996}). 
  We have marked a point A on each where the slope of the spiral arm
  clearly changes - this can be seen as the kink (also marked A) in
  Fig.~5. }
\end{figure*}

\subsection{$R-\theta$ plots of spiral arms}

In Fig.~17, we show the column density of gas (left) and stars (right)
on a polar plot, at times of 180, 210, 240, 270 and 300 Myr. Over
time, the spiral arms become shallower, in both stars and gas, again
indicating the pattern is becoming more tightly wound.

We also overplot the positions of the peaks determined from the
Gaussian fittings on Fig.~17. From the general shape of the arms, and
the positions of the peaks, it is apparent that in many cases, the
spiral arms cannot be easily fitted by straight lines, and are
therefore not logarithmic. Rather the gradient changes with
radius. Sometimes there is a clear, even sharp change in the slope of
the spiral arms.  For example at the time of 300 Myr, one arm becomes
shallower at larger radii, becoming more tightly wound. At 240 Myr,
one arm displays the opposite behaviour, becoming more open at larger
radii. The figure also shows that the two arms are often asymmetric,
displaying different gradients. At 270 Myr the structure of the gas is
very messy, hence the disparity between positions of the gas and
stellar arms seen in Fig.~14.

We compare in more detail the gas column density with a H$\alpha$ plot
from \citet{Shetty2007} in Fig.~18. Our results display considerably
more structure than the observations. This may simply be a consequence
of higher resolution and/or a consequence of the simplicity of our
treatment of the ISM. Also for both the simulations and observations
\citep{Shetty2007}, it is difficult to pick out the spiral arms at low
($ R<3$ kpc) radii.

There is a specific example of a deviation from logarithmic behaviour
at the point marked A. Here there is a kink in the spiral arm, and the
gradient becomes shallower. This in fact corresponds to the point
marked A in Fig.~6, a feature apparent in both the model and
observations. At this point, if the inner arm continued, it would be
much more open than observed. Instead, rather than continue with the
same pitch angle, the arm becomes more tightly wound. We postulate
again that the inner arm is longer lived (from the first crossing of
the perturber) whilst the outer spiral arm is much more recent
(Fig.~8), thus it is not surprising the sections have different pitch
angles, and different pattern speeds.

Interestingly there is also a bifurcation at the point marked A in
Fig.~18. This is the branch marked B in Fig.~6, which is actually
turning radially inwards. Although there is no such corresponding
feature in the observations, there is nevertheless what appears to be
a similar structure below A (marked B), on the next spiral arm, which
also moves radially inwards.

\subsection{Radial velocities}

Here we calculate the radial velocities of the gas. \citet{Shetty2007}
plot the average radial velocity versus azimuth, and find large radial
motions in the disc, and furthermore large net radial motions at
different radii. If the disc obeys standard density wave theory, we would
expect no large radial motions, and certainly no net radial motions.

In Fig.~19 we show the radial velocity versus $\psi$, the angle
between spiral arms. This formalism follows \citet{Shetty2007}, so
$\psi$=0 is located approximately on a spiral arm. We also plot the
density in Fig.~19, and both are averaged over a 1 kpc band placed at
$r=2$ (upper), and $r=4$ kpc (lower) at times of 180 and 300 Myr. The
radial velocities do not show particularly good agreement with
\citet{Shetty2007}. In \citet{Shetty2007}, the radial velocities tend
to be positive, whereas ours are a mixture of positive and
negative. This could indicate that the orbit of NGC~5195 is too
tightly bound in our simulations. The magnitudes of our radial
velocities are very high, even compared to \citet{Shetty2007}, and
although the velocity tends to dip at high density, the profiles are
more complex than would correspond to standard density wave theory.

In Fig.~20, we plot the net radial velocity versus radius, at
different times. At all times there are substantial net radial
velocities, particularly at larger radii. The figure also indicates
that the radial motion fluctuates between inwards and outwards,
indicating the disc is compressing and expanding.  The net radial
velocity in \citet{Shetty2007} varies from 0 to 30 km s$^{-1}$ for
radii $R <$ 4 kpc, but varies according to the assumed position
angle. Our results typically show half this range for $R<4$ kpc, and
tend to show a net negative velocity, which again may suggest that
NGC~5195 is too gravitationally bound in our simulations.

\section{Discussion}

We have modelled the galaxy M51 and its interaction with its companion
NGC~5195, focusing primarily on the dynamics of the gas, and secondly
the stellar disc. The tidal interaction produces spiral arms in the
stars and in the gas. The resulting spiral structure shows excellent
agreement with that of M51, and we have even successfully reproduced
individual kinks and branches identified in M51. We also capture the
larger scale structure, including the extensive tail of HI.

\subsection{Density `waves'}

The spiral structure found in interacting, and other, galaxies is
often interpreted in terms of `density wave theory'.  There is,
however, much ambiguity in the literature as to what is meant or
understood by the phrases `density waves' or `density wave
theory'. Before proceeding it is important to clarify the underlying
physical concepts and also the terminology we employ in this paper
with regard to the spiral structure we see in galaxies and in our
simulations. What we say here is a simplification of the more thorough
and detailed analysis to be found in Chapter 6 of \citet{Binney}.  
Our discussion will focus on the structure apparent in the
stellar disc as it is in the stellar disc that the structures are
excited and maintained. The gas for the most part merely responds to
the time-- and space--varying potential of the stars. Because the
orbits of the gas elements cannot intersect without producing shocks,
the gas acts as an amplifier for identifying the underlying stellar
gravitational potential. In addition the gaseous response is easier to
observe.

\subsubsection{Material spiral arms}

The simplest concept is that of {\it material spiral arms}. Consider a
galaxy in which all the stars in the disc are on circular orbits, and
ignore the self-gravity of the stars. That is, we assume the
contribution to the potential in which the stars are moving comes from
the bulge and the halo. Suppose at a particular instant one attaches a
flag to each of the stars which lie along a diameter in this galaxy. Then
at a later time, because the rotation rate $\Omega(R)$ is a decreasing
function of radius $R$, the flags will form a pattern in the shape of
a trailing spiral arm. This is a material spiral arm and will wind up
locally at a rate $ |d \Omega / d \ln R |$. Because $d \Omega / dR <
0$ the arms are trailing spirals. Because typically the gas in the
disc of a galaxy is moving highly supersonically, radial pressure
gradients are negligible and the angular velocity of the gas is almost
identical to that of the stars. Thus to a first approximation the gas
comoves with the arms. An example of material arms is to be found in
flocculent spirals, where each small segment of spiral arms forms
locally through self-gravity (e.g. Dobbs and Bonnell, 2008). The density
enhancement essentially comoves with those stars which form the
locally self-gravitating entity. The entity then becomes wound up as a
trailing spiral which is then dispersed on a timescale of order $ 1 /
| d \Omega /d \ln R | $. 
\begin{figure}
\centerline{
\includegraphics[bb=50 350 550 780,scale=0.24]{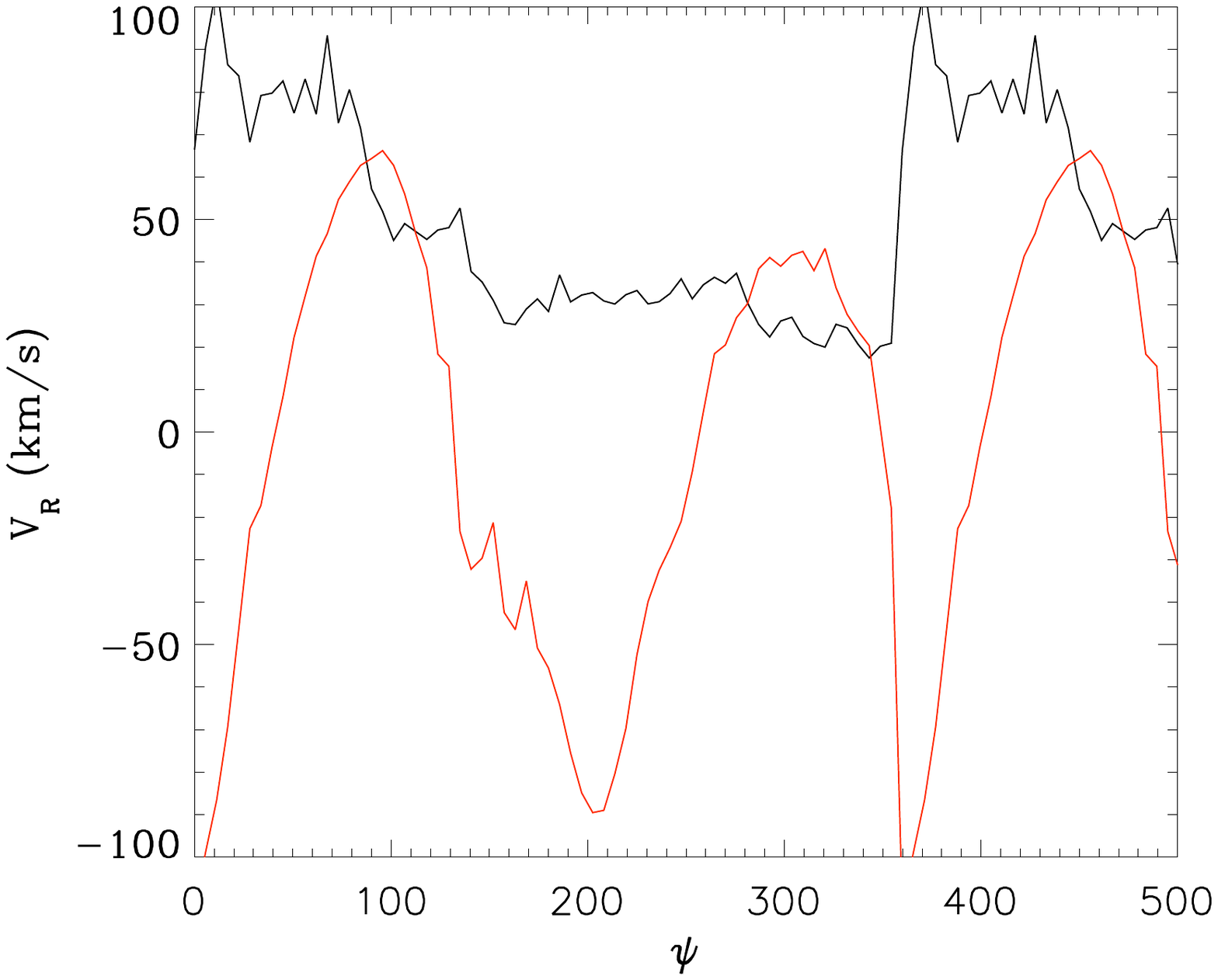}
\includegraphics[bb=50 350 550 780,scale=0.24]{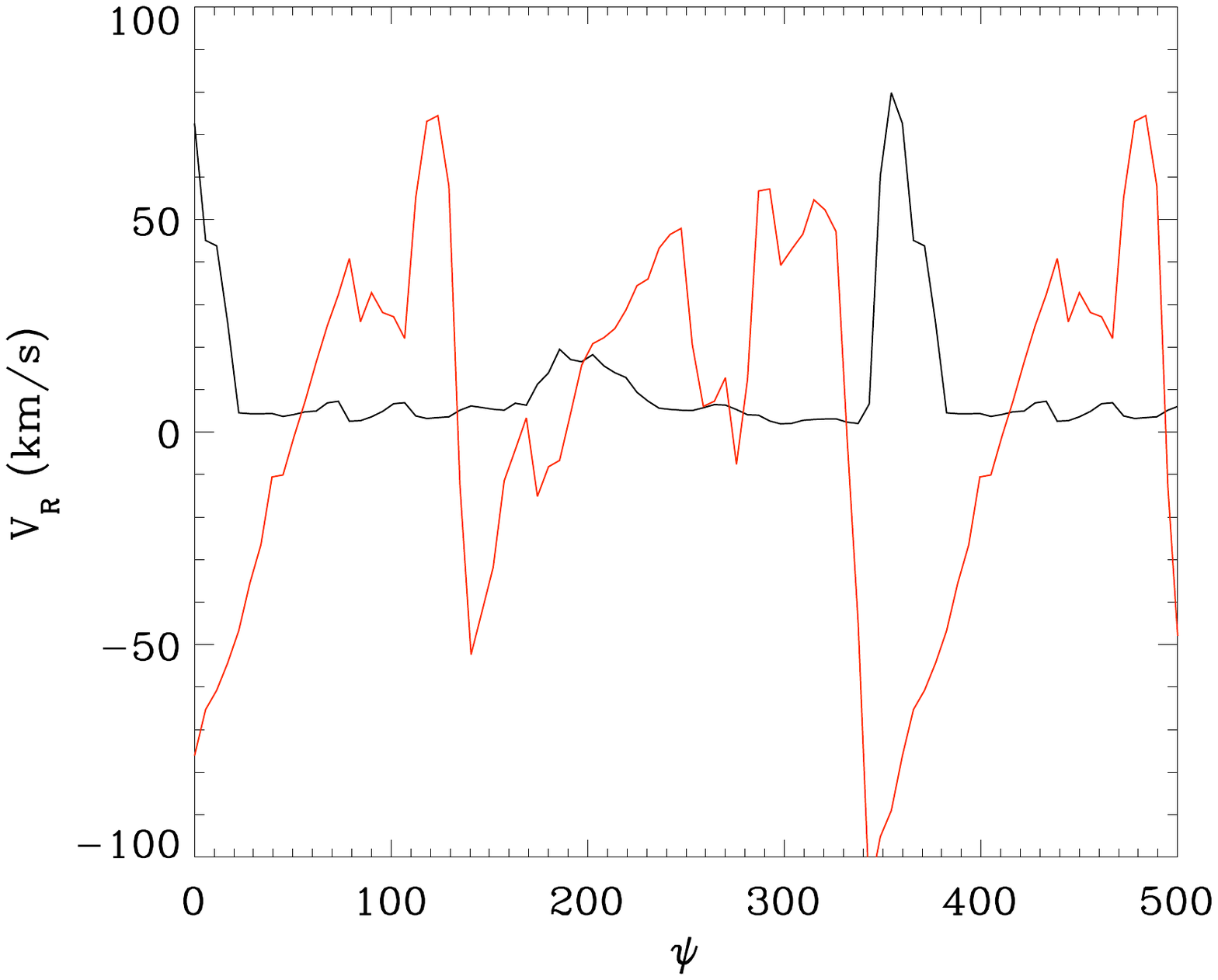}}
\centerline{
\includegraphics[bb=50 350 550 780,scale=0.24]{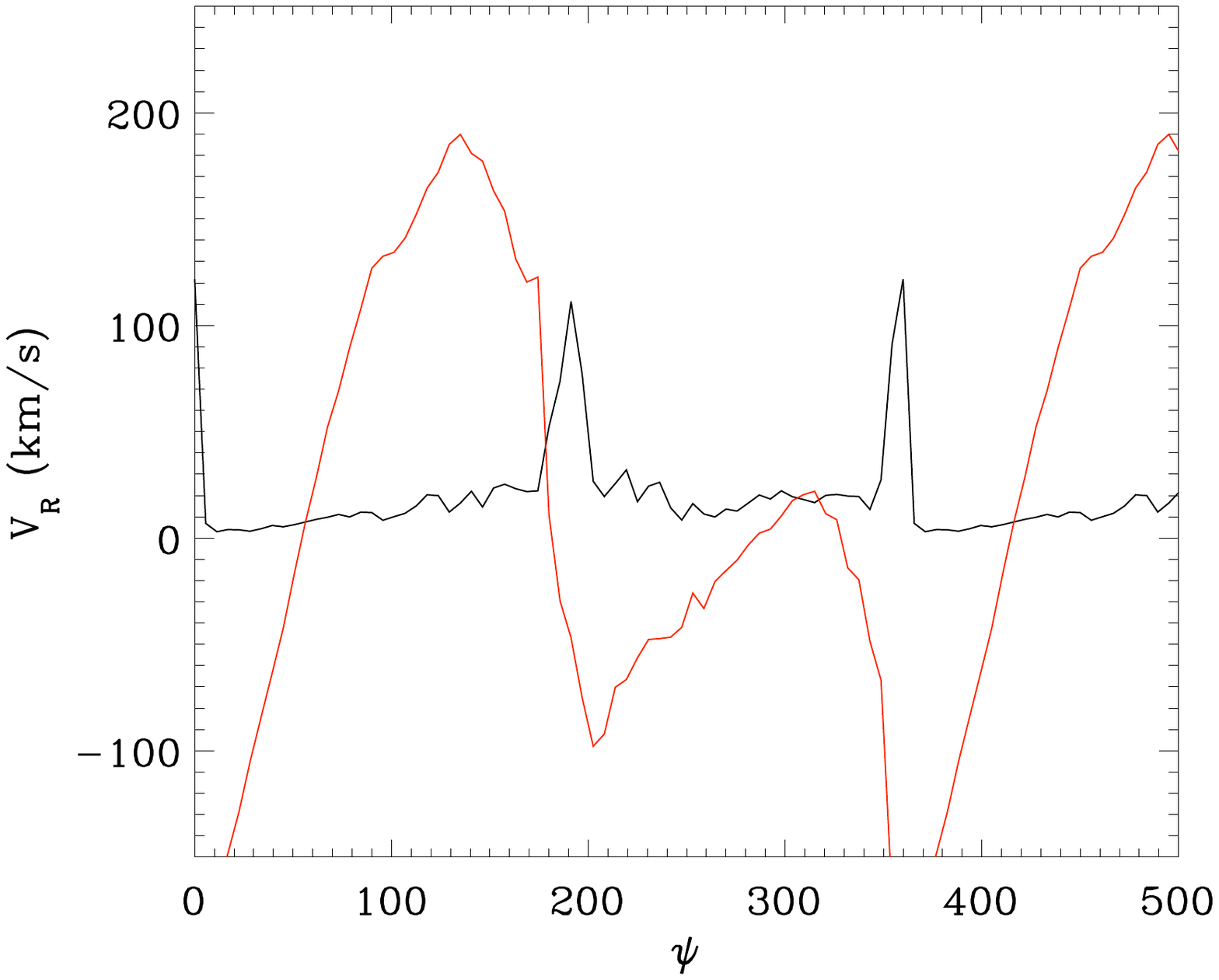}
\includegraphics[bb=50 350 550 780,scale=0.24]{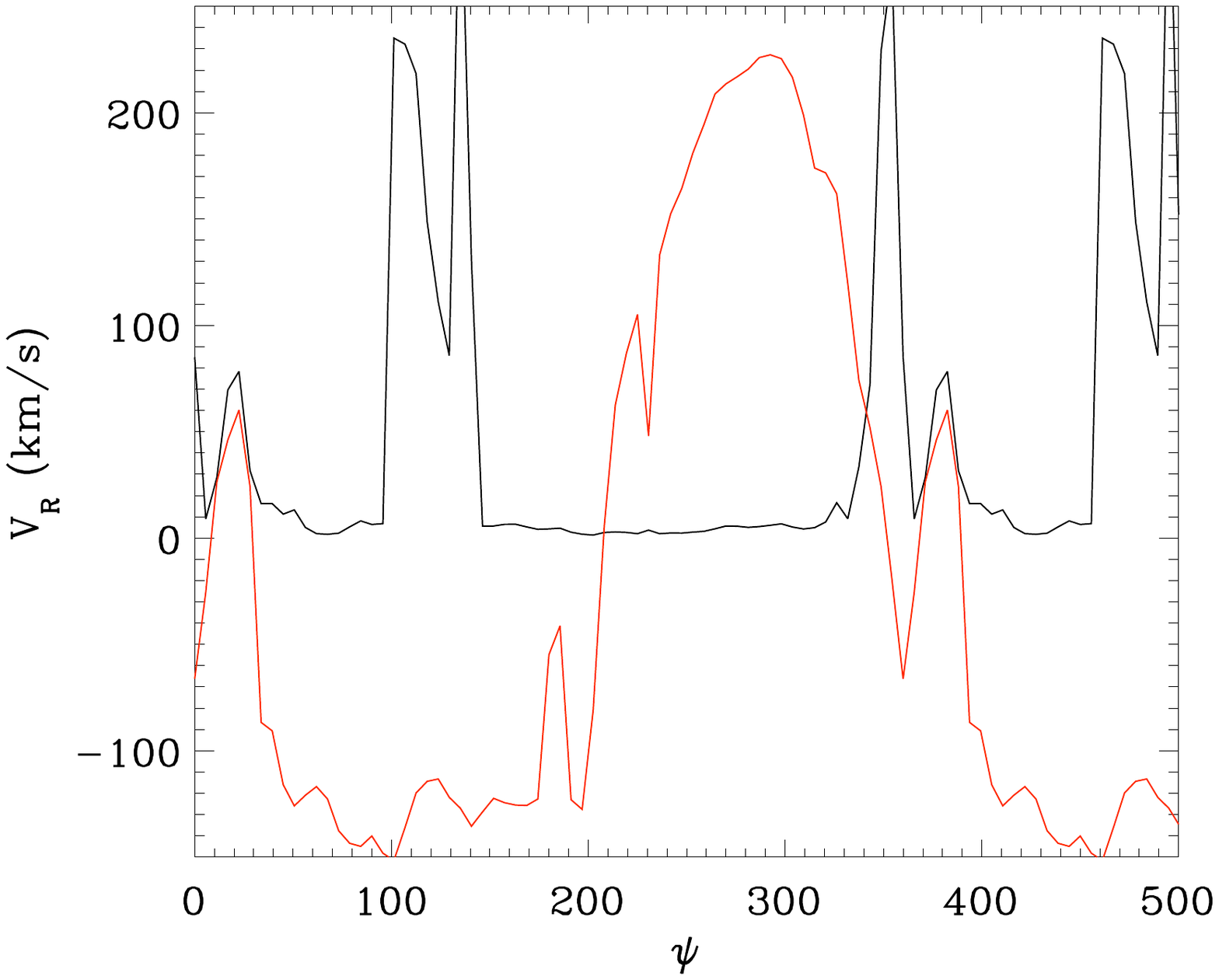}}
\caption{We show the (mass weighted) radial velocity (red) plotted versus azimuth at
  times of 180 Myr (left) and 300 Myr (right), and at radii of 2 (top)
  and 4 (lower) kpc. The density, in arbitrary units is also shown in
  black. The radial velocities are much larger than expected from
  density wave theory.}
\end{figure}

\begin{figure}
\centerline{
\includegraphics[bb=50 350 550 780,scale=0.45]{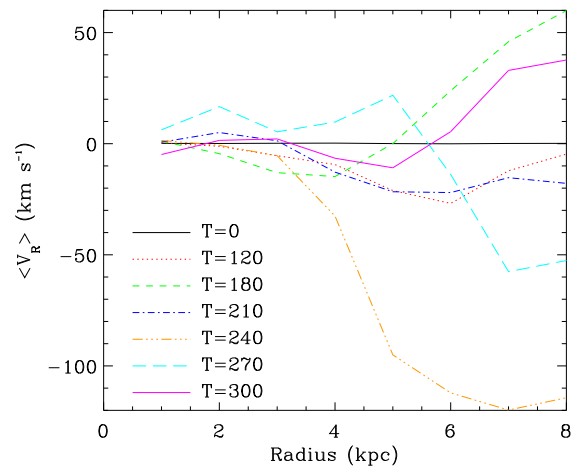}}
\caption{This figure shows the net radial velocity in 1 kpc bands,
  versus radius at different times during the simulation. The net
  radial velocity fluctuates from positive to negative values,
  indicating compression and expansion of the disc.}
\end{figure}

\subsubsection{Kinematic density waves}

The next simplest concept is that of {\it kinematic density waves}
(\citealt{Binney}, Section 6.2). These occur when, as we
assumed above, the self-gravity of the stars is negligibly small. The
nomenclature is misleading because they are not actually waves; there
is no oscillatory behaviour, no restoring forces and they do not
propagate (the group velocity is zero). Consider the galaxy described
above, and consider a star in a circular orbit at radius $R$. Give
this star a small radial impulse. Then, to linear order, the stellar
orbit is an ellipse, with one of its foci at the centre of the galaxy
(an $m=1$ perturbation). The star orbits the centre of the galaxy with
frequency $\Omega(R)$ and its distance from the centre of the galaxy
in the radial direction oscillates with frequency $\kappa(R)$ where
$\kappa$ is the epicyclic frequency, given by
\begin{equation}
\kappa^2 (R) = R \frac{d \Omega^2}{d R} + 4 \Omega^2 ,
\end{equation}
and in general $\kappa > \Omega$.  Thus in the inertial, non-rotating
frame, the orbit is not closed, but appears to precess (retrogradely)
at a rate $\Omega - \kappa$. However, in a frame which rotates
(progradely) at a rate
\begin{equation}
\Omega_{\rm ILR}(R) = \Omega(R) - \frac{1}{2} \kappa(R),
\end{equation}
the orbit is closed. In this frame the orbit, to linear order, takes
the form of an ellipse with centre coincident with the centre of the
galaxy (an $m=2$ perturbation). 

Since this star's orbit is non-circular, by conservation of angular
momentum it slows down when it is furthest from the centre. By
considering this star and its near neighbours, we can see that this
means that its neighbours get closer to it at this point, or, in other
words, the stellar density is enhanced there. To illustrate this,
imagine that we attach a light to this star, and make it flash each
time the star reaches its greatest orbital extent. Thus when the light
is on, it represents a region of enhanced density. Then in the frame
rotating with angular velocity $\Omega_{\rm ILR}$ the light flashes
with period $2 \pi/ \kappa$ but at points in this frame which are
fixed either side of the centre.

Now consider a set of stars which are initially uniformly distributed
around the circular orbit at radius $R$, and which are subject to a
perturbation by an object orbiting at some larger radius, thus
the perturbing object has some angular velocity $\Omega_{\rm pert} <
\Omega$. To simplify matters, suppose that the effect of the
perturbation is to give each star a small radial impulse at the moment
when it is closest to the perturber (an $m=1$ perturbation~\footnote{Note that for a tidal interaction, as we have in this paper, an m=2 perturbation is dominant. In this case,
we would introduce an impulse to the star closest to, and the star furthest from the, perturbed galaxy. Thus an
$m=2$ pattern again occurs, but induced at both the nearest and opposite side from the perturber simultaneously.}).
Also give each star a light which flashes when its orbit is at its greatest extent.
Then in the frame rotating with angular velocity $\Omega_{\rm ILR}$ we would see a
succession of flashes, either side of the centre, whose positions
advance in azimuth at a rate $\Omega_{\rm pert} - \Omega_{\rm
ILR}$.
Suppose the perturber has an angular frequency
 $\Omega_{\rm pert} = \Omega_{\rm ILR}$, then the succession of
flashes are all at the same fixed points in this frame, either side of the centre.
In other words, we see a steady light each side of the centre.
This would then represent a steady $m=2$ pattern of density which,
in the inertial frame, rotates with angular velocity $\Omega_{\rm ILR}$.

For a more general case, the perturber will induce a temporal Fourier component,
which coincides with $\Omega_{\rm ILR}$ at a given radius R.  This excites the
$m=2$ density pattern (of angular velocity $\Omega_{\rm ILR}$) as described above.
Extending over all radii, the effect of the perturbing galaxy is to induce a
{\it coherent} $m=2$ density enhancement across the disc, rotating with $\Omega_{\rm
ILR}(R)$. In general, it turns out that
\begin{equation}
\frac{ d \Omega_{\rm ILR}}{ d R} < 0,
\end{equation}
Thus, just as for the material spiral arms, these density enhancements
form a density structure in the shape of an $m=2$ trailing spiral
arm. At each radius the density maximum rotates with a local pattern
speed $\Omega_p(R) = \Omega_{\rm ILR}(R)$.

Such a structure is called a {\it kinematic density wave}. It appears
to be called a wave because the stars (and also therefore the gas)
move through the density maxima (indeed at a rate $\Omega -
\Omega_{\rm ILR}$), and so the particles forming the arms change with
time. But the term {\it wave} is misleading, because the structure so
formed does not propagate. The term {\it kinematic density pattern}
might be more appropriate.

The importance of these patterns is twofold. First, they are $m=2$ and
so are generically induced by tidal perturbations (just as the
moon produces two tides a day on the Earth)~\footnote{ The central
  bars in barred galaxies also produce $m=2$ perturbations which can
  set up similar patterns, but we do not consider those here.}. Second,
the spiral pattern so formed winds up at a rate $ | d
\Omega_{\rm ILR} / d \ln R| \ll | d \Omega / d \ln R |$, and so can be
comparatively long-lasting.

\subsubsection{Spiral density waves}

Now consider what happens to a kinematic density pattern if we allow
the stars which form it to be mildly self-gravitating. Then the spiral
density pattern gives rise to a corresponding spiral gravitational
potential. This means that the pattern, which was previously being
sheared out by differential precession, is now able to show some
coherence. For example, the stars in the gravitational maxima now feel
mutual attraction which counteracts the differential shear. This has
two effects: (i) the local rate of precession of the pattern is
modified, indeed now $\Omega_p > \Omega_{\rm ILR}$ and (ii) the
pattern now becomes a genuine wave in that it is able to propagate in
the radial direction. The radial group velocity, $V_{\rm g}$, of such
waves is given by (\citealt{Binney}, Chapter 6.2.4):
\begin{equation}
  \frac{V_{\rm g}}{R \Omega} = f \, \frac{2 \pi}{3.36} \, \frac{\Delta
    R}{R} \,  \frac{1}{Q},
\end{equation}
where $\Delta R = \sigma_R/\kappa$ is the radial excursion of stellar
orbits, given a radial velocity dispersion $\sigma_R$, $Q = (\sigma_R
\, \kappa)/(3.36 \,G \, \Sigma)$ is the usual Toomre parameter which
measures the strength of self-gravity, and $f$ is a dimensionless
factor which can be derived from the relevant dispersion relation.

\subsubsection{Density wave theory of spiral arms}

The aim of density wave theory, as propounded by \citet{Lin1964},
was to show that large-scale spiral structure within a galaxy can be
self-maintained in a quasi-steady state, without input from external
perturbations. If true, then this could account for grand design
spiral structure seen in galaxies without the need to appeal to
internal bars or external tidal interactions. Because, in the presence
of self-gravity, density waves propagate radially, it seemed a
reasonable proposition that the spiral pattern seen in grand design
spiral galaxies might correspond to a global, long-lasting,
self-sustaining spiral mode, with some pattern speed $\Omega_{\rm p}$
{\it independent of radius}. For such a fixed pattern speed, stellar
spiral density waves are able to propagate only at radii between the
inner Lindblad radius ($R_{\rm ILR}$) at which $\Omega_{\rm ILR} =
\Omega_p$ and the outer Lindblad radius ($R_{\rm OLR}$) at which $\Omega
+ \frac{1}{2} \kappa = \Omega_{\rm p}$. Between these two radii lies
the co-rotation radius ($R_{\rm CO}$) at which $\Omega = \Omega_{\rm
  p}$. In order to set up such a spiral density mode using spiral
density waves it is necessary for the waves to be able to propagate to
and fro between $R_{\rm ILR}$ and $R_{\rm OLR}$ (analogous to the
mechanism of setting up a vibrating mode on a violin string requiring
travelling waves to be able to communicate between the two ends of the
string). The major problem for this theory is that it has never been
shown how this can be achieved in practice. There are a number of
reasons for this. An example of one is that a major observational
advance since this theory was first propounded is that it is now known
that all observed spiral structures in galaxies consist of {\it
  trailing} spiral arms. Such structures propagate only inwards at
radii $R < R_{\rm CO}$ and outwards at radii $R > R_{\rm CO}$. Thus
using the {\it observed} spiral arms as density waves to set up a spiral
`mode' of the kind observed is not feasible. A further consequence of the 
lack of leading spiral arms is that the proposed feedback mechanism, by which global modes are maintained by the conversion of leading to trailing arms \citep{Mark1976}, does not appear to operate.
%A further consequence the 
%lack of leading spiral arms is that `swing amplification', which has
%been discussed as a possible energising process to maintain the
%putative global spiral modes, and which occurs at the moment when
%leading spiral arms are sheared by differential rotation to become
%trailing spiral arms does not appear to operate.

\subsubsection{Illustration}

An illustration of the response to be expected by a stellar galactic
disc to a tidal perturbation can be found in the paper by
\citet{Oh2008}. They consider the response of a mildly
self-gravitating disc of stars to a perturber which flies by on a
parabolic orbit. Although there are some limitations to their models (they 
adopt 2D thin disc models and fix the position of the perturbed galaxy), 
they find that the dominant response at late times is $m=2$, and
as can be seen from their Figure 14, the temporal behaviour of the
response at late times has a pattern speed equal to $\Omega_{\rm
  ILR}(R)$. This is true for their Model A2$^\ast$ in which stellar
self-gravity is set to zero, so that the response corresponds to a
kinematic density pattern and cannot propagate radially, and also for
their Model A2 which has mild stellar self-gravity and shows evidence
for some inward radial propagation.

\subsection{Our simulations}

The simulations of the interaction between M51 and NGC~5195 that we
have performed show that the response of the stellar and gas discs to
the interaction is strongly time-dependent and dynamic. Thus trying to
interpret the response in terms of full density wave theory, with
spiral modes of fixed pattern speeds (if they exist) is not a fruitful
way to proceed. From our detailed analysis, we find compelling
evidence that M51 does not fit the Lin-Shu hypothesis of spiral
density wave theory. This has been hinted at in recent observations
\citep{Shetty2007,Meidt2008} (see also \citealt{Buta2009}). Rather
than being a rigid pattern, the spiral pattern continuously evolves
throughout the interaction. The lengths, amplitudes, and azimuthal
separation of the arms change with time. Moreover, the spiral pattern
winds up over a timescale $\gtrsim$ 100 Myr (cf. \citealt*{Merrifield2006}). 
Thus there is no well defined global pattern speed for
the spiral arms; there is, rather, a radially decreasing pattern speed
of the spiral arms (Figure~15). There we see that the radial variation
in pattern speed is about $\Delta \Omega_p \approx 20$ km s$^{-1}$
kpc$^{-1}$ which corresponds to a wind up timescale of $2 \pi / \Delta
\Omega_p \approx 300$ Myr. Indeed what we are seeing is similar to the
findings of \citet{Oh2008}. The pattern speeds we find for the arms
(both stellar and gaseous) (Figure 15) lie slightly above the curve
$\Omega_{\rm ILR} (R)$ which is what we expect for a tidally induced
kinematic density pattern with the addition of mild self-gravity. The
angular frequency of the perturbing galaxy (NGC~5195) varies as it
proceeds along its orbit in the range 10 -- 30 km s$^{-1}$ kpc$^{-1}$,
with the highest frequencies corresponding to the closest approaches,
and so to the strongest tides. The response frequencies lie mainly in
this range.

We describe a highly chaotic picture of the dynamics of M51 and
NGC~5195. The companion orbits M51 twice, with two close non-coplanar
encounters producing a predominantly $m=2$ spiral response during each
passage. As seen from Fig.~17, the main spiral arms become more
difficult to distinguish with time, and show abrupt changes in pitch
angle. In addition dynamical interactions between the tidal impulses
and the arms and between the spiral arms themselves lead to apparent
bifurcations in the form of secondary arms, branches and spurs. We
show that these interarm branches are simply material shearing away
from the spiral arm. The simulations produce branches which fork away
from a spiral arm both inwards and outwards along the spiral arm.
From the detailed evolution of the disc, we find that features
including the kink along one spiral arm and large interarm branches
evolve relatively quickly (over 10's of Myr). They are not long
lasting features compared to the arms themselves. Thus, for example,
the interpretation of spiral arm bifurcations and other structure in
terms of 4:1 ultra-harmonic or other resonances
\citep{Elmegreen1989c,Chak2003} in such tidally driven structure is not
a fruitful exercise.

A number of authors have attempted to use the offsets between the
stellar spiral potential dips and the shocks in the gas, delineated
for example by molecular gas and/or star formation, as a means of
estimating the relative flow speed of the (stellar) spiral pattern and
the gas, often relying on the basic assumption that the spiral has a
fixed pattern speed (e.g. \citealt{Westpfahl1998,Gittins2004,
Kendall2008, Mart2009, Buta2009}). 
While this can be a useful procedure in barred galaxies \citep{Buta2009}, 
where one might expect there to be an overall
pattern speed driven by the rotation speed of the bar, it is likely to
be less so in interacting galaxies. We have already noted that in
flocculent galaxies it is expected that there is essentially no offset
between the stars and the gas \citep{DB2008}. From our
simulations of a strong tidal interaction we have found (Figures 11,
14 and 17) that there is little offset between the gaseous and stellar
arms, and that when offsets can be discerned they can have either
sign, and moreover the sign of the offset can vary with radius. This
comes about because in the case of M51 the tidal interaction is strong
and still on--going, and thus the responses both in the stars and in
the gas are still strongly dynamical.

\subsubsection{Swing amplification}
Previous simulations have highlighted the role of swing amplification in
maintaining spiral structure in interacting systems \citep{Hernquist1990,Donner1994}. With sufficient self gravity, spiral arms subject to shear become narrower and more pronounced. The stellar spiral arms in our simulation appear to be fairly broad (Fig.~13), suggesting that swing amplification is not operating. Ideally we would need to run calculations without self gravity to confirm whether the stellar spiral arms are essentially just kinematic features, or whether they are indeed amplified. However we can estimate the importance of swing amplification by calculating $X=\lambda_y/\lambda_{crit}$ \citep{Toomre1981}, where $\lambda_y$ is the wavelength of the perturbation and 
\begin{equation}
\lambda_{crit}=\frac{4 \pi^2 G \Sigma}{\kappa^2}.
\end{equation} 
Swing amplification becomes effective once $X<3$. 
For our simulations, $\lambda_y=2 \pi R/m$ where $m=2$ for a two armed spiral, hence
\begin{equation}
X=\frac{\pi R \kappa^2}{4 \pi^2 G \Sigma}.
\end{equation} 
We can rewrite this as
\begin{equation} 
X=\frac{3.36 Q}{4\pi}\frac{v_{c}}{\sigma}\frac{\kappa}{\Omega}
\end{equation} 
where $v_c$ is the circular velocity and $\sigma$ the stellar velocity dispersion. Taking $Q=1$ (Fig.~2),  $\kappa/\Omega\sim1.7$ and $v_c/\sigma \sim8$, typical for our simulations, gives $X=3.6$. More generally, these quantities vary across the disc, but $X$ lies between 3 and 4 for all but the innermost regions, where $X$ is higher. 

Thus there may be some swing amplification occurring in the spiral arms of our simulations, but it is unlikely to have a large effect. This is compatible with previous simulations which found that the mass of the disc needs to be about that of the halo for swing amplification to become important \citep{Sellwood1984,Byrd1989}, as well as the results of \citet{Oh2008}. Our simulations in fact suggest that swing amplification is not vital to producing well defined spiral arms, since the gas response naturally provides narrow, dense spiral arms (as we observe), even though the stellar arms are much broader. 

\subsection{Caveats}

Although the overall dynamics of the interaction seems to be well
accounted for by our simulations, the details require several
important caveats. 

As pointed out throughout the paper, the orbit differs from that found
by \citet{Theis2003}, since we adopt a live halo as opposed to a
static potential. In future, we should ideally use an orbit which
takes into account dynamical friction. However this significantly
increases the computational time for the \textsc{MINGA} code, and was
impractical when embarking on this project. This effect is likely
exacerbated by under-resolving the halo in our simulations, to reduce
computational time. Furthermore we only modeled the companion galaxy
as a point mass, again to concentrate computational resources to
modelling the disc of the M51 galaxy. We expect that modelling the
companion galaxy may also change the orbit of the two galaxies.

We used a very simple model of the interstellar medium (ISM) in
M51. We ignore the fact the in general the ISM is likely to be
multi-phase and interactive. The warm phase can condense to become
cool, the cool phase can give rise to star formation in dense regions,
and the resultant stellar feedback can recycle cool gas back to being
warm. Moreover we ignore pressure in the ISM due to magnetic fields
and cosmic rays, and dynamical feedback from supernovae etc. We
expect stellar feedback would change the distribution of the gas
primarily on smaller scales, though supernovae may well produce holes
and further substructure in the disc. Magnetic fields will act to
reduce the amount of substructure \citep{DP2008}, but may not be
sufficient to alter the larger scale branches seen in our
simulations. We note as well that in our main model, we assume a gas
temperature of $10^4$ K. The gas in M51 is predominantly molecular,
and will therefore tend to be significantly colder. In our model with
cold gas, the disc undergoes excessive fragmentation. It may well be
that including stellar feedback, and / or magnetic fields adds to
increase the local pressure of the gas, more similar to our $10^4$ K
model, thus reducing fragmentation and local collapse.

It also needs to be borne in mind that the relative positions of the
gaseous shock and the spiral potential minimum can depend on the sound
speed in the gas. For example, if the thermal energy in the gas is
comparable to the potential energy drop in the arm, it is possible for
the shock to be upstream of the potential minimum (Roberts, 1969),
whereas if the thermal energy is much less than the depth of the
gravitational potential of the arm, the shock is to be found
downstream of the potential minimum, where the gas is decelerated as
it climbs out of the potential well \citep{DBP2006}. 
We note, however, that we find from our models for
M51 differences in assumed sound speeds do not result in large
differences in the resulting large-scale gas distributions (Figure~8).

As a first attempt at modelling M51, we have only performed relatively
low resolution simulations. Even so we have been able only to carry
out one long timescale run. It is evident that a better model of the
internal structure of M51 would be required to obtain better agreement
between the observations and the simulations. Higher resolution
simulations will be required to analyse the details of molecular cloud
and star formation in M51. Also, for the current simulations we have only shown a scenario where the companion galaxy lies on a bound orbit and eventually merges with the main galaxy. A related question is what happens for a more generic case when the orbit is not bound, and how long does the spiral pattern (in the stars and gas) subsist?

\section*{Acknowledgments}

We thank Rob Kennicutt and Sarah Kendall for useful discussions. We would also like to thank an anonymous referee for a careful reading of the manuscript. We are also grateful to Alar Toomre for sending comments on an earlier draft of this paper, and making publicly available his 1981 conference proceedings. 

CLD thanks the University of Vienna for funding a visit to the
Institute for Astronomy. CLD also acknowledges support from the Institute of Astronomy 
(Cambridge) visitors' grant, and JEP similarly acknowledges support from the University of Exeter visitors' grant. 
This work, conducted as part of the award
`The formation of stars and planets: Radiation hydrodynamical and
magnetohydrodynamical simulations' made under the European Heads of
Research Councils and European Science Foundation EURYI (European
Young Investigator) Awards scheme, was supported by funds from the
Participating Organisations of EURYI and the EC Sixth Framework
Programme.

The calculations reported here were performed using the University of
Exeter's SGI Altix ICE 8200 supercomputer. CLD acknowledges Dave
Acreman for the maintenance and support of the Exeter
supercomputer. Figures included in this paper were produced using
\textsc{SPLASH}, a visualization package for SPH that is publicly
available from http://www.astro.ex.ac.uk/people/dprice/splash/
\citep{splash2007}.

\bibliographystyle{mn2e}
\bibliography{Dobbs}
\bsp
\label{lastpage}

\end{document}